\documentclass[a4paper,11pt]{report} 

\usepackage{amssymb}
\usepackage{graphicx}
\usepackage{amsmath}
\usepackage{amsthm}
\usepackage{hyperref}
\usepackage[T1]{fontenc} 
\usepackage[ansinew]{inputenc} 
\usepackage{bbm}
\usepackage{epsfig}
\usepackage{microtype}

\newcommand{\id}{\mathbbm{1}}
\newcommand{\tr}{\textnormal{Tr}}
\newcommand{\beq}{\begin{eqnarray}}
\newcommand{\eeq}{\end{eqnarray}}
\newcommand{\bra}[1]{\ensuremath{\langle #1 |}}
\newcommand{\ket}[1]{\ensuremath{| #1 \rangle}}
\newtheorem{theorem}{Theorem}
\newtheorem{definition}{Definition}
\newtheorem{prf}{Proof}
\newcommand{\btm}{\begin{theorem}}
\newcommand{\etm}{\end{theorem}}
\newcommand{\bdf}{\begin{definition}}
\newcommand{\edf}{\end{definition}}
\newcommand{\bpf}{\begin{prf}}
\newcommand{\epf}{\end{prf}}

\begin{document}

\begin{titlepage}

\vspace*{2cm}
\begin{center}

\Huge\textsc{Detection and Characterisation\\
\small of\\
\Huge Multipartite Quantum Entanglement}\\

\vspace{2cm}

\large Andreas Gabriel\\

University of Vienna, Faculty of Physics, Boltzmanngasse 5, 1090 Vienna, Austria

\end{center}
\end{titlepage}

\
\newpage
\
\newpage

\begin{center}\Huge \textbf{Abstract} \\ \vspace{1cm}\end{center}
The research field of quantum entanglement theory is comparatively new. While a basic understanding of the most simple systems in question (i.e. bipartite systems) has been established over the past few decades, multipartite entanglement still holds many unsolved questions and intriguing riddles. In particular, it is completely unclear how several concepts from the bipartite case can be generalised in a meaningful way to multipartite scenarios.\\
In this work, the main issues of multipartite entanglement detection, characterisation and classification are discussed.  The differences and similarities between the bipartite and the multipartite situation are reviewed, various possible generalisations are presented and results are obtained in several areas.\\
The focus of this work particularly lies on a formalism - the so called HMGH framework, which has been developed and expanded for the past several years - which allows for construction of very specific separability criteria, capable of discriminating between different kinds of multipartite entanglement. By means of these criteria, the questions of partial separability, genuine multipartite entanglement and - ultimately - multipartite entanglement classification (which appear to contain the most striking differences to bipartite entanglement) are adressed and discussed.\\
In order to illustrate the theoretical conclusions in these respects, several examples are given from different (and differently closely related) fields, showing the capabilities, strengths and weaknesses of the HMGH framework as well as giving insights into the current status of research in multipartite entanglement theory as a whole.
\newpage

\begin{center}\Huge\textbf{Zusammenfassung} \end{center}
Das Forschungsgebiet der Theorie der Verschränkung von Quantensystemen ist vergleichsweise jung. Ein grundlegendes Verständnis der elementarsten solcher Systeme (i.e. Zweiteilchensysteme) wurde in den letzten Jahrzehnten erreicht, doch Verschränkung in Mehrteilchensystemen birgt nach wie vor viele offene Fragen und Mysterien. Insbesondere ist es bis dato völlig unklar, wie spezielle Konzepte aus dem Zweiteilchenfall am sinnvollsten auf die Mehrteilchensituation verallgemeinert werden können.\\
Diese Arbeit behandelt die zentralen Aspekte und Fragestellungen der Detektion, Charakterisierung und Klassifikation von Mehrteilchenverschränkung. Unterschiede und Gemeinsamkeiten zwischen Zwei- und Mehrteilchenszenarien werden erläutert, verschiedene mögliche Verallgemeinerungen präsentiert und in einigen Bereichen werden Resultate erarbeitet.\\
Der besondere Schwerpunkt dieser Arbeit liegt bei einem Formalismus - dem sogenannten HMGH-Framework, der im Lauf der letzten Jahre entwickelt und erweitert wurde - der die Konstruktion sehr spezifischer Separabilitätskriterien ermöglicht, welche fähig sind, zwischen verschiedenen Arten von Mehrteilchenverschränkung zu unterscheiden. Mit Hilfe dieser Kriterien werden die Problemstellungen der teilweisen Separabilität (partial separability), der genuinen Mehrteilchenverschränkung (genuine multipartite entanglement) und - schließlich - der Klassifikation von Mehrteilchenverschränkung (die die den fundamentalsten und kritischsten Unterschiede zur Zweiteilchenverschränkung beinhalten) behandelt und diskutiert.\\
Um die so erhaltenen Resultate zu verdeutlichen werden mehrere Beispiele aus verschiedenen (und unterschiedlich nah verwandten) Themenbereichen präsentiert, die die Fähigkeiten, Stärken und Schwächen des HMGH-Frame\-works aufzeigen und Einblicke in den gegenwärtigen Status der Forschung auf dem Gebiet der Mehrteilchenverschränkungstheorie im Gesamten geben.

\newpage

\begin{center}\Huge \textbf{List of Publications}\\
\Large in reverse-chronological order\end{center}
\vspace{1cm}
\begin{itemize}
	\item A. Gabriel and B. C. Hiesmayr\\
	\textsl{Macroscopic Observables Detecting Genuine Multipartite Entanglement in Many Body Systems}\\
  ePrint arXiv:1203.1512 (Submitted), 2012
  \item B. C. Hiesmayr, A. Di Domenico, C. Curceanu, A. Gabriel, M. Huber, J.-A. Larsson and P. Moskal\\
  \textsl{Revealing Bell's Nonlocality for Unstable Systems in High Energy Physics}\\
Eur. Phys. J. C 72, 1856 (2012)
  \item Ch. Spengler, M. Huber, A. Gabriel and B. C. Hiesmayr\\
\textsl{Examining the Dimensionality of Genuine Multipartite Entanglement}\\
Accepted for publication in Quant. Inf. Proc. (2012), ePrint arXiv:1106.5664
  \item A. Di Domenico, A. Gabriel, B. C. Hiesmayr, F. Hipp, M. Huber, G. Krizek, K. Mühlbacher, S. Radic, Ch. Spengler and L. Theussl\\
  \textsl{Heisenberg's Uncertainty Relation and Bell Inequalities in High Energy Physics}\\
Foundations of Physics 42, 6, 778-802 (2012)
  \item A. Gabriel, M. Huber, S. Radic and B. C. Hiesmayr\\
  \textsl{Computable Criterion for Partial Entanglement in Continuous Variable Quantum Systems}\\
Phys. Rev. A 83, 052318 (2011)
  \item Z.-H. Ma, Z.-H. Chen, J.-L. Chen, Ch. Spengler, A. Gabriel and M. Huber\\
  \textsl{Measure of genuine multipartite entanglement with computable lower bounds}\\
Phys. Rev. A 83, 062325 (2011)
  \item M. Huber, P. Erker, H. Schimpf, A. Gabriel and B. C. Hiesmayr\\
  \textsl{Experimentally feasible set of criteria detecting genuine multipartite entanglement in n-qubit Dicke states and in higher dimensional systems}\\
Phys. Rev. A 83, 040301(R) (2011)
  \item M. Huber, H. Schimpf, A. Gabriel, Ch. Spengler, D. Bruß and B. C. Hiesmayr\\
  \textsl{Experimentally implementable criteria revealing substructures of genuine multipartite entanglement}\\
Phys. Rev. A 83, 022328 (2011)
  \item M. Huber, N. Friis, A. Gabriel, Ch. Spengler and B. C. Hiesmayr\\
  \textsl{Lorentz invariance of entanglement classes in multipartite systems}\\
Eur. Phys. Lett. 95, 20002 (2011)
  \item A. Gabriel, B. C. Hiesmayr and M. Huber\\
  \textsl{Criterion for k-separability in mixed multipartite systems}\\
Quantum Information \& Computation 10, 9 \& 10, 829-836 (2010)
  \item M. Huber, F. Mintert, A. Gabriel and B. C. Hiesmayr\\
  \textsl{Detection of high-dimensional genuine multi-partite entanglement of mixed states}\\
Phys. Rev. Lett. 104, 210501 (2010)\\
Accepted for the Virtual Journal of Quantum Information
\end{itemize}
\newpage \
\newpage

\tableofcontents

\newpage \

\chapter{Introduction}
Ever since its theoretical discovery in 1935 \cite{schroedinger}, quantum entanglement has increasingly witnessed attention from the scientific community. After first being considered an ''odd phenomenon`` of no real physical concern, it grew to be seen as one of the central and most fundamental mysteries of quantum physics, giving rise to a whole new field of research (quantum information theory) and, during the last few decades, even to several new kinds of technology which would not have been imaginable classically.\\
Entanglement theory is a very modern and dynamical field of research which after several decades of extensive studies has brought forward at least as many new questions as answers. While bipartite entanglement is slowly beginning to be understood quite well (despite some rather counter-intuitive aspects which still remain puzzling), multipartite entanglement theory is only at the very beginning of being investigated and has already proven to be a much more complex field, holding both the possibilities for even more sophisticated new technologies as well as whole new problems and complications.\\
The main problem in multipartite entanglement theory is the ambiguity of how to generalise results of bipartite entanglement theory (as simple generalisations of such often do not appear naturally). Unlike in the latter, multipartite entanglement can exhibit various different forms, which are not only hard to distinguish from one another, but are even extremely difficult to identify and properly define in the first place.\\
A recently introduced mathematical framework allows for investigation of these questions in a novel way, as it contains the possibility of constructing criteria for arbitrary kinds of entanglement which can be used both experimentally and theoretically to classify given entangled states.\\
\\
The aim of this work is to give a compact and precise overview over multipartite entanglement theory, focussing on the contributions to this field by the author (i.e. Refs.~\cite{hmgh,ghh,lorentzclasses,huberbruss,huber_dicke,ghrh,gme-conc,spengler_gmd,gabriel_gmegap}). While specific details can be found in referenced articles, this work is rather meant to be comprehensible and illustrative than complete (as a summary of an entire field as complex as entanglement theory is far beyond the scope of a single PhD thesis). The choice of focus-topics reflects the research performed during this course of PhD study.\\
For sake of completeness, the author's work which is not directly related to multipartite entanglement is mentioned in the appendix.\\
\\
This work is organised as follows. After giving a brief introduction into the mathematical background and terminology in chapter \ref{sec_math}, an overview over the most important and fundamental facts on bipartite and multipartite entanglement will be established (chapters \ref{sec_bip} and \ref{sec_multip}, respectively). Then, the HMGH-framework will be thoroughly introduced and explained in chapter \ref{sec_hmgh}. Finally, several open problems of multipartite entanglement theory will be discussed with special emphasis on their connection to the HMGH-framework, in particular the problem of multipartite separability properties and partial separability (chapter \ref{sec_multisep}) and the question of classification of multipartite entanglement (chapter \ref{sec_multiclasses}). As an illustration of the results obtained in the previous sections, several examples and applications will be given in chapter \ref{sec_apps} before the work is concluded.\\

\chapter{Mathematical Basics, Notation and Terminology\label{sec_math}}
In order to properly discuss entanglement in multipartite systems, firstly the mathematical background has to be introduced, which forms the basis of its description. All symbols used in this chapter will retain their definitions and meanings throughout this work (unless explicitly stated otherwise).\\
\\
\section{Hilbert Spaces and States}
Quantum systems are mathematically described by Hilbert spaces $\mathcal{H}$, which in the multipartite case possess a tensor product structure, i.e. are composed of several Hilbert spaces $\mathcal{H}_i$, describing the respective subsystems, such that
\beq \mathcal{H} = \mathcal{H}_1 \otimes \mathcal{H}_2 \otimes \cdots \otimes \mathcal{H}_n \eeq
where $n$ is the number of subsystems comprising the complete considered system. Instead of enumerating the subsystems, it is also customary to label them by A for Alice, B for Bob, C for Charlie, et cetera. Often, a state is labelled in order to clarify which subsystems it describes, e.g. $\ket{\Psi^{ABC}}$ is a tripartite state on $\mathcal{H}^A\otimes\mathcal{H}^B\otimes\mathcal{H}^C$.\\
The elements of this Hilbert space are called state vectors or pure states and are denoted by ket-vectors $\ket{\Psi}$. Since however, in general, pure states do not suffice to describe realistic situations, mixed states have to be considered. These are mathematically represented by density matrices $\rho$ (also known as density operators), which are elements of the Hilbert-Schmidt space $\mathcal{H}^S$ associated with the respective Hilbert space $\mathcal{H}$. For sake of simplicity, Hilbert spaces and the (uniquely) associated Hilbert-Schmidt spaces are often denoted synonymously by $\mathcal{H}$. Density matrices are of the form
\beq \rho = \sum_i p_i \ket{\Psi_i}\bra{\Psi_i} \label{eq_mixedstate} \eeq
where the $p_i$ form a probability distribution, i.e.
\beq p_i \geq 0 \quad \mathrm{and} \quad \sum_i p_i = 1 \eeq
Density matrices $\rho$ by definition satisfy
\beq \rho^\dagger = \rho \quad \quad \quad \tr(\rho) = 1 \quad \quad \quad \rho \geq 0 \eeq
Note that pure state decompositions of the form (\ref{eq_mixedstate}) are not unique, in the sense that any mixed state has infinitely many pure state decompositions $\{p_i, \ket{\Psi_i}\}$, while a density matrix of a pure state $\rho = \ket{\Psi}\bra{\Psi}$ unambiguously corresponds to a state vector $\ket{\Psi}$ (up to a global phase, which is of no physical relevance). The maximally mixed state is uniquely given by $\rho = \frac{1}{d} \id$, where $d$ is the dimension of the respective Hilbert space.\\

\section{QuBits, QuDits and Dimensions}
In quantum information theory, mainly finite-dimensional quantum systems are of concern, such that the Hilbert spaces associated with the individual subsystems are of the form $\mathcal{H}_i = \mathbbm{C}^{d_i}$, where $d_i$ is the dimension of the Hilbert space. Consequently, $d = d_1 \times d_2 \times \cdots \times d_n$ is the dimension of the whole composite Hilbert space. Although this work is mostly concerned with mixed states, and therefore the associated Hilbert Schmidt spaces are more important in this context, the dimensionality of a system by convention always refers to the complex dimension of the Hilbert space of state vectors (unless explicitly stated otherwise).\\
In analogy to the terminology of classical information theory, a quantum system of dimension $d$ is called a quantum dit, or qudit. In particular, it is called a qubit if $d=2$ and a qutrit if $d=3$. The standard (computational) basis of a qudit-system is given by
\beq \{\ket{i}\} \quad \mathrm{with} \quad 0 \leq i \leq (d-1) \quad \mathrm{where} \quad \bra{i}j\rangle = \delta_{ij} \eeq
For composite systems, the short hand notations
\beq \ket{a \ b \ \cdots n} \equiv \ket{a}\ket{b}\cdots\ket{n} \equiv \ket{a}\otimes\ket{b}\otimes\cdots\otimes\ket{n} \eeq
is customarily used for pure states. Such a state, which can be written as a tensor product of states on each subsystem is called a product state.\\

\section{Multipartite Operations}
The inverse operation of composing Hilbert spaces via the tensor product is given by the partial trace $\tr_x(\rho)$, where $x$ represents a subspace of $\mathcal{H}$. By partially tracing over the density matrix of the complete Hilbert space, reduced density matrices are obtained, which are states of the remaining part of the quantum system, e.g.
\beq \rho^{AC} = \tr_B(\rho^{ABC}) := \sum_{i=0}^{d_B - 1} \bra{i_B}\rho^{ABC}\ket{i_B} \eeq
where the scalar product is taken on $\mathcal{H}^B$.\\
A partition $\gamma$ of the $n$-partite Hilbert space $\mathcal{H}$ is given by a number of non-empty sets $\gamma_i$ which satisfy
\beq \bigcup_i \gamma_i = \{1,2,3,\cdots,n\} \quad \mathrm{and} \quad \gamma_i\cap\gamma_j = \{\} \ \forall \ i\neq j\eeq
This corresponds to a splitting of the quantum system, in which each $\gamma_i$ (i.e. the set of all subsystems whose labels are elements of $\gamma_i$) represents one split part. Partitions are also often denoted by $\{\gamma_1|\gamma_2|\cdots|\gamma_k\}$.
In particular, a $k$-partition is a partition of the Hilbert space into exactly $k$ nontrivial parts $\gamma_i$.\\

\section{Further Terminology}
Although it should be evident from the context, sets are always referred to by symbols in brackets, so that they are clearly distinguishable from scalars (e.g. $\{a\}$ is a set, while $a$ is a number). If $\{a\}$ is a set, then $a_i$ is the $i$-th element of $\{a\}$ and $|\{a\}|$ is its cardinality, i.e. the number of elements in $\{a\}$.\\
Complex conjugation is denoted by a$^\ast$, i.e. the complex conjugate of $a$ is $a^\ast$.\\

\chapter{Bipartite Entanglement\label{sec_bip}}
\begin{figure}[htp!]\centering\includegraphics[width=7cm]{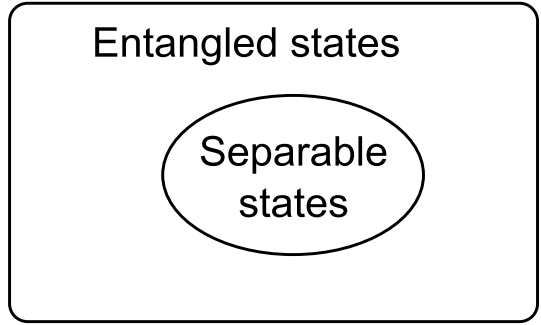}\caption[Geometry of separable and entangled states]{Illustration of the geometry of separable and entangled states. The set of separable states is convexly embedded within the set of entangled states, which extends to the border of the Hilbert space.}\label{fig_sepent}\end{figure}

Although bipartite entanglement is much less complex than multipartite entanglement, it offers a good starting point for investigation of the latter, since many basic principles and building pieces are common. It therefore seems sensible to start by defining and briefly discussing bipartite entanglement, such that these results can then form a basis on which to study multipartite entanglement.\\
\bdf A pure bipartite quantum state $\ket{\Psi} \in \mathcal{H} = \mathcal{H}_A \otimes \mathcal{H}_B$ is called separable, iff it can be written as a product of two unipartite states $\ket{\Psi_A} \in \mathcal{H}_A$ and $\ket{\Psi_B} \in \mathcal{H}_B$:
\beq \ket{\Psi} = \ket{\Psi_A}\otimes\ket{\Psi_B} \eeq
A mixed bipartite quantum state $\rho$ is called separable iff it can be decomposed into pure separable states, i.e. iff
\beq \rho = \sum_i p_i \ket{\Psi_i}\bra{\Psi_i} \eeq
where $\{p_i\}$ is a probability distribution (i.e. $p_i \geq 0$ and $\sum_i p_i = 1$) and all $\ket{\Psi_i}$ are separable (note however, that such a state may also have decompositions into entangled states).\\
Any state is called entangled iff it is not separable. \label{def_bipent}\edf
\noindent Since separability in mixed states is defined via the convex hull of separable pure states, the set of all separable states is always a convex and closed set, which is surrounded by entangled states (as illustrated in fig.~\ref{fig_sepent}).\\

\section{Detecting Bipartite Entanglement}
One of the main tasks in bipartite entanglement theory is the detection of entanglement in mixed states, which in general is a rather challenging task. To this end, various necessary separability criteria have been introduced (see e.g. Refs.~\cite{horodecki_separability, doherty, horodecki_reduction, rudolph}). Since these criteria are satisfied for all separable states, violation directly implies entanglement, while non-violation does not make any statement about presence or absence of entanglement. Due to the lack of a closed direct definition of entangled states (as opposed to the definition as not separable), no necessary criteria for entanglement could be formulated until now. Thus, the border between the sets of separable and entangled states can only be approached by these means from one side (namely from the set of entangled states inwards).\\
\\
The probably most prominent criterion for separability is the Peres-Horodecki-criterion, also known as the PPT-criterion \cite{peres_ppt}:

\btm \label{thm_ppt}  If a bipartite state $\rho$ is separable, it has to stay positive semidefinite under partial transposition (PPT), i.e.
\beq \rho^{T_A} = (T\otimes\id)(\rho) = \sum_{i,k=1}^{d_1}\sum_{j,l=1}^{d_2} \bra{i,j}\rho\ket{k,l} \ket{k,j}\bra{i,l} \geq 0 \eeq
where $T$ denotes the transposition operator. Conversely, a state which is non-positive under partial transposition (NPT) has to be entangled.\etm

\bpf For a separable state $\rho$, the partially transposed density matrix is
\beq \rho^{T_A} = && (T\otimes\id)(\rho) = (T\otimes\id)\left(\sum_i p_i \ket{\Psi_A^i}\bra{\Psi_A^i} \otimes \ket{\Psi_B^i}\bra{\Psi_B^i}\right) \nonumber \\ 
= && \sum_i p_i (\ket{\Psi_A^i}\bra{\Psi_A^i})^T \otimes \ket{\Psi_B^i}\bra{\Psi_B^i} \eeq
which is a positive semidefinite operator, since it is a convex sum of products of positive semidefinite operators. \qed \epf

\noindent While the effect of the partial transposition and thus also the partially transposed density matrix $\rho^{T_A}$ depend on the chosen basis, its eigenvalues do not. Therefore, this criterion requires no optimisation and can be computed quite simply, given a density matrix. It has also turned out to be one of the strongest and most effective separability criteria for bipartite systems so far and is therefore often used as a measure by comparison for other separability criteria.\\
\\
Another very important tool in entanglement detection is the entanglement witness theorem \cite{horodecki_separability}.
\begin{figure}[htp!]\centering\includegraphics[width=8cm]{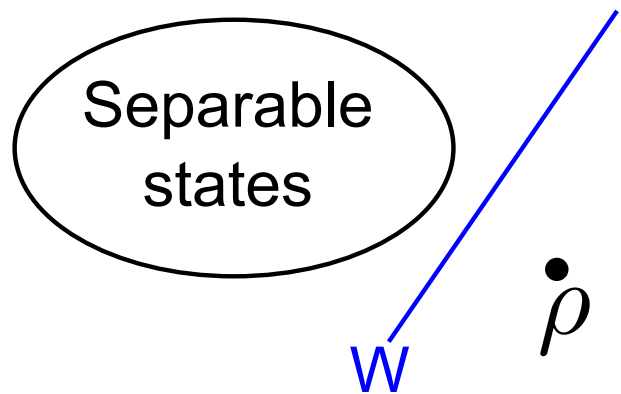}\caption[Working principle of entanglement witnesses]{Illustration of the working principle of entanglement witnesses. An entanglement witness $W$ is visualised as the hyperplane characterised by $\tr(W\omega) = 0$. All states $\omega$ on one side of this hyperplane have a positive expectation value $\tr(W\omega)>0$ and all states on the other side have a negative expectation value $\tr(W\omega) < 0$. As by definition, all separable states are located on the positive-valued-side, all states on the negative-valued-side can be identified as being entangled by means of the operator $W$.}\label{fig_witness}\end{figure}
\btm \label{thm_witness} For each entangled state $\rho$, there is an entanglement witness $W$ which detects this state, i.e. a hermitian Operator $W$ with $\tr(\rho W) < 0$ and $\tr(\sigma W) \geq 0$ for all separable states $\sigma$. \etm
\bpf The entanglement witness theorem is a direct consequence of the Hahn-Banach-theorem, which states the following: Given two disjoint convex sets, at least one of which is closed, then there exists a functional which assumes nonnegative values for all elements of the closed set and negative values for all elements of the second set. As both the set of separable states and the set containing the single (entangled) state $\rho$ are convex and closed, this implies the entanglement witness theorem. The Hahn-Banach-theorem and its proof can be found in most textbooks on functional analysis, e.g. in \cite{reedsimon}.\qed \epf

\noindent Although the entanglement witness theorem is hard to apply to a specific given problem (since it is in general very difficult to find a suitable entanglement witness for an arbitrary given state), it still is a very valuable and useful tool due to its generality. In particular, many other separability criteria can be reformulated in terms of entanglement witnesses (see e.g. \cite{horodecki_separability,zyczkowski}).\\
\\
A typical task in bipartite entanglement detection is usually of the form: Given a state $\rho_{\{\alpha_i\}}$ which depends on a number of parameters $\alpha_i$. For which values of these parameters is the state entangled, and for which is it separable?\\
Since a full cartography of the considered Hilbert space in this fashion is in most cases neither feasible nor useful (since the structure of high dimensional spaces can seldom be fully visualised or even imagined), one often resorts to investigating simplices of special states (i.e. lower dimensional subspaces which often exhibit high degrees of symmetry).\\

\section{Measuring Bipartite Entanglement}
While detection of entanglement can give a first rudimentary idea of the structure of a state space or of the properties of a certain state, it can never fully grasp the entanglement properties of an entangled state. In order to get a finer and more detailed picture of these properties, a straightforward approach is to quantify entanglement. The task in this context is not only to decide whether a state is entangled or not, but also if so, how much it is entangled. Evidently, this includes the detection of entanglement and is therefore in general a much more complex task.\\
To this end, several measures of bipartite entanglement have been introduced (see e.g. \cite{vedral_geomeasure, vidal_robustness, vidal_negativity}). Before some of the more prominent shall be presented here, observe that a proper entanglement measure should satisfy several conditions.\\
\bdf An entanglement measure $E(\rho)$ is a real-valued function $\mathcal{H} \rightarrow \mathbbm{R}$ which should ideally satisfy the following criteria \cite{bruss_character}:
\begin{itemize}
	\item[M1] $E(\rho) = 0 \Leftrightarrow \rho$ is separable.
	\item[M2] $\rho$ is maximally entangled $\Leftrightarrow E(\rho) = \max_{\omega \in \mathcal{H}} E(\omega)$
	\item[M3] $E(\rho)$ should not increase under any local operations and classical communications ($LOCC$): $E(\rho) \geq E(\Lambda^{LOCC}(\rho))$ (since $LOCC$ is often defined in slightly different ways, and it does not play a central role in this work, no precise mathematical definition of this concept shall be presented here).
	\item[M4] $E(p \rho_1 + (1-p) \rho_2) \leq p E(\rho_1) + (1-p) E(\rho_2) \ \ \ \forall \ 0\leq p \leq 1.$
\end{itemize}
\edf

\noindent There are several other conditions which may (and often are) demanded from an entanglement measure (such as additivity or continuity). However, these four will be sufficient for the discussion of entanglement quantification in this work.\\
Condition M1 guarantees that entangled states and separable states are indeed characterised as such by the measure.\\
While condition M2 sets the range of the measure, it only makes sense along with a proper definition of 'maximally entangled'. Since entanglement can be interpreted as information which exists apart from (or in between) the two parties individually, and since information about a quantum state corresponds to its purity, a maximally entangled bipartite state can be meaningfully defined as a pure state whose reduced density matrices are maximally mixed.\\
Entanglement cannot be created (or increased) by local operations and classical communication. This fact should be respected by any sensible measure of entanglement, which is stated in condition M3. Note that this implies invariance under local unitary transformations, i.e.
\beq E(\rho) = E(U_1\otimes U_2 \rho U_1^\dagger\otimes U_2^\dagger) \quad \forall \ U_i \in \mathbf{U}(d_i), \ i = 1,2 \eeq
where $\mathbf{U}(d_i)$ is the group of unitary $d_i \times d_i$ matrices.\\
Condition M4 means that $E(\rho)$ has to be a convex function. This stems from the definition of separable states via convex sums. The entanglement in mixture of two states can never be greater than the weighted averaged entanglement of these two states, while it may very well be lower (since e.g. the maximally mixed state can be decomposed into maximally entangled pure states, although it is separable itself, as it can also be decomposed into pure separable states).\\
In general it is not possible to compute an entanglement measure for an arbitrary state in a feasible way. Therefore, in order to be of actual use, an entanglement measure also should have computable and tight bounds, in addition to satisfying the above conditions.\\
\\
As two examples, consider two entanglement measures which were the first to be formulated historically: the entanglement of formation and the entanglement of distillation \cite{bennett_eof}.\\

\subsection{Entanglement of Formation}
\bdf The entanglement of formation $E_F$ of a pure bipartite state $\ket{\Psi}$ is defined as the von Neumann entropy $S(\rho)$ of either of its two reduced density matrices $\rho_A$ and $\rho_B$:
\beq E_F(\ket{\Psi}) = S(\rho_A) = S(\rho_B) \quad \mathrm{with} \quad S(\rho) = - \tr(\rho \ln \rho) \eeq
For a mixed state $\rho$, the entanglement of formation is defined via a convex roof construction, i.e. as the infimum over all pure state decompositions of $\rho$:
\beq E_F(\rho) = \inf_{\{p_i, \ket{\Psi^i}\}} \sum_i p_i E_F(\ket{\Psi^i}). \eeq
\edf

\btm The entanglement of formation of any general bipartite quantum state $\rho$ equals its entanglement cost, i.e. the number of maximally entangled states which are required to produce this state by means of the most effective conversion procedure, in the asymptotic limit of many copies of the state.\etm

\bpf See Ref.~\cite{hayden}. \epf

\noindent While the entanglement of formation of pure states is quite easy to evaluate, for mixed states it can in general not be computed, since the convex roof construction implies nontrivial optimisation. However, a remarkable method allows for its exact and analytical computation for bipartite qubit systems (see Ref.~\cite{wootters_concurrence}). Also, there exist several bounds and computation methods for special classes of states (see e.g. Refs.~\cite{terhal_eofiso, mintert_eofbound, audenaert}).\\

\subsection{Entanglement of Distillation}
As will be illustrated in more detail in section \ref{sec_bipdistill}, it is possible to convert a large number of weakly entangled states into a smaller number of more highly or even maximally entangled states. This procedure is called entanglement distillation.

\bdf The entanglement of distillation $E_D$ of a bipartite state $\rho$ is defined as the optimal conversion ratio of distilled maximally entangled states per copy of the input state $\rho$, in the asymptotical limit of many copies. \edf

\noindent Although the entanglement of distillation is defined in what appears to be a rather simple way, there is at present no way to compute it for a general state, since this would involve optimisation over all possible distillation protocols. Since at present there is no closed formulation of the latter, only bounds on this measure can be obtained. The value for any fixed distillation protocol clearly gives a lower bound, while the entanglement of formation always gives an upper bound. Only in special cases it is possible to exactly determine the entanglement of distillation, e.g. for pure states it coincides with the entanglement of formation \cite{horodecki_limits}, while e.g. for all PPT states it is zero (regardless of the state's being entangled or not, as will be discussed in more detail in section \ref{sec_bipdistill}).\\

\subsection{Properties of Bipartite Entanglement Measures}
The respective physical interpretations of the entanglement of formation and the entanglement of distillation lead to the conclusion, that any sensible bipartite entanglement measure $E(\rho)$ should satisfy
\beq E_D(\rho) \leq E(\rho) \leq E_F(\rho) \eeq
in order to be interpretable physically in a similar way \cite{horodecki_limits}, since a state can never possess more entanglement than what is needed to obtain it, nor less entanglement than can be distilled out of it.\\
\\
By considering the above two examples, it becomes apparent that a single entanglement measure can never fully characterise the entanglement properties of a bipartite state. Each of the two quantities measures entanglement in a physically meaningful way, yet they in general are not directly connected to one another. They are sensitive to different aspects of entanglement and thus capable of revealing different kinds of information, which can never be fully contained in a single quantity.\\

\section{Distillation and Distillability of Bipartite Entanglement \label{sec_bipdistill}}
As mentioned above, there are protocols to convert a large number of weakly entangled states into a smaller number of more highly entangled states. In principle, this is not surprising, since e.g. any state can always simply be projected onto a maximally entangled one with nonzero success probability. The key feature of entanglement distillation however is, that it can be implemented by means of local operations and classical communications ($LOCC$). That is, both parties can, through combined effort, achieve distillation only by manipulating their respective particles locally and coordinating these operations via classical communications. This possibility is quite nontrivial, as it is not possible to create entanglement by means of $LOCC$ (i.e. without transmitting quantum systems)\begin{figure}[htp!]\centering\includegraphics[width=10cm]{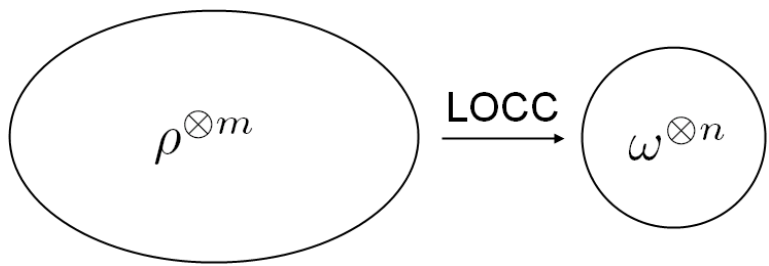}\caption[Illustration of bipartite entanglement distillation.]{Illustration of a distillation protocol. The two parties Alice and Bob can increase the entanglement present in some of their particle-pairs, by sacrificing the entanglement in the other pairs. Thus, many copies of a weakly entangled state ($\rho^{\otimes n})$ are transformed into fewer copies of a more strongly entangled state $(\omega^{\otimes m})$ (and a number of copies of separable or less entangled states). The overall entanglement does not increase during this procedure.}\label{fig_distill}\end{figure}.\\
As an example, consider a moderately simple distillation protocol, the so called BBPSSW protocol, which historically was the first such protocol to be suggested \cite{bbpssw1, bbpssw2} and is named after its authors Bennett, Brassard, Popescu, Schumacher, Smolin and Wootters. Without going into detail too much, it works as follows:
\begin{enumerate}
	\item Alice and Bob share a number $n$ of copies of a non-maximally entangled state $\rho$ ($\rho^{\otimes n}$).
	\item They transform each of the pairs into a standard form (called the Werner state \cite{werner}) by an operation called twirling (which consists of random unitary operations applied locally to both subsystems).
	\item Each party applies a certain operation - an XOR (exclusive OR) gate \cite{deutsch_computer} - to their respective parts of two pairs.
	\item Both then perform a certain measurement on one of these two particles. Depending on the outcome, the second involved pair of particles is either kept or discarded (while the measured particle pair is discarded in any event). In the former case, the entanglement of the state has been increased through the performed operations.
	\item The previous two steps can be repeated until the required or desired amount of entanglement per state is achieved.
\end{enumerate}
In the context of entanglement characterisation, the question arises whether different states behave differently in distillation. In particular, can all states be distilled? Evidently, this is not the case, since separable states can never be distilled. The much more interesting (and subtle) question therefore is: Can all entangled states be distilled? As it turns out, this is not the case; undistillable entangled - so-called bound entangled - states do exist \cite{horodecki_be}. In fact, a state $\rho$ can be distilled if and only if
\beq \bra{\Psi}\rho^{T_A}\ket{\Psi} < 0 \eeq
for some state $\ket{\Psi}$ with Schmidt rank 2 (to be defined below). As a consequence of this relation, PPT states (i.e. states not violating the Peres-Horodecki-criterion) can never be distilled and are therefore always bound entangled as soon as they are entangled. It is not entirely clear, whether the converse statement also holds, i.e. whether all NPT states are distillable. Although this is a controversially discussed question, there is much evidence pointing towards the existence of NPT bound entanglement (see e.g. \cite{pankowski, bandyopadhyay}).\\

\section{Classification of Bipartite Entanglement}
In the previous sections, several possible properties of entangled states have been discussed. These give rise to a classification scheme for bipartite entangled states. Each state can unambiguously be assigned a value for each property. A state may for example be NPT, thus entangled, having a certain entanglement of formation and entanglement of distillation. However, this classification scheme fails to grasp a central property of entangled states: the number of degrees of freedom involved in the entanglement.\\
The problem of describing this property is usually addressed by means of Schmidt numbers, which in turn are defined via Schmidt ranks \cite{terhal_schmidt}.\\
\btm For each pure bipartite state $\ket{\Psi}$ there exist local orthonormal bases $\{\ket{a_i} \in \mathcal{H}_1\}$ and $\{\ket{b_i} \in \mathcal{H}_2\}$ such that the state can be written as
\beq \ket{\Psi} = \sum_{i=1}^k c_i \ket{a_i}\otimes\ket{b_i} \eeq
for some $k \leq \min (d_1, d_2)$. The lowest possible $k$ for a given state is called this state's Schmidt rank. \etm

\bpf The theorem is known as Schmidt's theorem. The proof can be found in most linear algebra textbooks or e.g. in \cite{peres_qt}. \epf

\bdf The Schmidt number $r$ of a general bipartite state $\rho$ is defined as the maximal Schmidt rank that is at least necessary in order to construct the state, i.e. the minimal number $r$ such that there is no decomposition of $\rho$ into pure states of Schmidt ranks strictly smaller than $r$. \edf

\begin{figure}[htp!]\centering\includegraphics[width=8cm]{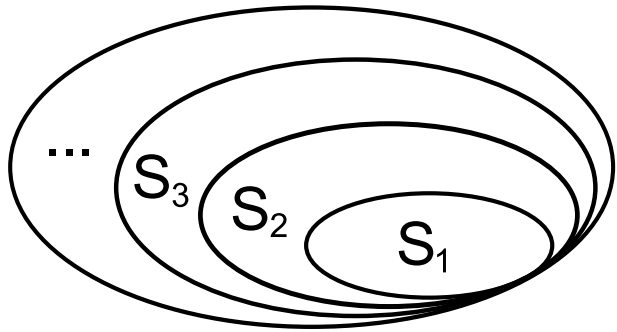}\caption[Illustration of the geometry of bipartite Schmidt numbers.]{Illustration of the geometry of the sets $S_i$ of all states with certain Schmidt numbers $i$. Each set is convexly embedded within the next without being a subset.}\label{fig_schmidt}\end{figure}

\noindent The Schmidt number is the number of degrees of freedom which are entangled. It ranges from $1$ to $\min(d_1, d_2)$, where a Schmidt number of 1 corresponds to a separable state, while a maximally entangled state necessarily has full Schmidt number (i.e. $r = \min(d_1, d_2)$). It is also conjectured that bound entangled states may always have non-maximal Schmidt number \cite{sanpera_schmidt}.\\
As a direct consequence of its definition, the Schmidt number is convex, i.e. the set of all states with Schmidt number 1 is convexly embedded within the set of all states with Schmidt number 2, et cetera. In other words, the unification $\bigcup_{j=1}^{i} S_j$ is a convex set for all $i$, while $S_i \cap S_j = \{ \} \forall i \neq j$, where $S_i$ is the set of all states with Schmidt number $i$ (as illustrated in fig. \ref{fig_schmidt}). Consequently, local operations and classical communications can only lower the Schmidt number, but never increase it.\\
\\
By combining all previously discussed classification properties, a composite characterisation scheme of bipartite entanglement can be obtained. Although this still does not allow for a complete characterisation of the entanglement present in a given state, it does give rise to a scheme of classification, which describes bipartite entanglement in a practical and useful way, and may be adapted to given situations and requirements by implementing further elements (such as different entanglement measures).\\

\chapter{Multipartite Entanglement\label{sec_multip}}
In order to investigate entanglement in general situations, one has to go beyond bipartite entanglement and rather consider multipartite scenarios (which of course contain the bipartite situation as a special case). Wherever entanglement is present - from quantum informational technologies to its appearance in nature -  multipartite systems offer more possibilities and in general a more suitable description of the respective situation.\\
In principle, multipartite entanglement can be approached by the same means as bipartite entanglement: it can be described by separability properties, entanglement measures, distillability, et cetera, each of which has more or less straightforward generalisations from the bipartite case (for which they were introduced and discussed in the previous chapter) to multipartite situations. However, these generalisations hold several subtleties and ambiguities which make multipartite entanglement a much more complex research field than bipartite entanglement.\\
In this chapter, only a brief summary of these generalisations will be given, along with a short discussion of their problematic implications. The respective aspects of multipartite entanglement are then investigated in detail in the following chapters.\\
\\
\section{Partial Separability\label{sec_partsep}}
Similarly to the study of bipartite entanglement, the first and most fundamental question concerning a multipartite state is: Is the state entangled? While in the bipartite case, the answer to this question was either ''yes`` or ''no``, the situation is significantly more complex for multipartite systems. Here, some subsystems may be entangled with each other, while others may be separable from them. Furthermore, this partial separability (or partial entanglement) can be defined in different (inequivalent) ways \cite{horodecki_qe}.\\

\bdf A pure multipartite quantum state $\ket{\Psi}$ is called $k$-separable (where $1 \leq k \leq n$), iff it factorises into $k$ states $\ket{\psi_i}$, each of which describes either one or several subsystems:
\beq \ket{\Psi} = \ket{\psi_1}\otimes\ket{\psi_2}\otimes\cdots\otimes\ket{\psi_k} \eeq
Equivalently, $\ket{\Psi}$ is called $k$-separable iff it is separable with respect to any $k$-partition of the respective Hilbert space.\\
A mixed multipartite state $\rho$ is called $k$-separable, iff it has a decomposition into $k$-separable pure states.\\
An $n$-separable $n$-partite state is called fully separable and a state which is not 2-separable (biseparable) is called genuinely multipartite entangled.\edf

\bdf A general multipartite state $\rho$ is called $\gamma_k$-separable, iff it is separable w.r.t. any $k$-partition, i.e. iff it can be written as
\beq \rho = \sum_i p_i \rho_1^i\otimes\rho_2^i\otimes\cdots\otimes\rho_k^i \eeq
where each $\rho^i_j$ is a state of one or several subsystems and $\{p_i\}$ is a probability distribution. \edf

\noindent In order to illustrate these definitions, consider the following tripartite example-states:
\beq \begin{array}{c} \ket{\Psi} = \alpha \ket{000} + \beta \ket{011} = \ket{0} \otimes (\alpha \ket{00} + \beta \ket{11}) \\
\rho_1 = a \ket{\Phi^+}\bra{\Phi^+}\otimes\ket{0}\bra{0} + b \ket{0}\bra{0}\otimes\ket{\Phi^+}\bra{\Phi^+} \\
\rho_2 = c \ket{0}\bra{0}\otimes\ket{\Phi^+}\bra{\Phi^+} + d \ket{1}\bra{1}\otimes\ket{\Phi^-}\bra{\Phi^-} \end{array} \eeq
where
\beq |\alpha|^2+|\beta|^2 = a + b = c + d = 1 \eeq
and
\beq \ket{\Phi^\pm} = \frac{1}{\sqrt{2}}(\ket{00}\pm\ket{11}) \eeq
are two entangled bipartite states.\\
While all three states $\ket{\Psi}$, $\rho_1$ and $\rho_2$ are biseparable (since $\ket{\Psi}$ can be written as a product of two state vectors and since both $\rho_i$ have decompositions into pure states with the same property), only $\ket{\Psi}$ and $\rho_2$ are also $\gamma_2$-separable, since they both are separable under a specific bipartition (namely $\{A|BC\}$).\\

\begin{figure}[htp!]\centering\includegraphics[width=8cm]{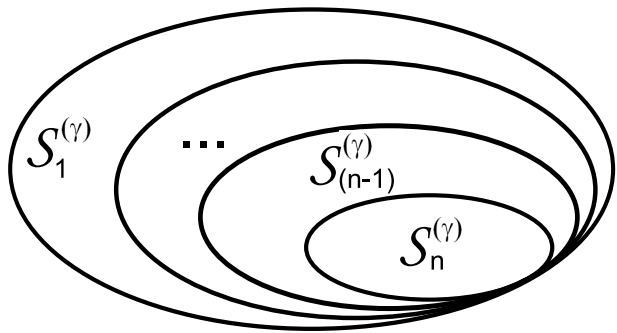}\caption[Illustration of the geometry of partial separability.]{Illustration of the geometry of the sets $\mathcal{S}_k^{(\gamma)}$ of $k$-separable ($\gamma_k$-separable) states. Each set is convexly embedded within the next. Note that in spite of the similarity to the visualisation of Schmidt numbers in fig. \ref{fig_schmidt}, there are significant differences. In particular, here each set is a subset of the next and convex.}\label{fig_ksepsets}\end{figure}

\noindent It follows directly from the definitions of $k$-separability and $\gamma_k$-separability, that they coincide for pure states and for $k = n$, while for mixed states and $k < n$, $\gamma_k$-separability is a stronger condition, i.e. a $\gamma_k$-separable state is always also $k$-separable, while the converse statement does not hold.\\
Both the set $\mathcal{S}_k$ of $k$-separable states and the set $\mathcal{S}^\gamma_k$ of $\gamma_k$-separable states form convex nested structures (as illustrated in fig. \ref{fig_ksepsets}), i.e.
\beq \mathcal{S}_n \subset \mathcal{S}_{n-1} \subset \cdots \subset \mathcal{S}_1 \quad \mathrm{and} \quad \mathcal{S}_n^\gamma \subset \mathcal{S}_{n-1}^\gamma \subset \cdots \subset \mathcal{S}_1^\gamma \eeq
As a consequence of the convex structure of the sets of $k$-separable and $\gamma_k$-separable states, a state which is $k$-separable ($\gamma_k$-separable) is always also $(k-l)$-separable ($\gamma_{(k-l)}$-separable) for all $0 \leq l \leq (k-1)$. In particular, all states are $1$-separable and $\gamma_1$-separable, therefore the definition of $k$-separability and $\gamma_k$-separability is only meaningful for $2 \leq k \leq n$.\\

\section{Multipartite Entanglement Measures \label{sec_multimeasures}}
The complexity of multipartite entanglement detection implies that also its quantification is a much more involved task than in the bipartite case. In particular, each kind of entanglement (in terms of partial separability) may be assigned an own entanglement measure, in order to be capable to specify how much of which sort of entanglement is contained in which subsystems. For example, a quantity sensitive to all kinds of entanglement is not able to distinguish between different kinds of entanglement, while it gives useful information about the overall entanglement contained in a present state. On the other hand (representing the opposite extreme situation), a measure for e.g. genuine multipartite entanglement has to be zero for most entangled states, while it can be optimally used in applications where only genuine multipartite entanglement is of concern. Also, it may be important for each individual party to quantify how much entanglement their respective subsystem shares with any one or several (or all) of the other parties. Considering that each of these kinds of entanglement may require more than one quantity for description (e.g. in analogy to the entanglement of formation and the entanglement of distillation of a bipartite state), in order to get a picture as complete as in the bipartite case, a vast number of measures would be needed.\\

\section{Multipartite Entanglement Distillation}
\begin{figure}[htp!]\centering\includegraphics[width=12cm]{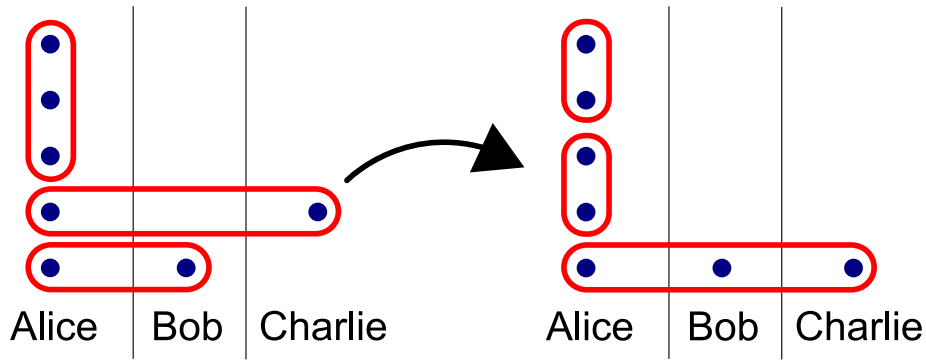}\caption[Distilling bipartite entanglement to multipartite entanglement]{Illustration of how multipartite entanglement can be distilled from bipartite entanglement. The red lines depict entanglement between the enclosed particles, represented by blue dots. After locally preparing a multipartite entangled state, its subsystems can be distributed to other parties via bipartite quantum teleportation.}\label{fig_gmedistill}\end{figure}

Entanglement distillation is one of the few features of entanglement which is not significantly more complex in the multipartite than in the bipartite case. Although due to the ambiguity of maximally entangled states there are different kinds of distillation (i.e. distillation protocols aiming at distilling different kinds of multipartite states, see e.g. \cite{multidist_ghz, multidist_w}), the principle is very similar (only the explicit distillation protocols differ). In fact, any multipartite entangled state can be distilled from mere bipartite entanglement, if the latter is shared between all involved parties (in several particle pairs). One of the parties can simply prepare the desired multipartite state locally and then use the bipartite entanglement to teleport its parts to the other parties \cite{teleport}, thus establishing the multipartite entangled state between them (as illustrated in fig.~\ref{fig_gmedistill}).\\

\section{Equivalence Classes of Multipartite Entanglement}
As a consequence of the different (equally sensible) kinds of multipartite entanglement measures, even the term ''maximally entangled``, which was a basic building piece for bipartite entanglement measures, becomes ambiguous in this case for two reasons. Firstly, because pure states with maximally mixed reduced density matrices are no longer equivalent to one another, and secondly, because there even are kinds of entanglement, which may be maximal for states which do not have maximally mixed reduced density matrices.\\
In several ways, the maximally entangled state is considered to be the Greenberger-Horne-Zeilinger ($GHZ$)-state, which for $n$-qudit-systems is defined as \cite{ghz}
\beq \label{eq_ghzstate} \ket{GHZ} = \frac{1}{\sqrt{d}} \sum_{i=0}^{d-1} \ket{i}^{\otimes n}\eeq
This state possesses the maximal amount of genuine multipartite entanglement, without containing any other form of entanglement (since all its reduced multipartite density matrices are separable, while it is a pure state with maximally mixed unipartite reduced density matrices). This property makes $GHZ$-states the preferred resource for several applications, such as e.g. quantum secret sharing protocols (as will be discussed in section \ref{sec_app_qss}).\\ 
Another well-known genuinely multipartite state is the $W$-state of $n$ qubits \cite{duer_w}
\beq \label{eq_wstate} \ket{W} = \frac{1}{\sqrt{n}} \sum_{i=1}^n \ket{w_i} \quad \mathrm{where} \quad \ket{w_i} = \ket{0}^{\otimes (i-1)} \otimes \ket{1} \otimes \ket{0}^{\otimes (n-i)} \eeq
Besides containing a certain amount of genuine multipartite entanglement, this state contains maximal bipartite entanglement (distributed evenly among all parties) \cite{fortescue}. The $W$-state belongs to the family of Dicke states \cite{dicke}, which are genuinely multipartite entangled states of $n$ qubits with a parameter $m$ (a natural number between $1$ and $n-1$):
\beq \ket{D_m^n} = \left(\begin{array}{c} n\\m\end{array}\right)^{-\frac{1}{2}} \sum_{|\{\alpha\}|=m} \ket{d_{\{\alpha\}}} \label{eq_dickestate} \eeq
where the sum runs over all $\{\alpha\} \subset \{1, 2, \cdots, n\}$ which are sets of $m$ different integers between $1$ and $n$, and $\ket{d_{\{\alpha\}}}$ is the product state with $\ket{1}$ in all subsystems whose numbers are contained in $\{\alpha\}$ and $\ket{0}$ else. Since the sum runs over all sets $\{\alpha\}$ with cardinality $m$, this is the equally weighted superposition of all $n$-qubit product states with $m$ $\ket{1}$s and $(n-m)$ $\ket{0}$s. For $m=1$, the Dicke state coincides with the $W$-state. For example, for $n=4$ and $m=2$, the state reads
\beq \ket{D_2^4} = \frac{1}{\sqrt{6}}\left(\ket{0011}+\ket{0101}+\ket{0110}+\ket{1001}+\ket{1010}+\ket{1100}\right) \eeq
\\
Dicke states and variations thereof - e.g. phased Dicke states (i.e. Dicke states with nonzero relative phases between some of the superposed product states) - appear in crystals and spin-chains \cite{krammerbruss} and can be used for various quantum informational tasks, such as e.g. quantum secret sharing or open-destination teleportation \cite{prevedel}. Furthermore, Dicke states are a rich resource for all kinds of quantum informational applications, since their reduced density matrices can exhibit different types of entanglement.\\
\\
Because of the reasons discussed above, entanglement classification is much more complex a task in multipartite systems than in bipartite ones, which is why until now no even nearly complete classification scheme for general multipartite quantum systems could be developed (results have only been obtained for specific low-dimensional systems, see e.g. \cite{duer_3qubits, verstraete_4qubits}). There are several frequently used approaches towards this problem, which will be discussed in chapter \ref{sec_multiclasses}.\\

\chapter{The HMGH-Framework\label{sec_hmgh}}
Quite recently, a framework for constructing various kinds of separability criteria for multipartite systems was introduced \cite{hmgh}. This framework (which shall be referred to as the HMGH-framework -- after the authors of its first introductory publication \cite{hmgh}, Huber, Mintert, Gabriel and Hiesmayr) allows for construction of very general and versatile separability criteria, based on convex inequalities of density matrix elements. Its advantages over other separability criteria are numerous. Apart from being the first systematic approach for characterising multipartite entanglement and its different aspects in general $n$ qudit systems, the criteria obtained can comparatively easily be implemented experimentally. Also, it should be emphasised that the framework is not only applicable to separability problems, but also to more specific tasks, such as multipartite entanglement classification or multipartite entanglement quantification.\\
In order to properly present the HMGH-framework with all its features and capabilities, the formalism upon which it is based has to be introduced.\\

\section{Definitions, Terminology and Formulation}
The first mathematical concept which is of grave importance to the HMGH-framework is the notion of convexity. Convexity plays a very important role throughout entanglement theory, since most relevant sets of states are convex sets or complements thereof, e.g. the set of $k$- or $\gamma_k$-separable states, the set of PPT states, et cetera. Furthermore, mixed states are convex combinations of pure states, therefore problems of mixed states can be reduced to comparatively simple problems of pure states by means of convex functions.\\
\bdf A function $f(\rho)$ is called convex iff
\beq p f(\rho_1) + (1-p) f(\rho_2) \geq f(p\rho_1+(1-p)\rho_2) \quad \forall \ \rho_i, \ \forall \ 0 \leq p \leq 1 \eeq \edf
\btm If a convex inequality of the form $f(\rho) \leq 0$ is satisfied for all pure states $\rho$, then it is also satisfied for all mixed states $\rho$. \etm
\bpf The theorem follows immediately from the definition of convexity since any mixed state $\rho$ is a convex combination of pure states $\ket{\psi_i}\bra{\psi_i}$:
\beq f\left(\sum_i p_i \ket{\psi_i}\bra{\psi_i}\right) \leq \sum_i p_i f\left(\ket{\psi_i}\bra{\psi_i}\right) \leq \sum_i 0 = 0 \eeq \qed \epf

\noindent The above theorem obviously also holds if the considered mixed and pure states do not constitute the whole Hilbert space, but only a part of it (as long as the considered mixed states always have decompositions into the respective pure states). For example, if such an inequality is satisfied for all $k$- or $\gamma_k$-separable pure states, it is also satisfied for all $k$- or $\gamma_k$-separable mixed states, respectively, since the latter are defined as having decompositions into the former.\\
\btm \label{thm_conv} All functions that can be written as sums of terms of the following forms are convex in $\rho$:
\beq c_1 |\bra{\phi_1}\rho\ket{\phi_2}| \label{eq_conv_1} \eeq
\beq - c_2 \sqrt[l]{\prod_{i=1}^l \bra{\phi_i}\rho\ket{\phi_i}}  \label{eq_conv_2} \eeq
where the $c_j$ are arbitrary positive real numbers and $l$ is an arbitrary positive integer. \etm
\bpf First, observe that any sum of convex functions is convex itself. Thus, it is sufficient to prove that the expressions (\ref{eq_conv_1}) and (\ref{eq_conv_2}) are convex individually.\\
For the first expression, this follows from the triangle inequality (as any transition element $\bra{\phi_1}\rho\ket{\phi_2}$ is a complex number $z$):
\beq p |z_1| + (1-p) |z_2| = |p z_1| + |(1-p) z_1| \geq |p z_1 + (1-p) z_2| . \eeq
For the second expression, abbreviate $\bra{\phi_i}\rho\ket{\phi_i} = r_i$ (and observe that these are nonnegative real numbers). Now, the convexity of the expression (\ref{eq_conv_2}) is equivalent to
\beq \sum_{i=1}^2 \prod_{j=1}^l a_i^j \leq \prod_{j=1}^l \sqrt[l]{\sum_{i=1}^2 (a_i^j)^l} \eeq
with $a_1^j = \sqrt[l]{p r_{j,1}}$ and $a_2^j = \sqrt[l]{(1-p) r_{j,2}}$. By defining
\beq A_i^j = \frac{a_i^j}{\sqrt[l]{\sum_{k=1}^2 (a_k^j)^l}} \eeq
it follows that
\beq \sum_{i=1}^2 \prod_{j=1}^l A_i^j = \sum_{i=1}^2 \prod_{j=1}^l \sqrt[l]{(A_i^j)^l} \leq \sum_{i=1}^2 \frac{\sum_{j=1}^l (A_i^j)^l}{l} = 1 \eeq
where the inequality follows from the fact that the geometric mean is always lower than or equal to the arithmetic mean for nonnegative numbers. Inserting the above definition for the $A_i^j$, one arrives at
\beq  \frac{\sum_{i=1}^2 \prod_{j=1}^l a_i^j}{\prod_{j=1}^l \sqrt[l]{\sum_{i=1}^2 (a_i^j)^l}} \leq 1 \quad \Leftrightarrow \quad \sum_{i=1}^2 \prod_{j=1}^l a_i^j \leq \prod_{j=1}^l \sqrt[l]{\sum_{i=1}^2 (a_i^j)^l} \eeq \qed \epf

\noindent Most criteria constructed from the HMGH-framework (in particular the most basic ones) are formulated via certain permutation operators acting on the two-fold copy Hilbert space of states. In order to understand the working principle of the framework, one needs to thoroughly define these permutation operators.
\bdf \label{def_permute} The permutation operator $\mathcal{P}_i$ acting on an element $\ket{\Phi} = \ket{\phi_1}\otimes\ket{\phi_2}$ of the two-fold copy Hilbert space $\mathcal{H}^{\otimes 2}$, where $\ket{\phi_i} \in \mathcal{H}$, is defined via its action as
\beq \label{eq_permute} \mathcal{P}_i \ket{\Phi} && = \mathcal{P}_i \ket{a_1, a_2, \cdots, a_n}\otimes\ket{b_1, b_2, \cdots, b_n} \\  \nonumber
&& = \ket{a_1, a_2, \cdots, a_{i-1}, b_i, a_{i+1}, \cdots, a_n}\otimes\ket{b_1, b_2, \cdots, b_{i-1}, a_i, b_{i+1}, \cdots, b_n} \eeq
where $n$ is the number of subsystems of $\mathcal{H}$ and the $a_j$ and $b_j$ are the contributions of the $j$-th subsystem from $\ket{\phi_1}$ and $\ket{\phi_2}$, respectively. That is, the permutation operator $\mathcal{P}_i$ swaps the $i$-th subsystems of the two single-copy Hilbert spaces.\\
The permutation operator $\mathcal{P}_{\{\alpha\}}$, where $\{\alpha\}$ is a set of integers between $1$ and $n$, is defined as
\beq \mathcal{P}_{\{\alpha\}} = \prod_{j \in \{\alpha\}} \mathcal{P}_j \eeq
Note that all $\mathcal{P}_j$ commute, therefore the order of the above product is irrelevant. Thus, $\mathcal{P}_{\{\alpha\}}$ swaps several subsystems at once between the two copies of the Hilbert space, namely all subspaces with labels $i \in \{\alpha\}$.\\
The permutation operator $\underline{\mathcal{P}}$ is defined as the operator permuting all $n$ subsystems of both copies of the Hilbert space:
\beq \underline{\mathcal{P}} = \prod_{i=1}^n \mathcal{P}_i \eeq
i.e.
\beq \underline{\mathcal{P}}\ket{a_1, a_2, \cdots, a_n}\otimes\ket{b_1,b_2,\cdots,b_n} = \ket{b_1,b_2,\cdots,b_n}\otimes\ket{a_1, a_2, \cdots, a_n}. \eeq \edf

Observe that all $\mathcal{P}_i$ (and products thereof) are hermitian and unitary, i.e.
\beq \mathcal{P}_i^\dagger = \mathcal{P}_i^{-1} = \mathcal{P}_i \eeq

\section{Working Principle}
\begin{figure}[htp!]\centering\includegraphics[width=12cm]{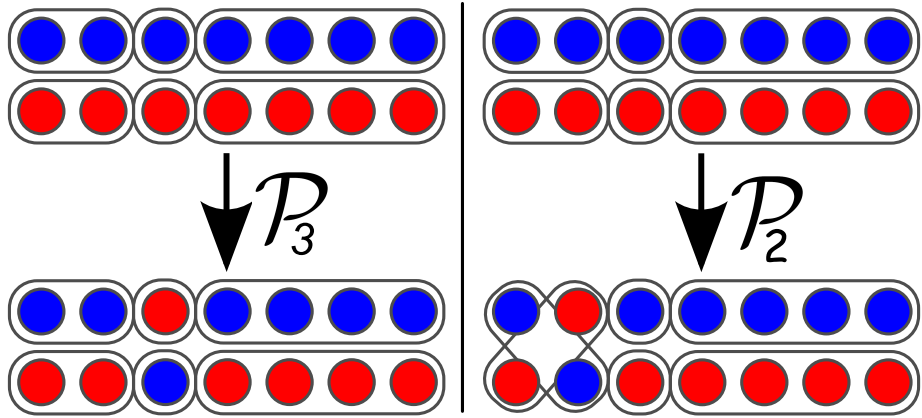}\caption[Illustration of the action of the permutation operators $\mathcal{P}_i$]{Illustration of the effect of the permutation operators $\mathcal{P}_i$ (or $\mathcal{P}_{\{\alpha\}}$) on two copies of a pure state (in this example, a seven-partite triseparable state). If all permuted subsystems (in the above examples, the one permuted subsystem, i.e. system 3 and 2, respectively) are separable from the rest of the system (as in the left hand picture), the two copies of the state remain unchanged and, in particular, separable from each other. Contrary to this, as illustrated in the right hand picture, if the permutation operator does not match the partition under which the given state is separable, the resulting state does not factorise into two copies any more. Using this, the permutation operators $\mathcal{P}_i$(or, more generally, the permutation operators $\mathcal{P}_{\{\alpha\}}$) can be used to detect and characterise different forms of multipartite entanglement.}\label{fig_permute}\end{figure}

The separability criteria constructed within the HMGH-framework via the permutation operators $\mathcal{P}_i$ share a common working principle. First of all, they are formulated via convex inequalities. Thus, it is sufficient to prove their validity for pure states and validity for mixed states is guaranteed. Now, if a pure state $\ket{\Psi}$ is separable with respect to any certain $k$-partition $\gamma_k$, there are $k$ different permutation operators $\mathcal{P}_{\{\alpha\}}$ which leave two copies of the state invariant (see left part of fig.~\ref{fig_permute})

\beq \mathcal{P}_{\{\alpha\}} \ket{\Psi} = \ket{\Psi} \quad\forall\ \{\alpha\}\in\gamma_k \eeq
In particular, these permutation operators conserve the product structure between the two copies of the state, while permutation operators $\mathcal{P}_{\{\alpha\}}$ with $\{\alpha\} \notin \gamma_k$ leave the two copies of the state entangled with each other (as illustrated in the right part of fig.~\ref{fig_permute}). In this way, problems of multipartite partial separability are effectively reduced to comparatively simple problems of bipartite separability between the two copies of the state.\\
\\
Some inequalities within the HMGH-framework do not rely on permutation operators like $\mathcal{P}_i$ (see section \ref{sec_class_excl}). These inequalities however implicitly contain the action of these operators. Since they do not possess the freedom of basis choice other inequalities (such as the ones discussed in Chapter \ref{sec_multisep}) possess, they can be written in a more final and explicit form. However, the working principle remains the same.\\

\section{A Simple Example\label{sec_hmgh_example}}
The most simple (and also historically first \cite{hmgh}) separability criterion which can be constructed via the HMGH-framework is a bipartite separability criterion. Although it is not particularly useful by itself (since there are several stronger bipartite separability criteria, such as e.g. the PPT criterion, theorem \ref{thm_ppt}), it serves as an illustrative example for how the framework can be used. Also, this criterion can be generalised to multipartite partial separability (as will be done in Chapter~\ref{sec_multisep}) in a very useful way.\\
\btm \label{thm_bipcrit} If a bipartite state $\rho$ is separable, it has to satisfy the inequality
\beq \label{eq_bipcrit} \sqrt{\bra{\Phi}\rho^{\otimes 2}\underline{\mathcal{P}}\ket{\Phi}} - \sqrt{\bra{\Phi}\mathcal{P}_1^\dagger\rho\mathcal{P}_1\ket{\Phi}} \leq 0 \eeq
for all fully separable states $\ket{\Phi}$ on the two-copy Hilbert space. \etm

\bpf Since the inequality is a convex function of $\rho$ (which follows from thm. \ref{thm_conv}), it is sufficient to prove it for pure states and its validity for mixed states is guaranteed.\\
If $\rho$ is separable, then $\mathcal{P}_1^\dagger \rho^{\otimes 2} \mathcal{P}_1 = \rho^{\otimes 2}$. Using $\ket{\Phi}=\ket{\phi_1}\otimes\ket{\phi_2}$, the inequality therefore assumes the form
\beq |\bra{\phi_1}\rho\ket{\phi_2}| - \sqrt{\bra{\phi_1}\rho\ket{\phi_1}\bra{\phi_2}\rho\ket{\phi_2}} \leq 0 \eeq
which follows from the positivite-semidefiniteness of $\rho$. \qed \epf

\noindent In other words, whenever the above inequality is violated by a state $\rho$ and a fully separable state $\ket{\Phi}$, $\rho$ is identified as being entangled, while non-violation of the inequality does not imply any statement about the separability properties of $\rho$. In particular, the inequality may be violated for some $\ket{\Phi}$, and satisfied for another (for the same $\rho$), which points out an important issue: How can one chose $\ket{\Phi}$ such that the detection power of the criterion is maximal?\\

\section{Optimisation \label{sec_bipopt}}
There are several (widely independent) conditions which an optimal $\ket{\Phi}$ should satisfy (for any given state $\rho$ under investigation) \cite{ghrh}:
\begin{itemize}
	\item[C1] $\ket{\Phi}$ should be fully separable, i.e. $\ket{\Phi} = \ket{\phi_1^A}\otimes\ket{\phi_1^B}\otimes\ket{\phi_2^A}\otimes\ket{\phi_2^B}$. 
	\item[C2] The two parts $\ket{\phi_1}$ and $\ket{\phi_2}$ of $\ket{\Phi}$ on the two copies of the Hilbert space should be orthogonal in each subsystem: $\left\langle \phi_1^X | \phi_2^X \right\rangle = 0$, where $X \in \{A, B\}$.
	\item[C3] $\ket{\phi_1}$ and $\ket{\phi_2}$ should be chosen such that $|\bra{\phi_1}\rho\ket{\phi_2}|$ is maximal.
\end{itemize}
Condition C1 can be split up into two weaker conditions which should be satisfied for different reasons. Firstly, $\ket{\Phi}$ should be separable with respect to the two-copy-partition (i.e. $\ket{\Phi} = \ket{\phi_1}\otimes\ket{\phi_2}$), since this is necessary for the proof to hold and thus for the inequality to make sense. Therefore, this may be seen as a technical requirement. Secondly, each of these $\ket{\phi_i}$ should itself be separable. Although this is not necessary for the criterion to be well-defined or even to detect entanglement, the state $\ket{\Phi}$ for which the criterion is strongest (i.e. for which the violation of the inequality is greatest) will always be fully separable. This can be seen by computing the criterion for an entangled state 
\beq \ket{\Phi} = \left(\cos\alpha_1 \ket{\phi_1^1} + \sin\alpha_1 \ket{\phi_1^2}\right)\otimes \left(\cos\alpha_2 \ket{\phi_2^1} + \sin\alpha_2 \ket{\phi_2^2}\right) \eeq
where all $\ket{\phi_i^j}$ are product states such that the $\ket{\phi_i} = \cos\alpha_i\ket{\phi_i^1} + \sin\alpha_i\ket{\phi_i^2}$ are entangled. Now, the first term of the inequality (\ref{eq_bipcrit}) of the criterion (theorem \ref{thm_bipcrit}) reads
\beq \left| \cos\alpha_1\cos\alpha_2 \bra{\phi_1^1}\rho\ket{\phi_2^1} + \cos\alpha_1\sin\alpha_2 \bra{\phi_1^1}\rho\ket{\phi_2^2} + \right. \nonumber\\  \left.
\sin\alpha_1\cos\alpha_2 \bra{\phi_1^2}\rho\ket{\phi_2^1} + 
\sin\alpha_1\sin\alpha_2 \bra{\phi_1^2}\rho\ket{\phi_2^2} \right| \eeq
This leads to a weakening of the detection quality of the criterion for several reasons. Not only is a superposition of this form very likely to have a comparatively small absolute value (and thus not to satisfy condition C3, which will be discussed below) due to interference between the individual terms. Also some of the terms might (depending on the $\ket{\phi_i^j}$) be expectation values (and not transition elements). Entanglement however is based on coherence and is thus indicated by transition elements (off-diagonal density matrix elements) rather than expectation values (density matrix diagonal elements). Due to the construction of the separability criterion, such contributions are automatically (at least) cancelled by similar contributions in the second term of the inequality. Furthermore, entangled $\ket{\phi_i}$ lead to contributions of the off-diagonal density matrix elements in the second term of the inequality, effectively reducing its violation for entangled states.\\
Condition C2 is a very central technical requirement. In order for the criterion to work at all, the states $\ket{\phi_1^X}$ and $\ket{\phi_2^X}$ need to be different, since otherwise the permutation operator $\mathcal{P}_X$ would leave the state $\ket{\Phi}$ invariant, leaving the inequality trivially satisfied for all states $\rho$. Since for any choice of $\ket{\phi_1^X}$ any $\ket{\phi_2^X}$ can be decomposed into the parallel and perpendicular contributions 
\beq \ket{\phi_2^{X \parallel}} = \ket{\phi_1^X}\bra{\phi_1^X}\phi_2^X\rangle \quad \mathrm{and} \quad \ket{\phi_2^{X \perp}} = (\id - \ket{\phi_1^X}\bra{\phi_1^X})\ket{\phi_2^X} \eeq
and only the latter are capable of yielding a violation of the criterion, it is evident that the optimal choice has to be such that $\bra{\phi_1^X}\phi_2^X\rangle = 0$ for $X = A, B$.\\
Condition C3 represents the optimisation of $\ket{\Phi}$ in dependence of the given state $\rho$. It is quite clear that the inequality can only be violated (i.e. yield a value greater than zero) if its positive term is as large as possible (while still satisfying all other conditions).\\

\section{Experimental Implementation\label{sec_experiment}}
A very advantageous feature of all separability criteria constructed within the HMGH-framework is their experimental implementability. All these criteria can be written as functions of density matrix elements, each of which can be expressed as a linear combination of local observables (e.g. Pauli operators for qubits). Furthermore, the number of density matrix elements needed for any of the criteria is much smaller than the total number of elements the density matrix has, in particular, very few (often just one) off-diagonal elements are needed. As a consequence, the number of observables required in order to experimentally implement one of the criteria is much smaller than the number of observables for a full quantum state tomography (in some cases, the former does not even grow exponentially with the system size, while the latter always does).\\
As an example, consider inequality (\ref{eq_bipcrit}) (theorem \ref{thm_bipcrit}) for an arbitrary two-qudit-state $\rho$ with $\ket{\Phi} = \ket{0011}$. To this end, consider the short hand notation
\beq [kl] := \mathrm{Tr}(\rho \ \sigma_k\otimes\sigma_l) \eeq
where the $\sigma_j$ are the Pauli operators on the two-dimensional spaces spanned by $\ket{\phi_i^A}$ and $\ket{\phi_i^B}$, respectively. For the above choice of $\ket{\Phi}$ they read
\beq \begin{array}{c} \sigma_0 = \ket{0}\bra{0}+\ket{1}\bra{1} \\ \sigma_1 = \ket{0}\bra{1} + \ket{1}\bra{0} \\ \sigma_2 = i \ket{0}\bra{1} - i \ket{1}\bra{0} \\ \sigma_3 = \ket{0}\bra{0} - \ket{1}\bra{1} \end{array} \eeq
The inequality now reads
\beq |\bra{00}\rho\ket{11}| - \sqrt{\bra{01}\rho\ket{01}\bra{10}\rho\ket{10}} = \frac{1}{4}|[11]-[22]+i([12]+[21])| - \nonumber \\  \frac{1}{4}\sqrt{([00]-[30]+[03]-[33])([00]+[30]-[03]-[33])} \leq 0 \eeq
By simple counting, one can see that it can be implemented by means of only eight local observables $\sigma_k\otimes\sigma_l$, as opposed to $(d^4-1)$ observables (i.e. for example fifteen for two qubits, eighty for two qutrits, et cetera) necessary for a full quantum state tomography (which is required for implementing most other separability criteria, such as e.g. the PPT criterion).\\

\section{Alternative Formulation}
In order to put the HMGH-framework into a more general context, the permutation operators $\mathcal{P}_i$ and their action on separable states can be formulated in different ways. In contrast to the rather mathematical formulation given in the previous sections, they will be formulated in terms more common in physics in this section.\\
To this end, observe that the permutation operator $\mathcal{P}_i$, as defined in definition \ref{def_permute} (eq. (\ref{eq_permute})) can, for fixed $\ket{\Phi} = \ket{\phi_1}\otimes\ket{\phi_2}$, also be written in terms of creation and annihilation operators as
\beq \mathcal{P}_i = a^{(i)\dagger}_{\phi_2^i} a^{(i)}_{\phi_1^i} b^{(i)\dagger}_{\phi_1^i} b^{(i)}_{\phi_2^i} \eeq
where the $a^{(i)}_\phi$ and $b^{(i)}_\phi$ are annihilation operators, annihilating the mode $\phi$ in the first and second copy of the $i-th$ subsystem, respectively, and the $a^{(i)\dagger}_\phi$ and $b^{(i)\dagger}_\phi$ are the corresponding creation operators.\\
Also, the condition for separability of a state $\rho$ under a $k$-partition $\gamma$ 
\beq \mathcal{P}_{\{\gamma_i\}}^\dagger \rho \mathcal{P}_{\{\gamma_i\}} = \rho \quad \forall \ i \eeq
which is a basic building piece of the HMGH-framework, can be rewritten in a way more common in physics, namely via commutators:
\beq [\mathcal{P}_{\gamma_i}, \rho] = 0 \quad \forall \ i \eeq
such that separability can be formulated as a symmetry property of the state $\rho$.\\

\chapter{Multipartite Separability Properties \label{sec_multisep}}
Although the question of full separability for multipartite states is very well studied (see e.g. \cite{horodecki_multifs, wocjan, yu, hassan}), the problem of partial separability has only quite recently started to be investigated thoroughly (e.g. \cite{seevinck_partsep}). The different kinds of multipartite partial separability ($k$- and $\gamma_k$-separability, as defined in section \ref{sec_partsep}) require different kinds of tools for investigation. While the task of detecting $\gamma_k$-(in)separability (i.e. separability under specific partitions) does in principle not require special tools, but can be dealt with by means of tools for bipartite entanglement, this is not the case for $k$-(in)separability. Although the two definitions coincide for the pure state case, significant differences arise when mixed states are considered.\\

\section{\texorpdfstring{$\gamma_k$}--Separability}
The question whether any given $n$-partite mixed state $\rho$ is $\gamma_k$-separable with respect to any given $k$-partition $\gamma$ can be addressed by means of common bipartite separability criteria. To this end, all subsystems of the $i$-th part of the partition $\gamma$ are considered as one single subsystem $A_i$, while all other subsystems together form a single subsystem $B_i$ (where $i = 1, 2, ..., k$). Now, it is easy to see that $\rho$ is $\gamma_k$-separable with respect to the partition $\gamma$ if and only if $A_i$ is separable from $B_i$ for all $i$.\\
Despite this somewhat simple possibility of characterisation, $\gamma_k$-separability still holds some remarkable and quite counterintuitive features. In particular, a mixed quantum state $\rho$ may be $\gamma_k$-separable with respect to different $k$-partitions $\gamma$, but not be $\gamma_{(k+1)}$-separable. In this sense, $\gamma_k$-separability is not unique.\\
For example, consider the so-called Smolin-state of four qubits \cite{smolinstate}
\beq \label{eq_smolinstate} \nonumber \rho = && \frac{1}{4} ( \ket{GHZ}\bra{GHZ} + \ket{GHZ_{(1,2)}}\bra{GHZ_{(1,2)}} \\ && + \ket{GHZ_{(1,3)}}\bra{GHZ_{(1,3)}} + \ket{GHZ_{(1,4)}}\bra{GHZ_{(1,4)}} ) \eeq
where $\ket{GHZ} = (\ket{0000}+\ket{1111})/\sqrt{2}$ is the usual four-qubit GHZ state, and the $\ket{GHZ_{(i,j)}}$ are the same state after application of a bit flip in the $i$-th and $j$-th subsystem (i.e. the same state in a different basis, e.g $\ket{GHZ_{(1,2)}}=\ket{GHZ_{(3,4)}}=(\ket{1100}+\ket{0011})/\sqrt{2}$). This state can also be decomposed as
\beq \label{eq_smolinstatebip} \nonumber \rho = && \frac{1}{4}( \ket{\Phi^+}\bra{\Phi^+}\otimes\ket{\Phi^+}\bra{\Phi^+} + \ket{\Phi^-}\bra{\Phi^-}\otimes\ket{\Phi^-}\bra{\Phi^-} \\ && + \ket{\Psi^+}\bra{\Psi^+}\otimes\ket{\Psi^+}\bra{\Psi^+} + \ket{\Psi^-}\bra{\Psi^-}\otimes\ket{\Psi^-}\bra{\Psi^-} ) \eeq
where
\beq \ket{\Psi^\pm} = \frac{1}{\sqrt{2}}\left( \ket{01} \pm \ket{10} \right) \quad \ket{\Phi^\pm} = \frac{1}{\sqrt{2}}\left( \ket{00} \pm \ket{11} \right) \eeq
are the two-qubit Bell states. Since $\rho$ is invariant under exchange of its subsystems (due to the symmetry of the decomposition (\ref{eq_smolinstate})), the decomposition (\ref{eq_smolinstatebip}) is possible with respect to any two-qubit-versus-two-qubit partition.\\
As can be seen from these different possible pure state decompositions, the Smolin state is $\gamma_2$-separable under all bipartitions into two sets of two subsystems each (i.e. the bipartitions $\{1,2|3,4\}$, $\{1,3|2,4\}$ and $\{1,4|2,3\}$). From this, one may naively conclude, that the state is actually fully separable, since each subsystem is separable from every other subsystem in some partition. This, however, is not true. In fact, the Smolin state is not separable under any one-versus-three-subsystem bipartition (as can be easily proven by means of bipartite separability criteria, e.g. the PPT-criterion).\\
This example illustrates the ambiguity of $\gamma_k$-separability. Although the maximal $k$ for which a given state is $\gamma_k$-separable is an absolute value, there need not be one unique partition with respect to which this state is $\gamma_k$-separable.\\

\section{\texorpdfstring{$k$}--Separability and Genuine Multipartite Entanglement}
\begin{figure}[htp!]\centering\includegraphics[width=9cm]{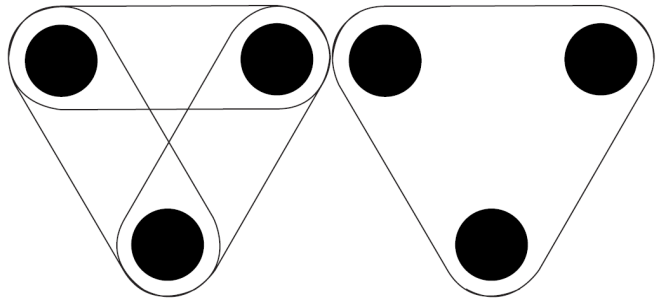}\caption[Conceptual problems of detecting $k$-separability]{Illustration of the problems in detecting $k$-separability. Each sphere represents a subsystem, while the entanglement between the subsystems is depicted by frames around the respective spheres. In this tripartite example, the left state is biseparable (since there is no genuine multipartite entanglement of all three subsystems present, e.g. a mixture of three states containing bipartite entanglement between different subsystems), while the state on the right hand side is genuinely multipartite entangled (e.g. a GHZ-state). However, neither of the two states is $\gamma_2$-separable, since both are fully entangled.}\label{fig_1sep2sep}\end{figure}

$\gamma_k$-separability is a stronger criterion than $k$-separability, in the sense that every $\gamma_k$-separable state is also always $k$-separable, but not vice-versa. Thus, $k$-(in)separability can also be detected by bipartite separability criteria in the way described in the previous section. However, most $k$-separable states are not $\gamma_k$-separable and can therefore not be identified as such by this method (as illustrated in fig.~\ref{fig_1sep2sep}). Such $k$-separable but $\gamma_k$-inseparable states arise from the fact that the $k$-separable pure states composing a $k$-separable mixed state can be separable with respect to different $k$-partitions, i.e. the resulting state is in general not separable under any particular partition.\\
The problem of deciding whether a general given state is $k$-separable or not has recently been studied increasingly intensively (mainly for $k=2$), and different tools and approaches have been developed (see e.g. \cite{guehne_gme1, guehne_gme2}). One of the most successful approaches, and the only systematic and fully analytic one so far, lies within the HMGH-framework (stronger results have only been obtained for very specific types of states \cite{graphgme1, graphgme2} or by semi-definite programming \cite{guehne_sdp}, i.e. not fully analytically).\\

\subsection{Genuine Multipartite Entanglement}
Since the case of $k=2$, i.e. the detection of genuine multipartite entanglement in a given state, is much more important to applications (in the context of quantum information technology) than other questions of partial separability, most of the research in this field is concentrated on this particular problem.\\
The conceptually most simple and straightforward approach utilises so-called fidelity witnesses. These are a special kind of entanglement witnesses of the form
\beq \label{eq_fidelitywitness} W = \max_{\sigma\in\mathcal{S}_2}\tr(\ket{\Psi}\bra{\Psi}\sigma)\id - \ket{\Psi}\bra{\Psi} = \alpha\id - \ket{\Psi}\bra{\Psi}\eeq
where $\alpha$ is a positive real number and $\ket{\Psi}$ is a state exhibiting the desired property (e.g. genuine multipartite entanglement). By optimising over all states $\sigma$ of a certain kind (e.g. biseparable states), this operator by construction can only have nonnegative expectation values for these states, thus, any state with a negative expectation value necessarily cannot be of this kind (i.e. has to be genuinely multipartite entangled). For example, witnesses for genuine multipartite entanglement near the three-qubit GHZ- and W-state are given by \cite{acin_3qubits}
\beq W_{GHZ} = \frac{3}{4}\id - \ket{GHZ}\bra{GHZ} \quad \mathrm{and} \quad W_W = \frac{2}{3}\id - \ket{W}\bra{W} \label{eq_fidelitywitnesses} \eeq
respectively.\\
Such witnesses are a good starting point for investigations, as their application does not require an extensive beforehand knowledge of the investigated system and the results often give important insights. However, fidelity witnesses are not practical for advanced studies of entanglement, because they involve (mostly numerical) optimisation procedures, only work in a small region of the considered state space (i.e. in the vicinity of the state $\ket{\Psi}$ they are constructed from) and thus have a rather low detection efficiency.\\
Some of the strongest and most versatile criteria for genuine multipartite entanglement can be derived within the HMGH-framework, the most simple of which is a straight-forward multipartite generalisation of the bipartite separability criterion (\ref{eq_bipcrit}) \cite{hmgh}, a less general version of which was also developed independently in Ref.~\cite{guehne_gme2}:
\btm The inequality
\beq  \sqrt{\bra{\Phi}\rho^{\otimes2}\underline{\mathcal{P}}\ket{\Phi}} - \sum_{\{\gamma\}}\sqrt{\bra{\Phi}\mathcal{P}_{\{\gamma_1\}}^\dagger\rho^{\otimes2}\mathcal{P}_{\{\gamma_1\}}\ket{\Phi}} \leq 0 \label{eq_gme_ghz} \eeq
is satisfied for all biseparable states $\rho$ and for all fully separable states $\ket{\Phi}$, where the sum runs over all bipartitions $\{\gamma\}$. \etm
\bpf In analogy to the proof of thm. \ref{thm_bipcrit}, firstly observe that the left hand side of the inequality is a convex function. Therefore, it suffices to prove the inequality for arbitrary pure states $\rho = \ket{\Psi}\bra{\Psi}$ and its validity for mixed states is guaranteed. Now, any biseparable pure state $\ket{\Psi}$ is separable under a specific bipartition $\tilde{\{\gamma\}}$, for which
\beq \mathcal{P}_{\{\tilde{\gamma}_1\}}^\dagger\rho^{\otimes2}\mathcal{P}_{\{\tilde{\gamma}_1}\} = \rho^{\otimes2} \eeq
The inequality now reads
\beq \sqrt{\bra{\Phi}\rho^{\otimes2}\underline{\mathcal{P}}\ket{\Phi}} - \sqrt{\bra{\Phi}\rho^{\otimes2}\ket{\Phi}} -  \sum_{\{\gamma\}\neq\{\tilde{\gamma}\}}\sqrt{\bra{\Phi}\mathcal{P}_{\{\gamma_1\}}^\dagger\rho^{\otimes2}\mathcal{P}_{\{\gamma_1\}}\ket{\Phi}} \leq 0 \eeq
It follows from the positivity of $\rho$ that the first two terms combined are non-positive, as they can be rewritten as
\beq \sqrt{\bra{\Phi}\rho^{\otimes2}\underline{\mathcal{P}}\ket{\Phi}} - \sqrt{\bra{\Phi}\rho^{\otimes2}\ket{\Phi}} = |\bra{\phi_1}\rho\ket{\phi_2}| - \sqrt{\bra{\phi_1}\rho\ket{\phi_1}\bra{\phi_2}\rho\ket{\phi_2}} \eeq
using that by definition $\ket{\Phi} = \ket{\phi_1}\otimes\ket{\phi_2}$. Since the remaining sum is also non-positive, this proves the theorem.\qed\epf

\noindent As this inequality is a straightforward generalisation of inequality (\ref{eq_bipcrit}) and therefore is based on the same working principle, the optimal choice of $\ket{\Phi}$ is given by the conditions discussed in section \ref{sec_bipopt}.\\
\begin{figure}[htp!]\centering\includegraphics[width=6cm]{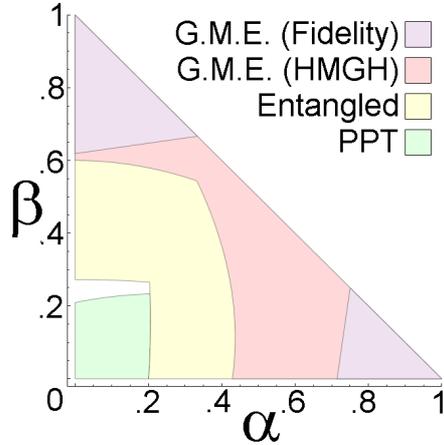}\caption[Detection quality of fidelity witnesses and ineqs. (\ref{eq_gme_ghz}), (\ref{eq_bipcrit})]{Illustration of the genuine multipartite entangled states of the form (\ref{eq_ghzw}) detected by inequality (\ref{eq_gme_ghz}) in comparison to those detected by the fidelity witnesses for GHZ- and W-states (\ref{eq_fidelitywitnesses}), the entanglement detected by the bipartite separability criterion inequality (\ref{eq_bipcrit}) and the PPT criterion. For both inequalities, the two choices $\ket{\Phi} = \ket{000111}$ and $\ket{\Phi} = \ket{+++---}$ (where $\ket{\pm} = \frac{1}{\sqrt{2}}(\ket{0}\pm\ket{1})$) are used (in accordance with the conditions discussed in section \ref{sec_bipopt} for the cases $\beta = 0$ and $\alpha = 0$, respectively). The purple area is detected to be genuinely multipartite entangled by the fidelity witnesses, the red and purple areas are detected to be genuinely multipartite entangled, since they violate inequality (\ref{eq_gme_ghz}), the yellow, red and purple areas are detected to be entangled (not to be fully separable) by violation of inequality (\ref{eq_bipcrit}). The green area is the set of PPT states, i.e. all states outside are necessarily entangled.\newline The comparison to the fidelity witnesses demonstrates that the detection power of inequality (\ref{eq_gme_ghz}) is indeed quite satisfactory. However, the gap between the PPT-area and the area detected by the bipartite separability criterion (\ref{eq_bipcrit}) indicates that this criterion (and, by generalisation, also the criterion for genuine multipartite entanglement which is based upon it) is not capable of optimally detecting W-type entanglement (while $GHZ$-type entanglement is indeed detected optimally in this sense).}\label{fig_gmeineq}\end{figure}

\noindent Figure \ref{fig_gmeineq} illustrates the detection quality of the criterion (\ref{eq_gme_ghz}) in comparison to the detection range of the fidelity witnesses for GHZ- and W-states, as well as to the bipartite separability criterion inequality (\ref{eq_bipcrit}) and the PPT criterion for three-qubit states of the form
\beq \label{eq_ghzw} \rho = \alpha \ket{GHZ}\bra{GHZ} + \beta \ket{W}\bra{W} + \frac{1-\alpha-\beta}{8}\id \eeq
where $\ket{GHZ}$ and $\ket{W}$ are given by (\ref{eq_ghzstate}) and (\ref{eq_wstate}), respectively. This illustration suggests, that while inequality (\ref{eq_gme_ghz}) optimally detects GHZ-like genuine multipartite entanglement, it is not optimally suited to detect W-type entanglement. Mathematically, this is due to the number of significant off-diagonal density matrix elements in the respective states. Since the criterion only utilises a single off-diagonal element, it cannot be optimal for a state whose entanglement is described by several such elements. Thus, different criteria are necessary in order to optimally detect different types of entanglement using the HMGH-framework.\\
Such criteria can be constructed conceptually quite simply by using more complex generalisations and extensions of the bipartite separability criterion (\ref{eq_bipcrit}):
\begin{enumerate}
	\item Add up the absolute values of all characteristic off-diagonal density matrix elements of the state under investigation.
	\item Subtract square roots of products of two density matrix diagonal elements each, such that the whole expression is strictly lower than or equal to zero for biseparable states. 
	\item To do so, use the bipartite separability criterion (\ref{eq_bipcrit}) for estimations. The bound has to be proven for pure states only, since by construction, the constructed criterion is formulated as a convex function.
	\item In order to formulate the criterion more elegantly, all density matrix elements can optionally be expressed via permutation operators on two-fold copies of states.	
\end{enumerate}
This concept can in principle be applied to all different kinds of multipartite entangled states. Consider for example the $n$-qubit $m$-Dicke state (\ref{eq_dickestate}). The above method yields the following criterion \cite{huber_dicke}
\btm The inequality
\beq
\sum_{\{\sigma\}} \left( |\bra{d_{\{\alpha\}}}\rho\ket{d_{\{\beta\}}}| - \sqrt{\bra{d_{\{\alpha\}}}\otimes\bra{d_{\{\beta\}}}\mathcal{P}_{\{\alpha\}}^\dagger\rho^{\otimes2}\mathcal{P}_{\{\alpha\}}\ket{d_{\{\alpha\}}}\otimes\ket{d_{\{\beta\}}}}\right) \nonumber \\ - m(n-m-1) \sum_{\{\alpha\}} \bra{d_{\{\alpha\}}}\rho\ket{d_{\{\alpha\}}} \leq 0 \label{eq_gme_dicke} \eeq
is satisfied for all biseparable states, where the sum runs over all sets $\{\sigma\} = \{\{\alpha\},\{\beta\}\}$ which satisfy $|\{\alpha\}| = |\{\beta\}| = m-1$ and $|\{\alpha\}\cap\{\beta\}| = (m-1)$, and where
\beq \ket{d_{\{\alpha\}}} = \bigotimes_{i\in\{\alpha\}} \ket{1}_i \otimes \bigotimes_{j\notin\{\alpha\}} \ket{0}_j \eeq
By construction, for each $n$ and $m$ the maximal value $m$ of the inequality is attained for the corresponding Dicke state $\ket{D_m^n}$ (for $m \leq n/2$).\etm
\bpf Since the left hand side of the inequality is a convex function of $\rho$ (as a consequence of thm. \ref{thm_conv}), it is sufficient to prove the inequality's validity for pure states, and validity for mixed states follows immediately. Since each biseparable pure state is separable under a specific bipartition, assume that the state is separable under the partition $\{A|B\}$. Now, denoting the first term in the inequality as $O_{\{\alpha\},\{\beta\}}$, the second term as $P_{\{\alpha\},\{\beta\}}$ and the third term as $D_{\{\alpha\}}$, the following relations hold:
\beq O_{\{\alpha\},\{\beta\}} \leq \left\{ \begin{array}{cc} 
P_{\{\alpha\},\{\beta\}} & \mathrm{if} \ x \in A, y\in B \ \mathrm{or} \ x \in B, y \in A \\
\frac{1}{2}(D_{\{\alpha\}} + D_{\{\beta\}}) & \mathrm{if} \ x,y \in A \ \mathrm{or} \ x,y \in B \end{array}\right. \eeq
where $x = \{\alpha\}\backslash\{\beta\}$ and $y = \{\beta\}\backslash\{\alpha\}$. The first statement follows from thm. \ref{thm_bipcrit} (since $\mathcal{P}_{\{\alpha\}}$ acts as $\mathcal{P}_A$ in this case), while the second statement follows from the positivity of the density matrix $\rho$. Counting the maximal number of necessary $D_{\{\alpha\}}$ elements for this estimation yields the coefficient $m(n-m-1)$ of the last term of the inequality.\qed\epf

\begin{figure}[htp!]\centering\includegraphics[width=8cm]{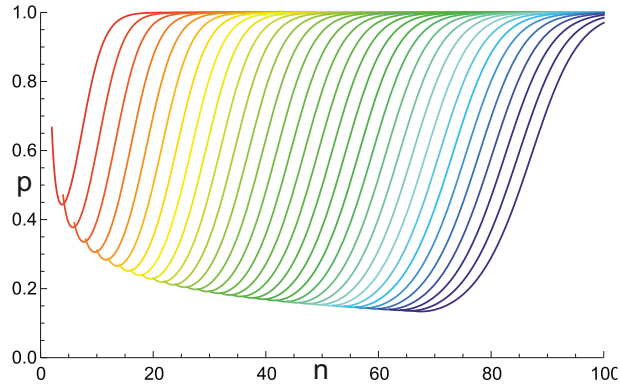}\caption[Detecting genuine multipartite entanglement in Dicke states]{Illustration of the noise resistance of the criterion eq. (\ref{eq_gme_dicke}) for different $n$-qubit $m$-Dicke states mixed with white noise (i.e. states of the form $\rho = p \ket{D^n_m}\bra{D^n_m} + \frac{1-p}{2^n}\id$). The lines represent the detection threshold $p$ for different $m$, ranging from $m=1$ (red line) to $m=33$ (purple line).}\label{fig_dicke}\end{figure}

\noindent Such criteria, tailored specifically for special kinds of states by means of the HMGH-framework, are comparatively noise-resistant (as illustrated in fig.~\ref{fig_dicke}) and easily implemented experimentally (as discussed in section \ref{sec_experiment}), which makes them a very versatile tool for efficiently detecting genuine multipartite entanglement.\\

\subsection{General \texorpdfstring{$k$}--Separability}
Although the concept of general $k$-separability (i.e. for arbitrary $k$) and its detection in given states is of comparatively low importance in quantum informational applications, it is still crucial for establishing a full understanding of multipartite entanglement as a whole. So far, the HMGH-framework offers the only analytic solution to this problem, as the convex inequalities detecting genuine multipartite entanglement (i.e. 2-inseparability) can be generalised to detect $k$-inseparability for arbitrary $k$. Explicitly, the corresponding criteria have only been derived for the most simple case, i.e. as a generalisation of eq. (\ref{eq_gme_ghz}) \cite{ghh}:
\btm The inequality
\beq \sqrt{\bra{\Phi}\rho^{\otimes2}\underline{\mathcal{P}}\ket{\Phi}} - \sum_{\{\gamma\}}\prod_{i=1}^k\left(\bra{\Phi}\mathcal{P}_{\{\gamma_i\}}^\dagger\rho^{\otimes2}\mathcal{P}_{\{\gamma_i\}}\ket{\Phi}\right)^{\frac{1}{2k}} \leq 0 \label{eq_ksep} \eeq
is satisfied for all $k$-separable states $\rho$ and for all fully separable states $\ket{\Phi}$, where the sum runs over all $k$-partitions $\{\gamma\}$. \etm
\bpf In analogy to the proof of thm. \ref{thm_bipcrit}, firstly observe that the left hand side of the inequality is a convex function. Therefore, it suffices to prove the inequality for arbitrary pure states $\rho = \ket{\Psi}\bra{\Psi}$ and its validity for mixed states is guaranteed. Now, any $k$-separable pure state $\ket{\Psi}$ is separable under a specific $k$-partition $\tilde{\{\gamma\}}$, for which
\beq \mathcal{P}_{\tilde{\{\gamma\}}_i}^\dagger\rho^{\otimes2}\mathcal{P}_{\tilde{\{\gamma\}}_i} = \rho^{\otimes2} \eeq
The inequality now reads
\beq \sqrt{\bra{\Phi}\rho^{\otimes2}\underline{\mathcal{P}}\ket{\Phi}} - \sqrt{\bra{\Phi}\rho^{\otimes2}\ket{\Phi}} -  \sum_{\{\gamma\}\neq\tilde{\{\gamma\}}}\prod_{i=1}^k\left(\bra{\Phi}\mathcal{P}_{\gamma_i}^\dagger\rho^{\otimes2}\mathcal{P}_{\gamma_i}\ket{\Phi}\right)^{\frac{1}{2k}} \leq 0 \eeq
It follows from the positivity of $\rho$ that the first two terms combined are non-positive. Since the remaining sum is also non-positive, this proves the theorem.\qed\epf

\begin{figure}[htp!]\centering\includegraphics[width=10cm]{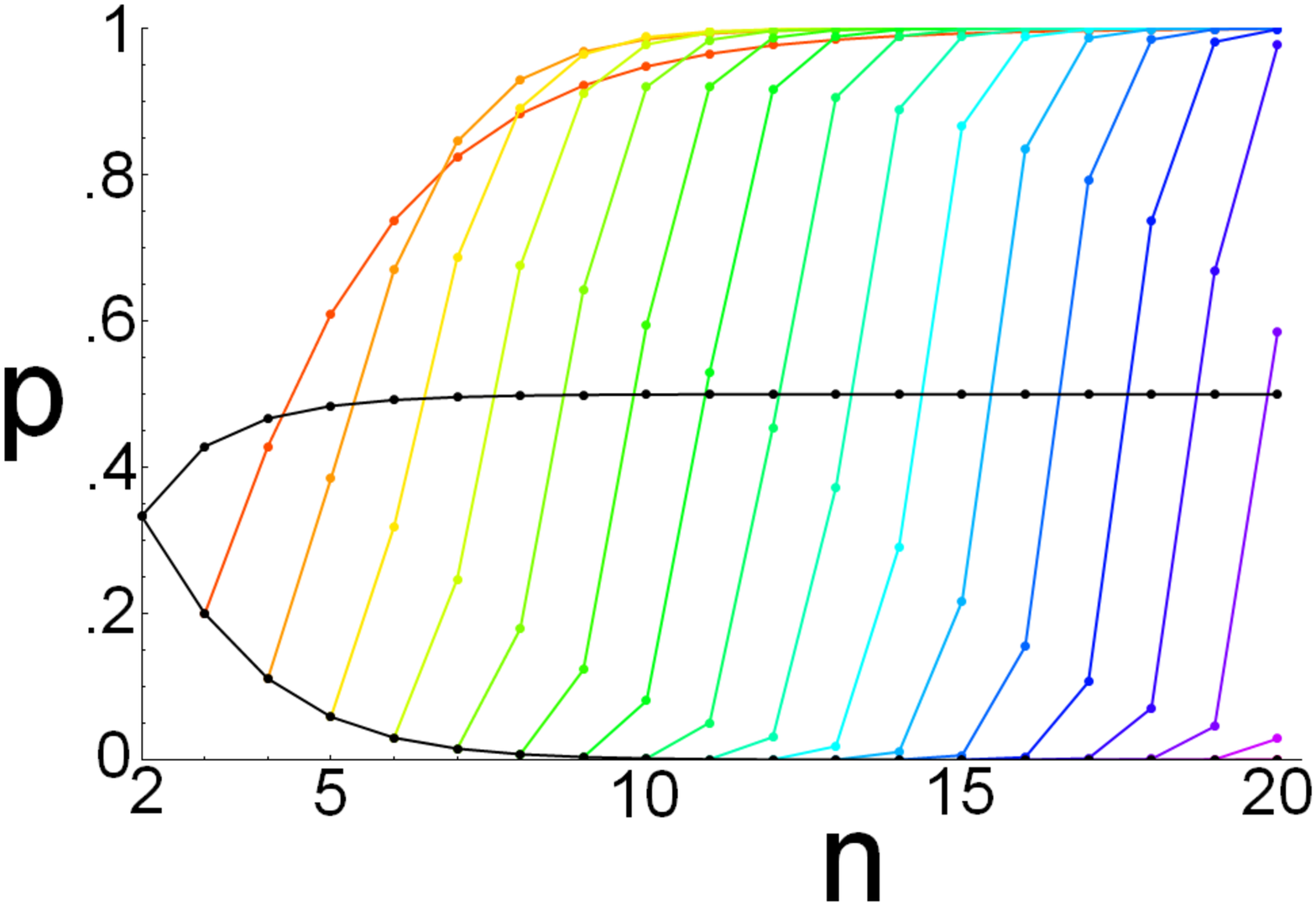}\caption[Problems of detecting $k$-inseparability with criterion (\ref{eq_ksep})]{Illustration of the problems in detecting $k$-inseparability using criterion (\ref{eq_ksep}) for the $n$ qubit state $\rho = p\ket{GHZ}\bra{GHZ} + \frac{1-p}{2^n}\id$. The lines correspond to the detection threshold for different $k$, ranging from $k=3$ (red line) to $k=19$ (purple line). The black lines correspond to $k=2$ (upper line) and $k=n$ (bottom line).\newline
While the detection quality for $k=2$ and $k=n$ is satisfactory (in fact, for $k = n$, the threshold is equal to the one yielded by the PPT criterion), this is not the case for other $k$. In particular, the order in which the state is detected to be $k$-inseparable for different $k$ becomes unsorted for $n \geq 5$, which indicates that the detection threshold in these cases is clearly far from optimal.}\label{fig_ksepproblem}\end{figure}

\noindent Although the mere possibility to analytically detect $k$-inseparability is a considerable advance in multipartite entanglement theory, this criterion (and, along with it, all other similar generalisations of separability criteria constructed within the HMGH-framework) has a serious weak point. For high numbers $n$ of parties and for $2 << k << n$, the detection power of the criterion is very low. In fact, it is not uncommon for this kind of criterion to detect a state to be $(k-1)$-inseparable before detecting it to be $k$-inseparable (although $(k-1)$-separability is a weaker condition than $k$-separability), as illustrated in fig.~\ref{fig_ksepproblem}. This problem is caused by the vast number of terms which are subtracted from a single positive term. The number of such terms is the number of inequivalent $k$-partitions of the $n$-partite system, which is given by the Sterling-number in the second kind \cite{huber_ichikawa}
\beq S(n,k) = \sum_{i=1}^k \frac{(-1)^{k-i}i^{n-1}}{(i-1)! (k-i)!} \label{eq_sterling} \eeq
which grows exponentially in $n$ and over-exponentially in $k$ (e.g. $S(4,2) = 7$, $S(10,3) = 9330$ and $S(20,8) \approx 1.5 \times 10^{13}$).\\
As a solution for this problem has not been found so far, the problem of detecting $k$-(in)separability remains essentially unsolved for general $k$ and $n$.\\

\subsection{Measuring Genuine Multipartite Entanglement}
As discussed in section \ref{sec_multimeasures}, quantifying multipartite entanglement is a highly ambiguous task, since for a complete characterisation of the entanglement of any given multipartite state, different kinds of multipartite entanglement need to be measured. However, in specific applications only specific kinds of entanglement are relevant, thus it suffices to restrict the description of states to these types of entanglement in these cases.\\
Like in the bipartite case, it is quite easy to define sensible entanglement measures for pure states. Via convex roof constructions, these can even be extended to mixed states in a straightforward fashion. This however leads to effectively incomputable expressions. In order to be useful in practice, such measures thus require computable (and tight) bounds. A good example for such a measure is the gme-concurrence \cite{gme-conc}:
\bdf The gme-concurrence is defined as
\beq \label{eq_cgme} C_{gme}(\ket{\Psi}) = \min_{\{\gamma\}} \sqrt{2(1-\tr(\rho_{\gamma_1}^2))} \eeq
where the sum runs over all bipartitions $\{\gamma\}$ and $\rho_{\gamma_1}$ is the reduced density matrix of the first part of $\{\gamma\}$. For mixed states, the gme-concurrence is defined via the convex roof construction
\beq C_{gme}(\rho) = \inf_{\{p_i, \ket{\Psi_i}\}} \sum_i p_i C_{gme}(\ket{\Psi_i}) \eeq
where the infimum is taken over all pure state decompositions of $\rho$. \edf

\btm The gme-concurrence is a measure of genuine multipartite entanglement, i.e. it is nonzero iff the considered state is genuinely multipartite entangled. \etm
\bpf If a state $\rho$ is genuinely multipartite entangled, then each of its pure state decompositions contains a genuinely multipartite entangled pure state $\ket{\Psi}$ with $C_{gme}(\ket{\Psi}) > 0$ (since reduced density matrices of fully entangled pure states are always mixed). Therefore, $C_{gme}(\rho) > 0$ in this case.\\
Conversely, if $\rho$ is biseparable, then there is a decomposition into biseparable pure states $\ket{\Psi_i}$ with $C_{gme}(\ket{\Psi_i}) = 0$ (since the respective reduced density matrices are pure). Since $C_{gme}$ is nonnegative, such a decomposition will always be an infimum of the convex roof construction and therefore ensure $C_{gme}(\rho) = 0$. \qed\epf

\btm Computable lower bounds on the gme-concurrence are given by
\beq C_{gme}(\rho) \geq 2\left(\sqrt{\bra{\Phi}\rho^{\otimes2}\underline{\mathcal{P}}\ket{\Phi}} - \sum_{\{\gamma\}} \sqrt{\bra{\Phi}\mathcal{P}_{\{\gamma_1\}}^\dagger\rho^{\otimes2}\mathcal{P}_{\{\gamma_1\}}\ket{\Phi}}\right) \eeq
for arbitrary fully separable $\ket{\Phi}$, where the sum runs over all bipartitions $\{\gamma\}$. Note that the bound is similar to the expression in ineq. (\ref{eq_gme_ghz}).\etm

\bpf Since the bound is a convex function of $\rho$, it is sufficient to prove the inequality for pure states. This can be done for arbitrary (but fixed) dimension $d$ and number $n$ of subsystems, where the structure of the proof is always the same. Here, for sake of simplicity and comprehensibility, the proof shall be presented for the three-qubit case ($n=3$ and $d=2$).\\
The most general pure three-qubit state can be written as
\beq \ket{\Psi}&& = a\ket{000} + b\ket{001} + c\ket{010} + d\ket{011} \nonumber\\ &&+ e\ket{100} + f\ket{101} + g\ket{110} + h\ket{111} \eeq
For this state, the squared concurrences $C^2(\rho_\gamma) = 2(1-\tr(\rho_{\gamma_1}^2))$ (with respect to the three possible bipartitions $\{\gamma\}$) read
\beq C^2(\rho_{A|BC}) = 4|ah-de|^2 + F_1 \nonumber \\ C^2(\rho_{B|AC}) = 4|ah-cf|^2 + F_2 \\ \nonumber C^2(\rho_{C|AB}) = 4|ah-bg|^2 + F_3 \eeq
where the $F_i$ are nonnegative functions. It thus follows that
\beq C(\rho_{A|BC}) \geq 2(|ah|-|de|) \nonumber \\ C(\rho_{B|AC}) \geq 2(|ah|-|cf|) \\ \nonumber C(\rho_{C|AB}) \geq 2(|ah|-|bg|) \eeq
And therefore
\beq C_{gme}(\rho) = \max_{\{\gamma\}}(C(\rho_\gamma)) \geq 2(|ah|-|de|-|cf|-|bg|) \eeq
For the choice $\ket{\Phi} = \ket{000111}$, this expression coincides with the postulated bound, as both yield
\beq |\bra{000}\rho\ket{111}|&&-\sqrt{\bra{001}\rho\ket{001}\bra{110}\rho\ket{110}}-\sqrt{\bra{010}\rho\ket{010}\bra{101}\rho\ket{101}}\nonumber\\&&-\sqrt{\bra{100}\rho\ket{100}\bra{011}\rho\ket{011}} \eeq
Since $C_{gme}(\rho)$ is invariant under local unitary transformations, it follows that the bounds must hold irrespective of such transformations as well, i.e. for any fully separable $\ket{\Phi}$.\qed\epf

\noindent Since all criteria for partial separability built within the HMGH-framework utilise the same expression (implicitly or explicitly) that has been proven to be related to the GME-concurrence, it stands to reason that all these criteria give rise to different bounds on this genuine multipartite entanglement measure (or similar ones). Using the criterion ineq. (\ref{eq_ksep}), this approach could even be generalised to measuring $k$-inseparability for arbitrary $k$. All this however, has not yet been thoroughly investigated.\\
Measures for other kinds of multipartite entanglement are even less known and understood. In most cases, the best known approach is using fidelity witnesses as entanglement measures for specific kinds of states. Such measures can be defined as
\beq E_X(\rho) = -\max_{U}\ \tr(U\rho U^\dagger W_X) \eeq
where $X$ represents the kind of state in question, $W_X$ the corresponding fidelity witness given by (\ref{eq_fidelitywitness}) and the maximum is taken over all local unitary operations $U = U_1\otimes U_2\otimes\cdots\otimes U_n$. The main disadvantage of this method is (as in the mere entanglement detection problem) the lack of detection power (even if the optimisation should be performable in a feasible way). Since tools for multipartite entanglement detection and characterisation (which form the foundation for multipartite entanglement quantification) are being developed quite intensively at present, progress in this direction is to be expected in the near future.\\

\chapter{Classes of Multipartite Entanglement\label{sec_multiclasses}}
Classification of multipartite entanglement is one of the most challenging tasks in entanglement theory. While the situation for bipartite states is quite simple (as the issue can be resolved by the concept of Schmidt numbers), it is widely unclear how this solution can be generalised most sensibly to multipartite systems and whether such a concept is capable of giving rise to a complete characterisation scheme for arbitrary multipartite systems.\\

\section{Conditions for Classification Schemes\label{sec_classes}}
Although it is unclear, how a sensible classification scheme for multipartite entanglement can be constructed, there are several conditions such a scheme should satisfy. These mainly originate from two ideas:
\begin{itemize}
	\item[C1] The resulting classes should not be too coarse, in the sense that states which exhibit (qualitatively) different entanglement properties should not belong to the same class of states.
	\item[C2] The classes should also not be too fine, in the sense that equivalent states (i.e. states which exhibit qualitatively similar entanglement properties, particularly states which can be converted into one another via local operations and classical communications) should always belong to the same class.
\end{itemize}
In particular (as a consequence of C2), the classification of states should be Lorentz invariant (i.e. a state should be assigned to the same entanglement class from all inertial frames of reference). Although at present there is no closed consistent relativistic description of quantum information theory, an approach based on the Wigner rotation of state vectors (for an overview, see e.g. Ref.~\cite{friis_da}) is very successful in forming a foundation for a future development of a complete relativistic quantum information theory. In order to be Lorentz invariant (in the above sense), an entanglement classification scheme has to satisfy two conditions \cite{lorentzclasses}: 
\begin{itemize}
	\item[(1)] all classes have to be invariant under local-unitary transformations, and
	\item[(2)] all classes have to be convex (i.e. the mixture of two states belonging to some class should belong to the same class).
\end{itemize}
While the necessity of (1) also follows directly from statement C2, statement (2) is much less obvious (although it seems a reasonable requirement) and leads to several complications. For example, it implies that different classes must not be disjoint, but have to overlap. For example, certain mixed separable states (particularly, the maximally mixed state) have to belong to $all$ classes of entanglement, since they can be composed of all kinds of states. This requires entanglement classes to be arranged in a sort of hierarchy or direction, since certain kinds of operations (particularly local operations and classical communications) may only work in one way (leading e.g. from a class of entanglement to the set of separable states, but never back).\\
Due to all these complex requirements, it is very difficult to classify entangled states in a satisfactory manner, especially since notations and definitions are often used ambiguously throughout the scientific community, such that no common basis of terminology has been established yet. Nevertheless, several promising approaches have been developed and started to be studied.\\

\section{Classification via Tensor Ranks}
One possible generalisation of bipartite Schmidt numbers is given by the concept of tensor ranks (see e.g. \cite{chitambar}).
\bdf The tensor rank $r$ of a pure state $\ket{\Psi}$ is defined as the lowest possible number of product states $\ket{\phi^p}$, into a superposition of which $\ket{\Psi}$ can be decomposed, i.e. the lowest number $l$ such that
\beq \ket{\Psi} = \sum_{i=1}^l c_i \ket{\phi^p_i} \eeq
where $c_i \in \mathbbm{C}$ with $\sum_i |c_i|^2 = 1$.\\
For mixed states $\rho$, the tensor rank can be generalised as the lowest number $l$, such that $\rho$ has a decomposition into pure states $\ket{\psi_i}$ of tensor rank $r(\ket{\psi_i}) \leq l \ \forall \ i$. \edf

\noindent At first glance, the tensor rank appears to induce a sensible classification of multipartite entanglement, which is why it was hoped to hold the key to a complete classification scheme. One of the most obvious disadvantages of the tensor rank is the fact that it is quite difficult to determine for a given (even pure) state. Apart from this, the tensor rank seems to offer many desirable features in the context of entanglement classification \cite{chen_tensorranks}. In particular, for multipartite qubit systems, it induces a sensible structure of equivalence classes of states, since all the well-known genuinely multipartite entangled states have inequivalent tensor ranks (particularly, the $n$-qubit GHZ-state has a tensor rank of 2, and the $n$-qubit $m$-Dicke state has a tensor rank of $n-m+1$, where $m \leq n/2$). Furthermore, the tensor rank formalism contains a hierarchy of equivalence classes (which, as discussed in the previous section, is a requirement for a sensible classification scheme): A state $\ket{\phi_{in}}$ can only be converted into a state $\ket{\phi_{out}}$ by means of local operations and classical communications, if this operation does not increase the tensor rank, i.e. if $r(\ket{\phi_{in}}) \geq r(\ket{\phi_{out}})$.\\
Since there are still many aspects of the tensor rank that are not sufficiently investigated, it cannot yet be estimated how exactly a potential classification scheme on this basis might be structured. However, from the present data it is already clear that the tensor rank alone cannot yield a sufficiently fine classification scheme, as several inequivalent states possess the same tensor rank and thus would be assigned to the same class. This can be seen e.g. by considering an $n$-qutrit GHZ-state and a three-qubit W-state, which both have a tensor rank of 3, while they should not be equivalent due to the different number of parties (for $n\neq 3$) and the qualitatively different entanglement properties. Various other examples of this kind can easily be constructed, e.g. a two-qutrit maximally entangled state embedded within a biseparable three-qutrit state has tensor rank three, just like a three-qutrit GHZ-state or a three-qubit W-state embedded within a three-qutrit system (i.e. all three states have the same tensor rank, while exhibiting completely different entanglement properties).\\

\section{Classification via Dimensionality of Entanglement}
In contrast to the concept of tensor ranks, the Schmidt number of bipartite systems can be generalised to multipartite systems in a different way, which is based rather on the physical interpretation of the Schmidt number (i.e. the number of degrees of freedom involved in the entanglement) than its purely mathematical meaning (i.e. the number of superposed product states). This concept is called the genuine dimensionality of entanglement \cite{spengler_gmd}.
\bdf A pure quantum state $\ket{\Psi}$ is called $f$-dimensionally entangled (with $f \geq 2$), iff it contains $f$-dimensional entanglement between any of its subsystems, i.e. iff there exists a reduced density matrix $\rho_i$ with rank$(\rho_i)\geq f$.\\
The state is called genuinely $f$-dimensionally entangled, iff all of its subsystems are $f$-dimensionally entangled with each other, i.e. iff
\beq \mathrm{rank}(\rho_i) \geq f \ \forall \ i \eeq
A mixed state $\rho$ is (genuinely) $f$-dimensionally entangled iff there is an at least (genuinely) $f$-dimensionally entangled pure state in each of its pure state decompositions.\edf

\noindent At present, tools for deciding whether a given state is $f$-dimensionally entangled or not are being developed at a promising rate (see e.g. Refs~\cite{li_gmd, lim_gmd}). However, like the classification via tensor ranks, the concept of entanglement dimensionality does not fully grasp the entanglement properties of a state (for example, all entangled qubit systems are 2-dimensionally entangled and all genuinely multipartite entangled qubit states are genuinely 2-dimensionally entangled - regardless of the specific type of entanglement present). Still, a combination of different tools - particularly, the concepts of tensor rank, dimensionality of entanglement and partial separability - offers a promising approach, as examples for inequivalent states which are equivalently described by these three concepts are not easily found.\\
In this context, it would be most desirable to find a way of constructing tools which incorporate these three concepts at once. A first step in this direction can be taken via the HMGH-framework, which allows for construction of criteria sensitive to at least two of them, namely the dimensionality of entanglement and partial separability \cite{spengler_gmd}. By superposing criteria for genuine multipartite entanglement (such as (\ref{eq_gme_ghz}) or (\ref{eq_gme_dicke})) in different bases, one can construct criteria for genuinely $f$-dimensional genuine multipartite entanglement.\\
\btm Criteria for genuinely multipartite entangled $f$-dimensional entanglement can be formulated via the quantities
\beq \label{eq_q0} Q_0 = \sum_{k\neq l}^{d-1} \left(\sqrt{\bra{k,l}\rho^{\otimes2}\ket{k,l}}-\sum_{\{\gamma\}}\sqrt{\bra{k,l}\mathcal{P}_{\{\gamma_1\}}^\dagger\rho^{\otimes2}\mathcal{P}_{\{\gamma_1\}}\ket{k,l}}\right) \eeq
where $\ket{k,l} = \ket{k}^{\otimes n}\otimes\ket{l}^{\otimes n}$ and the sum over $\{\gamma\}$ represents a sum over all possible bipartitions, and
\beq \label{eq_qm} Q_m = \frac{1}{m} \left(\sum_{k,l=0}^{d-2}\sum_{\{\sigma\}}\left( |\bra{e^k_{\{\alpha\}}}\rho\ket{e^l_{\{\beta\}}}| - \sum_{\{\delta\}}\sqrt{\bra{e^k_{\{\alpha\}}}\bra{e^l_{\{\beta\}}}\mathcal{P}_{\{\delta\}}^\dagger\rho^{\otimes2}\mathcal{P}_{\{\delta\}}\ket{e^k_{\{\alpha\}}}\ket{e^l_{\{\beta\}}}}\right) \right. \nonumber \\
 \left. - (d-1)m(n-m-1) \sum_{l=0}^{d-2}\sum_{\{\alpha\}}\bra{e^l_{\{\alpha\}}}\rho\ket{e^l_{\{\alpha\}}}\right) \eeq
with $1 \leq m \leq \frac{n}{2}$, where $\ket{e_{\{\alpha\}}^l}$ is the product vector with $\ket{l+1}$ in the $m$ subsystems whose labels are contained in the set $\{\alpha\}$, and $\ket{l}$ in all $(n-m)$ other subsystems. The sum over $\{\sigma\}$ runs over all sets $\{\sigma\} = \{\{\alpha\},\{\beta\}\}$ such that $|\{\alpha\}\cap\{\beta\}| = m-1$, and $\{\delta\}$ is defined as
\beq \{\delta\} = \left\{ \begin{array}{cc} \{\alpha\} & \mathrm{if} \ k=l \\ \{\{\chi\}: |\{\chi\}\cap\{\alpha\}|>0 \wedge |\{\chi\}\cap\{\beta\}| = 0\} & \mathrm{if} \ k \neq l \end{array} \right. \eeq
These quantities $Q_i$ ($0 \leq i \leq \frac{n}{2}$) are bounded from above by $f-1$ for all at most genuinely $f$-dimensionally entangled states and are non-positive for all biseparable states. Conversely, if for a given state $\rho$ any $Q_i > f-2$ (for any integer $f \geq 2$), then this state is detected to be at least genuinely $f$-dimensionally genuinely multipartite entangled.\etm
\bpf Note that for $d=2$, the expressions (\ref{eq_q0}) and (\ref{eq_qm}) are equivalent to (\ref{eq_gme_ghz}) and (\ref{eq_gme_dicke}), respectively, and are thus by construction maximally violated by the corresponding $n$-qubit GHZ- and $m$-Dicke-states, with a value of $1$. Since the inequalities are also non-positive for biseparable states, this proves the theorem for $d=2$.\\
Validity for $d>2$ follows from the convexity of the inequalities, as any non-genuinely multi-dimensional entangled state which is embedded within more than two local dimensions (i.e. for which more than one term in the sums over $k$ and $l$ are nonzero) necessarily yields a value lower than or equal to one.\qed\epf

\begin{figure}[htp!]\centering\includegraphics[width=8cm]{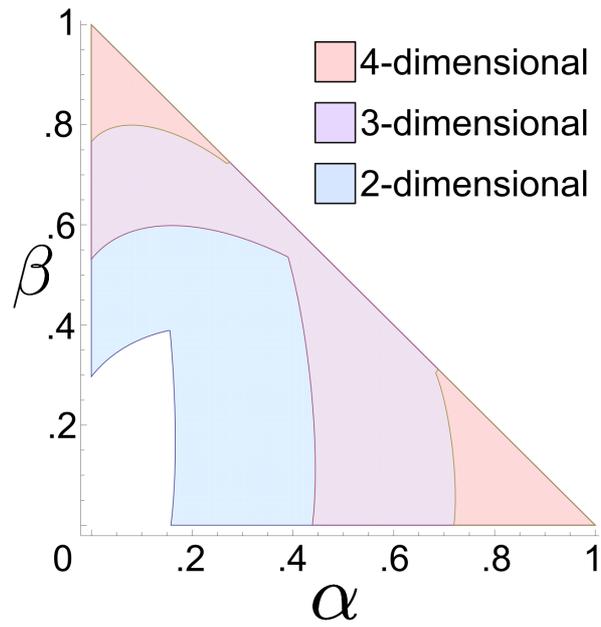}\caption[Detection quality of the criteria $Q_0$ and $Q_1$]{Illustration of the detection quality of the criteria $Q_0$ and $Q_1$ for the tripartite four-level state (\ref{eq_gmd_examplestate}). The red region (outmost in the corners) is detected to be genuinely four-dimensionally genuinely multipartite entangled, the purple region (middle one of the three coloured regions) is detected to be genuinely three-dimensionally genuinely multipartite entangled, and the green region (innermost, adjacent to the white region near the origin) is detected to be genuinely two-dimensionally genuinely multipartite entangled.}\label{fig_gmd_ghzw}\end{figure}

\begin{figure}[htp!]\centering\includegraphics[width=9cm]{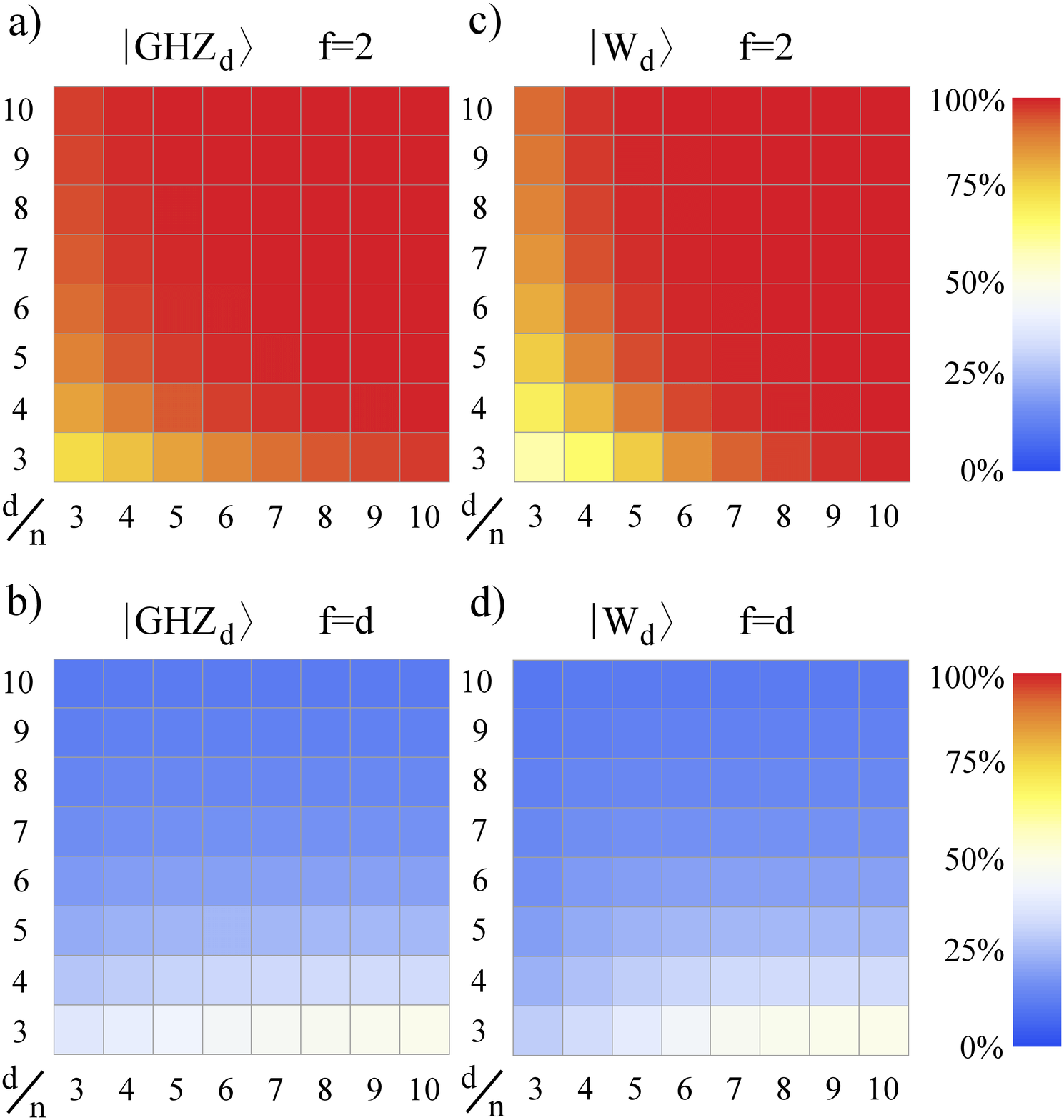}\caption[Noise resistance of the criteria $Q_0$ and $Q_1$]{Noise resistance (detection thresholds) of the criteria $Q_0$ and $Q_1$ for the states $\rho = p \ket{GHZ_d^n}\bra{GHZ_d^n} + \frac{1-p}{d^n}\id$ (a and b) and $\rho = p \ket{W_d^n}\bra{W_d^n} + \frac{1-p}{d^n}\id$ (c and d) for different $n$ and $d$ for $f=2$ and $f=d$.\newline It can be seen that for fixed $f$ (e.g. $f=2$, as depicted in a and c), the noise resistance increases with $n$ and $d$, while for $f=d$ (as depicted in b and d), it decreases with growing $d$ (but still increases with $n$).}\label{fig_gmd_temperature}\end{figure}

\noindent As an example, consider the $n$-qudit-states
\beq \ket{GHZ_d^n}=\frac{1}{\sqrt{n}}\sum_{i=0}^{d-1}\ket{i}^{\otimes n}\label{eq_quditghz} \eeq
and
\beq \label{eq_quditw} \ket{W_d^n} = \frac{1}{\sqrt{n(d-1)}}\sum_{i=0}^{d-1}\sum_{j=1}^n \ket{w_j^i} \eeq
where $\ket{w_j^i}$ is the product state of $\ket{i+1}$ for subsystem $j$ and $\ket{i}$ in all $(n-1)$ other subsystems, mixed with white noise, i.e.
\beq \label{eq_gmd_examplestate} \rho = \alpha \ket{GHZ_d^n}\bra{GHZ_d^n}+\beta\ket{W_d^n}\bra{W_d^n}+\frac{1-\alpha-\beta}{d^n} \id \eeq
In fig.~\ref{fig_gmd_ghzw}, the detection quality of the criteria $Q_0$ and $Q_1$ is illustrated, showing that they are capable of detecting large areas of genuine $d$-dimensional entanglement. Although the detection quality (i.e. the noise resistance) for fixed $f$ increases with growing $n$ and $d$, the detection quality for detecting genuine $d$-dimensional entanglement decreases with increasing $d$ (as illustrated in fig.~\ref{fig_gmd_temperature}).\\

\section{Classification via Exclusion\label{sec_class_excl}}
A quite different and rather intuitive approach, which can also be formulated within the HMGH-framework \cite{huberbruss}, works by excluding known states from the whole state space and defining new classes on the remaining set of states. Since this approach is based rather on intuition than on mathematical investigations, the resulting classification scheme is not unique and (in general) does not necessarily satisfy either of the two conditions $C1$ or $C2$ discussed in section \ref{sec_classes}. Nevertheless, it can be a very useful tool, as the requirements for its application -- in particular, the necessary before-hand knowledge about the considered state space, its symmetries and properties -- are very low, such that this classification scheme can ideally be used to obtain first results in an uncharted state space, which further study can then be based upon.\\
One possible way to define sensible equivalence classes for this kind of classification is as follows.
\bdf A class of states $\mathcal{C}(\ket{\Psi})$, represented by either one or several pure states $\ket{\Psi}$, is defined as the set of all local-unitary and permutational equivalents of these pure states, as well as the set of all mixtures of any such states:
\beq 
\mathcal{C}(\ket{\Psi}) = \left\{\rho \ | \ \rho = \sum_i p_i \ket{\phi_i}\bra{\phi_i}, \ \ket{\phi_i} = \Pi_i U_i^{local}\ket{\Psi}\right\}
\eeq
where $\Pi_i$ are subsystem-permutation operators and $U_i^{local}$ are local unitary transformations. \edf

\noindent By this definition, any given pure state induces an entanglement class. Thus, starting from a completely unknown state space (e.g. the set of all $n$ qudit states, for any fixed $n$ and $d$), one can find a classification scheme by applying the following steps:
\begin{enumerate}
	\item Find a pure state (or a set of pure states) which does not belong to any already defined class.
	\item Define the entanglement class associated with it (them).
	\item If there are states left which do not belong to a class already defined, go back to step 1. Otherwise, a complete classification scheme has been obtained and the task is accomplished.
\end{enumerate}
Since this procedure is not unique, there are several possible further restrictions, however, it is unclear which are favorable in the sense that they lead to a more satisfactory classification scheme. For example, it might be useful to restrict the choice of pure states in step 1 to a single family of states (for each class) of the form
\beq \ket{\Psi} = \sum_i c_i \ket{\psi_i} \quad \mathrm{where} \quad \sum_i |c_i|^2 = 1 \eeq
with variable $c_i \in \mathbbm{C}$ or $c_i \in \mathbbm{C}\backslash\{0\}$ (i.e. such that a pure states $\ket{\Psi}$ with different $c_i$ can compose a mixed state $\rho \in \mathcal{C}(\ket{\Psi})$).\\
Also, there might come a time, when there are only mixed states left unclassified, at which point there are several possibilities to proceed. All remaining states could, for example, be assigned one (or several) already existing class(es). Alternatively, a new class could be defined as the whole state space, such that automatically any state belongs to this class and most states also belong to (at least) one other class (which is an often-used strategy to obtain classification schemes, see e.g. \cite{acin_3qubits}).\\
Apart from the construction of entanglement classes, the problem of deciding which class a given state belongs to is of high interest. This question can be addressed by means of the HMGH-framework, as it allows for construction of separability criteria which are sensitive to this kind of classification \cite{huberbruss}. Given any representative pure state(s), an inequality can be constructed which is satisfied for all states belonging to the respective class, i.e. violation of the inequality implies that the investigated state does not belong to the considered class.\\
As an example, consider the $n$-qubit states
\beq \ket{\Psi^{(2)}} = c_1 \ket{0}^{\otimes n} + c_2 \ket{1}^{\otimes n} \quad \mathrm{with} \quad |c_1|^2+|c_2|^2=1 \label{eq_doublestate}\eeq
and
\beq \ket{\Psi^{(n)}} = \sum_{i=1}^n \lambda_i \ket{w_i} \quad \mathrm{with} \quad \sum_{i=1}^n |\lambda_i|^2 = 1 \label{eq_ntuplestate}\eeq
where $\ket{w_i} = \ket{0}^{\otimes (i-1)}\otimes\ket{1}\otimes\ket{0}^{\otimes (n-i)}$, which are generalisations of the GHZ-state and the W-state, respectively. Note that the definition of $\mathcal{C}(\ket{\Psi^{(x)}})$ is to be understood such that each pure state in a decomposition of a mixed state $\rho \in \mathcal{C}(\ket{\Psi^{(x)}})$ may have different $c_i$ or $\lambda_i$.\\
\btm
For all $n$-qubit-systems, the inequality
\beq \Re e && \left[\sum_{i\neq j} \left(\bra{w_i}\rho\ket{w_j} + (-1)^{n+1}\bra{\overline{w_i}}\rho\ket{\overline{w_j}}\right)\right] \nonumber \\
&& - (n-2)\sum_i \left(\bra{w_i}\rho\ket{w_i}+\bra{\overline{w_i}}\rho\ket{\overline{w_i}}\right) 
- \sum_{i\neq j} \left(\bra{d_{ij}}\rho\ket{d_{ij}} + \bra{\overline{d_{ij}}}\rho\ket{\overline{d_{ij}}}\right) \nonumber \\
&& - \frac{n(n-1)}{2}\left(\bra{0}^{\otimes n}\rho\ket{0}^{\otimes n} + \bra{1}^{\otimes n}\rho\ket{1}^{\otimes n}\right) \leq 0 \label{eq_doubleineq}\eeq
is satisfied for all states $\rho \in \mathcal{C}(\ket{\Psi^{(2)}})$, including all biseparable states. Here, $\ket{w_i}$ is defined as above, $\ket{d_{ij}}$ is the product state of $\ket{1}$ in the $i$-th and $j$-th subsystem and $\ket{0}$ in all other subsystems, and an overline denotes a bit flip in all subsystems, i.e. e.g. $\ket{\overline{001}} = \ket{110}$.\\
Furthermore, the inequality
\beq \Re e \left[ \bra{0}^{\otimes n}\rho\ket{1}^{\otimes n}\right] - \alpha\left(1-\bra{0}^{\otimes n}\rho\ket{0}^{\otimes n}-\bra{1}^{\otimes n}\rho\ket{1}^{\otimes n}\right) \leq 0 \label{eq_ntupleineq} \eeq
is satisfied for all $n$-qubit-states $\rho \in \mathcal{C}(\ket{\Psi^{(n)}})$, including all biseparable states, where $\alpha = \frac{3}{2}$ for $n=3$, $\alpha = 1$ for $n=4$ and $\alpha =\frac{1}{2}$ for $n > 4$.\etm
\bpf
The idea of the proofs is to show that the inequalities are satisfied for a pure state of the form (\ref{eq_doublestate}) or (\ref{eq_ntuplestate}) in an arbitrary basis (which also proves the validity for mixed states, by convexity of the inequalities). The proofs are mathematically quite simple, but since they are also rather cumbersome, they will not be presented here. The complete proofs can be found in Ref.~\cite{huberbruss}.\epf

\begin{figure}[htp!]\centering \includegraphics[width=8cm]{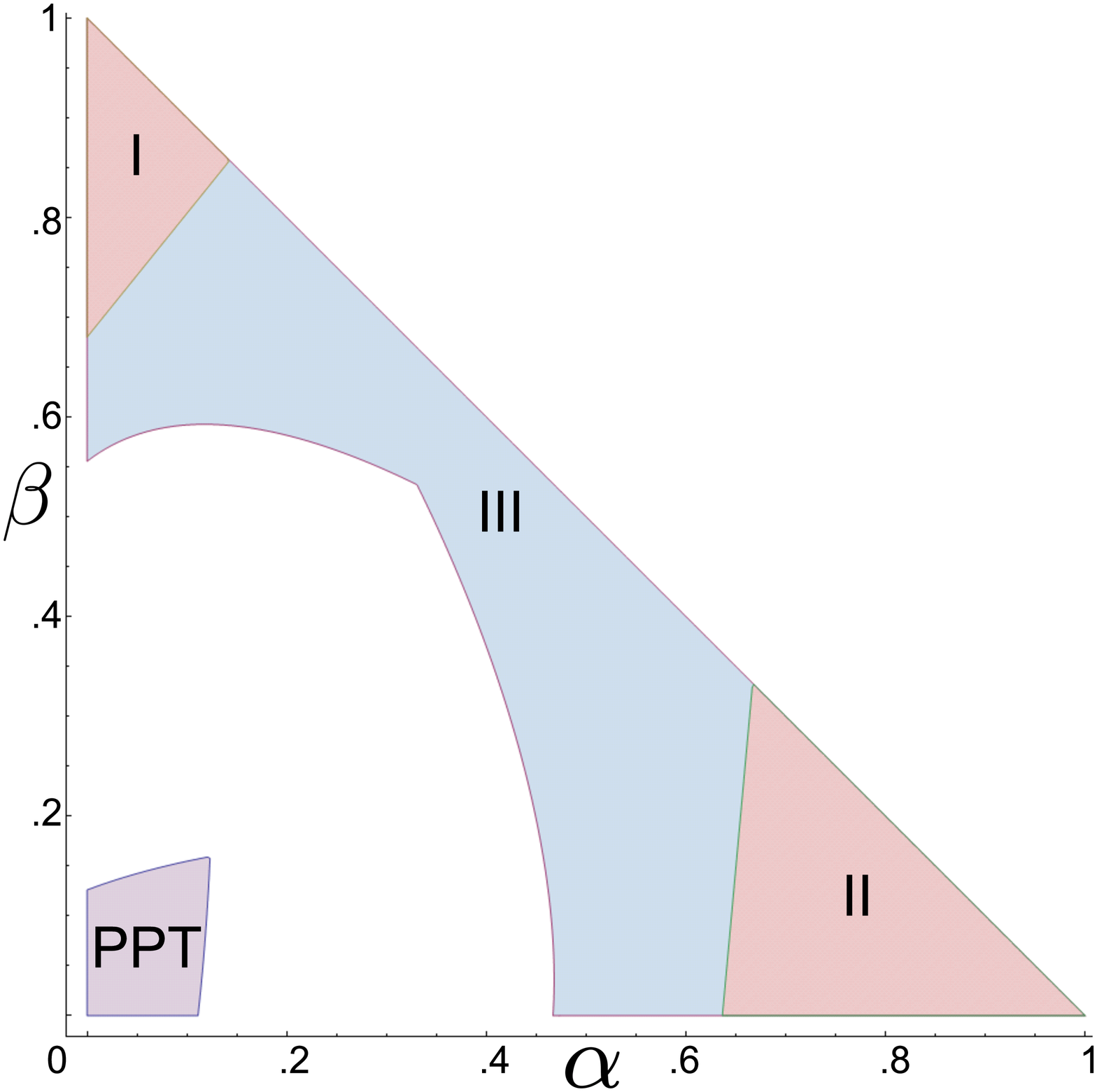}\caption[Detection quality of the classification inequalities (\ref{eq_doubleineq}), (\ref{eq_ntupleineq})]{Illustration of the detection quality of the two classification inequalities (\ref{eq_doubleineq}) and (\ref{eq_ntupleineq}) for the four-qubit state $\rho = \alpha \ket{GHZ_4}\bra{GHZ_4} + \beta \ket{W_4}\bra{W_4} + \frac{1-\alpha-\beta}{16} \id$. The areas I and II are detected by inequalities (\ref{eq_ntupleineq}) and (\ref{eq_doubleineq}), respectively, i.e. are detected not to belong to $\mathcal{C}(\ket{\Psi^{(4)}})$ and $\mathcal{C}(\ket{\Psi^{(2)}})$, respectively. For comparison, the areas III (genuine multipartite entanglement detected by (\ref{eq_gme_ghz}) and (\ref{eq_gme_dicke})) and PPT (the set of PPT-states) are also depicted.}\label{fig_classwghz}\end{figure}

\noindent Similar inequalities and class definitions can be conceived for arbitrary types of states, which allows for a successive cartography of any investigated state space. Although this kind of classification has several advantages over other classification schemes, there also are significant problems. Not only is it rather difficult to find a complete set of pure states which induce a complete classification. The associated detection inequalities also only barely have satisfactory detection quality (as illustrated in fig.~\ref{fig_classwghz}).

\begin{figure}[htp!]\centering\includegraphics[width=7cm]{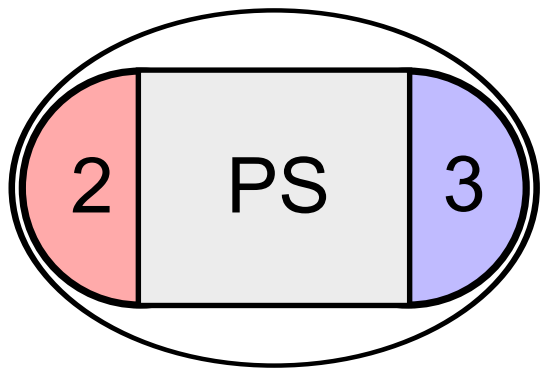}\caption[Classification of the set of three-qubit]{Qualitative illustration of the classification scheme induced by the states $\ket{\Psi^{(2)}}$ (denoted by '2') and $\ket{\Psi^{(3)}}$ (denoted by '3') for three qubit systems. The set labelled 'PS' depicts the set of partially separable states, which is a subset of both the sets '2' and '3'. Note that these two classes do not completely classify the space of three-qubit-states, and that a two-dimensional image can not fully grasp all properties of a high-dimensional space.}\label{fig_3qubitsclass}\end{figure}

\noindent A rough and qualitative illustration of the set of all three-qubit states is depicted as an example in fig. \ref{fig_3qubitsclass}. Note that a two-dimensional image can never grasp all properties of a high dimensional space such as this
.\\

\chapter{Examples and Applications\label{sec_apps}}

\section{GHZ-Type Isotropic States}
\begin{figure}[htp!]\centering\includegraphics[width=10cm]{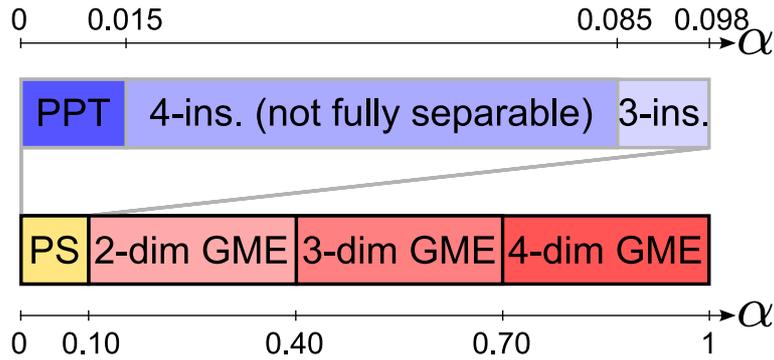}\caption[Parameter areas of the GHZ-type isotropic state]{Illustration of the different parameter areas of the GHZ-type isotropic state of four qudits with $d=4$ mixed with isotropic noise (\ref{eq_ghz44iso}). In the area labelled '4-dim GME', i.e. for $\alpha > \frac{149}{213}$, the state is detected to be genuinely 4-dimensionally genuinely multipartite entangled by criterion (\ref{eq_q0}), 3-dimensionally ('3-dim GME') for $\alpha > \frac{85}{213}$ and 2-dimensionally ('2-dim GME') for $\alpha > \frac{7}{71}$. Below this threshold, the state is not detected to be genuinely multipartite entangled, but appears to be partially separable ('PS'). In detail (as magnified in the upper part of the figure), different areas of partial separability can be identified by means of inequality (\ref{eq_gme_ghz}). For values of $\alpha$ above $\frac{3}{35}$, the state is detected not to be 3-separable ('3-ins'). For $\alpha > \frac{1}{65}$, it is still detected not to be fully separable, while below this threshold the state becomes PPT.}\label{fig_ghz_example}\end{figure}

The most well-known and deeply studied multipartite entangled state is the GHZ-state. Since also the most elementary criteria constructed from the HMGH-framework work best for this state, it constitutes an ideal example for illustrating how these different criteria complement each other. Consider the four-partite four-level GHZ state
\beq \ket{GHZ_4^4} = \frac{1}{2}(\ket{0000}+\ket{1111}+\ket{2222}+\ket{3333}) \eeq
mixed with isotropic noise, i.e. the state
\beq \rho = \alpha \ket{GHZ_4^4}\bra{GHZ_4^4} + \frac{1-\alpha}{256}\id \label{eq_ghz44iso} \eeq
In fig. \ref{fig_ghz_example}, the different areas of the parameter $\alpha$ are illustrated, where the state $\rho$ has different separability properties in terms of partial separability and genuine dimensionality of genuine multipartite entanglement, as detected by the criteria (\ref{eq_gme_ghz}) and (\ref{eq_q0}), respectively.\\

\section{Continuous Variable Systems}
After being the standard system for quantum informational considerations in the early days of quantum mechanics, the importance of continuous variable systems (e.g. position- or momentum-degrees of freedom of particles) dropped significantly, when the much more simple qudit-systems assumed this role. However, since a basic understanding of qudit-systems has been established, the much more complex continuous variable systems began to be investigated more thoroughly again, and today are known to enable technological applications, such as teleportation networks \cite{vloocknetwork}.\\
Unlike states of discrete systems, state vectors of continuous variable quantum systems are described by square-integrable functions $\ket{\Psi(x)} \in \mathcal{L}^2$. Mixed states on these systems have the form
\beq \rho = \int_{-\infty}^\infty d\alpha \ p(\alpha) \ket{\Psi(\alpha,x)}\bra{\Psi(\alpha,x')} \eeq
Unfortunately, most concepts and criteria developed for discrete quantum systems can not straightforwardly be generalised to continuous variable systems. Among the most prominent separability criteria, only the PPT criterion has been implemented for continuous systems \cite{simon_ppt} (although this is much more complex than applying it to qudit-systems). Most problems in continuous variable entanglement theory are approached by means of the covariance matrix formalism (for an overview, see e.g. \cite{adesso_cv}). Gaussian states, i.e. states with Gaussian profile, e.g. states the form
\beq \ket{\Psi(x)} \propto e^{-\frac{x^2}{\sigma}} \eeq
can be described by this framework with very few parameters and thus be characterised and detected very well. However, the covariance matrix approach fails to characterise non-Gaussian states, which makes the latter a major open problem in continuous variable entanglement.\\
Most criteria constructed from the HMGH-framework can also be applied to these infinitely dimensional systems \cite{ghrh}. In particular, they allow for detection of entanglement also in non-Gaussian states, as these criteria do not rely on the covariance matrix formalism. Consider e.g. the tripartite state
\beq \rho = p \ket{\omega}\bra{\omega} + (1-p) \rho_{mix} \eeq
with the non-Gaussian entangled state
\beq \ket{\omega} = \frac{1}{\alpha d^2} \int_{-d}^d dx (d-\alpha|x|) \ket{xxx} \eeq
and the noise function
\beq \rho_{mix} = \frac{1}{2\delta} \int_{-\delta}^\delta d^3x \ket{x_1}\bra{x_1}\otimes\ket{x_2}\bra{x_2}\otimes\ket{x_3}\bra{x_3} \eeq
For $d>\delta$, state is detected to be genuinely multipartite entangled by criterion (\ref{eq_gme_ghz}) for all $p>0$ and arbitrary values of $d$ and $\delta$, while for $d\leq\delta$ it is still detected if
\beq p > \frac{3 d^3 \alpha^2}{3 d^3 \alpha^2+2\delta} \eeq
The state is furthermore detected to be entangled (not fully separable) for 
\beq p > \frac{d^3 \alpha^2}{d^3 \alpha^2+2\delta} \eeq
Investigation of genuine multidimensional entanglement is not a sensible task for continuous variable systems, as e.g. the criterion (\ref{eq_q0}) would diverge for all states, where genuine multipartite entanglement is detected, if it was adapted to this system (i.e. any detected genuine multipartite entanglement is equivalent to detected genuine infinitely dimensional genuine multipartite entanglement).\\

\section{Many-Body Systems}
\begin{figure}[ht!]\centering\includegraphics[width=7cm]{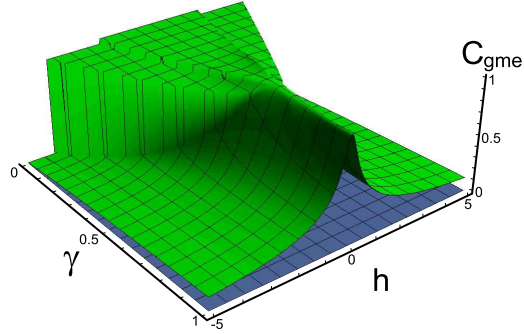}\caption[Genuine multipartite entanglement in many-body ground states]{Visualisation of the genuine multipartite entanglement content of the ground state of the investigated Hamiltonian for different values of the external magnetic field $h$ and the anisotropy parameter $\gamma$, measured by the gme-concurrence (\ref{eq_cgme}). While it is clear that the ground state becomes less and less genuinely multipartite entangled with increasing $|h|$ (as the state of lowest energy approaches the product state $\ket{0}^{\otimes n}$ or $\ket{1}^{\otimes n}$, depending on the sign of $h$), the amount of genuine multipartite entanglement also decreases drastically with increasing $\gamma$ (nevertheless, there always is an interval of $h$ in which the gme-concurrence is close to unity).}\label{fig_mbe_cgme}\end{figure}

\begin{figure}[ht!]\centering\includegraphics[width=9cm]{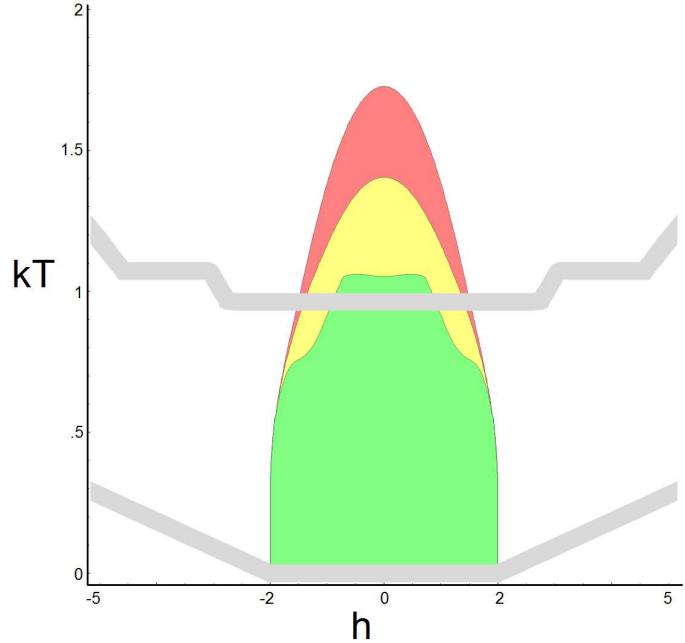}\caption[Multipartite entanglement in thermal many body states]{Illustration of the multipartite entanglement detected in thermal states of the discussed Hamiltonian by means of the $k$-entanglement-gap-witnesses (green, yellow and red for $k=2$, $k=3$ and $k=4$, respectively) and the detection inequalities (\ref{eq_gme_dicke}) for Dicke states (area between the gray lines), which in this case are the strongest criteria within the HMGH-framework. By construction, the $k$-entanglement-gap-witnesses can only detect multipartite entanglement in the area $|h|<2$, outside which the Hamiltonian's ground state becomes separable. In this area, both the GME-gap-witness and the HMGH-inequalities detect significant and about comparable amounts of genuine multipartite entanglement (with certain areas being detected by the witness, but not by the inequality, and other areas vice versa). In the area $|h|>2$ however, large areas are still detected by the inequality.}\label{fig_manybody}\end{figure}

Many-body systems, such as spin-chains or lattices, have been investigated in connection with entanglement very intensively in the past decades (see e.g. \cite{haselgrove_mbe, brukner_mbe, vedral_mbe, dowling_entgap}, or, for an overview, e.g. \cite{amico_mbesumm}). However, hardly any work has been done so far in this direction concerning partial separability or genuine multipartite entanglement. In this section, a common many-body system shall be studied in an exemplary fashion by means of the HMGH-framework as well as by special tools for many-body-entanglement, which can be adapted to detect $k$-inseparability (and, in particular, genuine multipartite entanglement). To do so, the latter need to be defined first.\\
Consider a many-body system described by a bounded Hamiltonian $\mathcal{H}$ (which for finite-dimensional Hilbert spaces, as are usually used to describe many-body systems, is always the case). Since the sets $\mathcal{S}_k$ of $k$-separable states are compact, there are certain minimal energy values
\beq E_{k-sep} = \min_{\sigma\in\mathcal{S}_k} \tr(\sigma \mathcal{H}) \label{eq_ksepen}\eeq
which a $k$-separable state can possess. If now the ground state of $\mathcal{H}$ is not $k$-separable (or, in case of a degenerate ground state, if there is no $k$-separable state in the ground state manifold), this implies
\beq E_{k-sep} > E_0 \eeq
where $E_0$ is the ground state energy of $\mathcal{H}$. Therefore, by definition of $E_{k-sep}$, all states with energies $\epsilon$ satisfying
\beq E_{k-sep} > \epsilon \geq E_0 \eeq
are necessarily $k$-inseparable, i.e. partially entangled (or, for $k=2$, genuinely multipartite entangled). Note that as a consequence of the definition (\ref{eq_ksepen})
\beq E_{n-sep} \geq E_{(n-1)-sep} \geq \cdots \geq E_{2-sep} \geq E_{1-sep} = E_0 \eeq
has to hold.\\
In this fashion, the Hamiltonian itself can effectively be used as an entanglement witness. This idea was first used for detecting generic entanglement, calling the detecting energy interval between the ground state energy and the minimal separable energy the entanglement gap \cite{dowling_entgap}. Later, it was generalised to partial separability and genuine multipartite entanglement in the form described above, denoting the corresponding energy interval between $E_0$ and $E_{k-sep}$ the $k$-entanglement gap and the interval between $E_0$ and $E_{2-sep}$ the GME-gap \cite{gabriel_gmegap}.\\
As an example, consider the Heisenberg model of a system of multiple spin-$\frac{1}{2}$-particles with nearest-neighbor interaction, given by the Hamiltonian
\beq \mathcal{H} = \frac{1}{2} \sum_{\langle i,j\rangle}\left(J_x \sigma_i^x\otimes\sigma_j^x + J_y \sigma_i^y\otimes\sigma_j^y+ J_z \sigma_i^z\otimes\sigma_j^z\right) + h \sum_{i = 1}^n \sigma_i^z \eeq
where $h$ is the external magnetic field, $\sigma_i^l$ is the $l$-th Pauli spin matrix acting on the $i$-th particle, the coefficients $J_l$ determine the spin-coupling strength and orientation in the three spatial dimensions and the sum over $\langle i,j\rangle$ runs over all index-pairs corresponding to adjacent particles (thus, this Hamiltonian can describe all kinds of lattices in arbitrarily many dimensions, depending only on the choice of these index-pairs).\\
Consider the example
\beq \begin{array}{c} J_x = 1 \\
J_y = 1-\gamma \\
J_z = 1-2\gamma \end{array} \eeq
with $0 \leq \gamma \leq 1$, interpolating between the isotropic antiferromagnetic Heisenberg XXX model ($\gamma = 0$) and the anisotropic Heisenberg XZ model ($\gamma=1$). Note that the maximal magnitude of the $J_i$ can be chosen to unity without loss of generality, by measuring all energies in units of $J_x$.\\
In many-body physics, one is usually concerned with thermal states (or, as a special case thereof, ground states) of given Hamiltonians, thus, this example will also deal with this class of states. A thermal state of a Hamiltonian $\mathcal{H}$ is given by
\beq \rho = \frac{1}{Z} \sum_i e^{-\frac{E_i}{kT}} \ket{E_i}\bra{E_i} \eeq
where $E_i$ are the eigenvalues of $\mathcal{H}$ corresponding to the eigenvectors $\ket{E_i}$, $kT$ is the temperature multiplied by Boltzmann's constant and
\beq Z = \sum_i e^{-\frac{E_i}{kT}} E_i \eeq
is a normalisation constant called the partition function.\\
In fig.~\ref{fig_mbe_cgme}, the genuine multipartite entanglement content of the above Hamiltonian's ground state (as measured by the gme-concurrence (\ref{eq_cgme})) is depicted for different values of $\gamma$ and $h$
. In this context, the most interesting case appears to be the isotropic case $\gamma = 0$, on which the rest of this example will thus be focussed.\\
In fig.~\ref{fig_manybody}, the genuine multipartite entanglement detection ranges of the GME-gap-witness and of the detection inequalities of the HMGH-framework are compared for the thermal state of the above Hamiltonian with $\gamma = 0$. While for magnetic fields of low magnitude, the GME-gap-witness detects slightly higher temperatures to still be genuinely multipartite entangled, in general the HMGH-inequalities detect a much larger state space area.

\section{Quantum Secret Sharing\label{sec_app_qss}}
Quantum secret sharing \cite{hillery_qss}, a multipartite form of quantum cryptography \cite{gisin_crypto}, currently is one of the most important technological applications of multipartite entanglement. It is the solution to the following problem: Assume that Alice wants to share a secret message with $(n-1)$ other parties Bob, Charlie, Daisy, et cetera. However, some of these parties might not be trustworthy, such that the message should only be readable by all recipients together. What can Alice do to distribute the message, such that only all $(n-1)$ other parties together can read it, while each individual party has no information on it whatsoever.\\
Quantum secret sharing works as follows. Alice supplies each party (including herself) with one of $n$ particles which together are in an $n$-qubit GHZ-state
\beq \ket{GHZ}=\frac{1}{\sqrt{2}}(\ket{0}^{\otimes n}+\ket{1}^{\otimes n}) \eeq
Since the GHZ-state contains maximal $n$-partite entanglement and no lower entanglement (in the sense that all of its reduced density matrices are separable), this guarantees that the $(n-1)$ recipient parties can only act together. Now, each party randomly measures their respective qubit either in the x- or in the y-basis, given by the respective sets of eigenstates
\beq \ket{x\pm} = \frac{1}{\sqrt{2}}(\ket{0} \pm \ket{1}) \quad\mathrm{and}\quad \ket{y\pm} = \frac{1}{\sqrt{2}}(\ket{0}\pm i\ket{1}) \eeq
By comparing their respective measurement results, the $(n-1)$ recipient parties can uniquely predict the state Alice's qubit is left in before the measurement (as can be seen by rewriting the GHZ-state in the x- or y-basis, and as is explicitly given in table \ref{table_qss} for the case $n=3$).\\
\begin{table*}[htp!]\centering\begin{tabular}[t]{r|cccc}
			         & \ket{x+} & \ket{x-} & \ket{y+} & \ket{y-} \\ \hline
			\ket{x+} & \ket{x+} & \ket{x-} & \ket{y+} & \ket{y-} \\
			\ket{x-} & \ket{x-} & \ket{x+} & \ket{y-} & \ket{y+} \\
			\ket{y+} & \ket{y-} & \ket{y+} & \ket{x-} & \ket{x+} \\
			\ket{y-} & \ket{y+} & \ket{y-} & \ket{x+} & \ket{x-}
		\end{tabular}\caption[Quantum secret sharing]{Tripartite quantum secret sharing. The rows represent Bob's different possible measurement outcomes, and the columns represent Charlie's. By revealing their respective measurement results to one another, Bob and Charlie can uniquely predict the state of Alice's qubit before the measurement. Whenever Alice choses the 'right' basis for her measurement, Bob and Charlie can thus predict her measurement outcome with certainty. For example, if Bob measures $\ket{x-}$ and Charlie measures $\ket{y+}$, Alice's qubit is left in the state $\ket{y+}$.}\label{table_qss}
\end{table*}
\noindent Next, all $n$ parties (including Alice) publicly announce their choice of basis. In half of the cases, Alice will have chosen the 'wrong' basis (i.e. not the basis corresponding to the state of her qubit, such that her measurement result is entirely random), in which case all $n$ qubits are discarded and the procedure is repeated. In all other cases, the recipient parties can determine Alice's measurement outcome if and only if they pool their knowledge. If even one party is missing (or contributing false data), the prediction fails. Alice's measurement outcome can therefore be used as a key to securely encrypt messages.\\
\\
One of the crucial loopholes in this scheme is the distribution of the GHZ-state. A potential eavesdropper could intercept one or several of the sent qubits and manipulate or replace them, thus compromising the security of the secret. The loophole can be closed by verifying that indeed a genuinely $n$-partite state is shared between the $n$ parties. Since the desired state is a GHZ-state, inequality (\ref{eq_gme_ghz}) offers an ideal criterion to do so (as discussed in Ref.~\cite{schauer}).\\
Rewriting the inequality e.g. for the three-qubit-case with $\ket{\Phi}=\ket{000111}$ and the short-hand notation $\rho_{ijklmn} = \bra{ijk}\rho\ket{lmn}$
\beq \label{eq_qssverify} |\rho_{000111}|-\sqrt{\rho_{001001}\rho_{110110}}-\sqrt{\rho_{010010}\rho_{101101}}-\sqrt{\rho_{100100}\rho_{011011}} \leq 0 \eeq
in expectation values of Pauli operators (see section \ref{sec_experiment}), 
\beq \begin{array}{c}
\rho_{000111}=\frac{1}{8}([111]-[221]-[212]-[122]-i([222]-[112]-[121]-[211])) \\
\rho_{001001}=\frac{1}{8}([000]-[003]+[030]+[300]+[330]-[303]-[033]-[333]) \\
\rho_{110110}=\frac{1}{8}([000]+[003]-[030]-[300]+[330]+[303]-[033]+[333]) \\
\rho_{010010}=\frac{1}{8}([000]+[003]-[030]+[300]-[330]+[303]-[033]-[333]) \\
\rho_{101101}=\frac{1}{8}([000]-[003]+[030]-[300]-[330]+[303]-[033]+[333]) \\
\rho_{100100}=\frac{1}{8}([000]+[003]+[030]-[300]-[330]-[303]+[033]-[333]) \\
\rho_{011011}=\frac{1}{8}([000]-[003]-[030]+[300]-[330]-[303]+[033]+[333]) \end{array} \eeq
it becomes apparent that half of the necessary measurements Bob and Charlie have to do (eight out of sixteen) for implementing this security check are measured already during the quantum secret sharing protocol itself, i.e. whenever Alice choses the 'wrong' measurement basis, the particles can be used to verify inequality (\ref{eq_qssverify}) instead of discarding them. In this way, the number of additional measurements required is kept comparatively low and the security check becomes rather cheap (in terms of entangled states).\\

\section{Error Estimation}
Since the criteria constructed from the HMGH-framework are comparatively easily implementable in experiments (as discussed in section \ref{sec_experiment}), it is desirable to also be able to control the error propagation from the measurements to the yield of these criteria. In order to give an illustrative example for this, consider the criterion (\ref{eq_ksep}). For reasons of simplicity, assume that all density matrix diagonal elements contain the same relative error $\delta$, and let the (absolute) error of the measured off-diagonal density matrix element (the first term in the inequality) be $o$. The total error $\Xi$ of the inequality's value can be determined by the Gaussian law of error propagation, which states that the total measurement uncertainty $\Xi$ of a function $f$ of several measured values $x_i$ is given by
\beq \Xi = \sqrt{\sum_i \left(\frac{\partial f}{\partial x_i} \xi_i\right)} \eeq
where $\xi_i$ is the respective measurement uncertainty of $x_i$. For the criterion under investigation, this yields \cite{ghrh}
\beq \Xi^2 && =o^2+\sum_{\{\alpha\},i} \left(\frac{1}{2k} \prod_{j=1}^{2k} \frac{(x_j)^{\frac{1}{2k}}}{x_i}\xi_i\right)^2 = o^2+\frac{1}{4k^2}\sum_{\{\alpha\},i}\prod_{j=1}^{2k} (x_j)^{\frac{1}{k}} \delta^2\nonumber \\
 && \leq o^2 + \delta^2 \frac{\gamma}{8k^3} \eeq
where the $x_i$ are the different density matrix diagonal elements used in the inequality, and $\gamma$ is the number of $k$-partitions of an $n$-partite system, given by the Sterling number in the second kind (\ref{eq_sterling}). For most practical cases, the second term is much smaller than the first, and thus $\Xi \approx o$.\\
Due to the simple form the criteria from the HMGH-framework assume when written in terms of density matrix elements, the overall error can be estimated rather easily (as compared to other separability criteria, which involve e.g. optimisation, eigenvalue computation, and other nontrivial functions).\\

\chapter{Summary and Conclusion}
While the phenomenon of bipartite quantum entanglement is already widely understood, the rather young research field of multipartite entanglement theory still holds many puzzles and mysteries. Results from the bipartite case often cannot be generalised in a straightforward way, which gives rise to various open problems, ranging from the only partially solved problem of mere entanglement detection, over its quantification in different ways, to problems like multipartite entanglement classification, which are still completely unanswered.\\
In order to address these issues, several different approaches have been developed over the past years, yielding results in different areas and starting to form a first rudimentary picture of multipartite entanglement as a whole. One of the most recent advances was made by the development of the HMGH-framework, a versatile tool which finds application in all these topics and already significantly improved on several previous results. It allows for construction of criteria for characterising entanglement in various ways and is very easily applicable (both theoretically and experimentally).\\
\\
In the present work, different problems of multipartite entanglement characterisation have been reviewed and discussed, with special emphasis on the contribution by the HMGH-framework. While this work is not claimed to be a complete discussion of multipartite entanglement theory, it is meant to give an overall insight into the problems and working principles of the topic.\\
The conceptually comparatively simple question of partial separability is essentially understood, however, tools for distinguishing between states with different separability properties still require further improvement for the problem of general $k$-separability. Nevertheless, genuine multipartite entanglement can already be detected and quantified quite satisfyingly.\\
The probably most important still widely open question in multipartite entanglement theory concerns the classification of multipartite entanglement. For general systems, very little is known at all, apart from the fact that there are inequivalent classes of multipartite entanglement which exhibit different entanglement properties. Since even the number, let alone the form, of these types of entanglement is generally unknown, this is a very difficult topic to address. First advances can be made from several directions, still even a coarse understanding of the problem seems not within range.\\

\newpage
\vspace*{4cm}
\begin{center}
\Large{\bf{Acknowledgements}}\\
\vspace{1cm}
\normalsize I would like to thank all the people who helped me make my way to where and who I am today -- both privately and professionally. In particular, I am deeply grateful to Tina Hinterleitner for enriching my life in so many ways, to Johanna and Ursula Gabriel for always being there when I needed them, and to Reinhold Bertlmann, Beatrix Hiesmayr and Marcus Huber (in chronological order) for guiding me and assisting me on my academical way.
\end{center}

\newpage \
\newpage

\listoffigures

\newpage \

\begin{appendix}
\chapter{Bipartite Separability Criteria in the HMGH-Framework}
The HMGH-framework (and generalisations of the elementary separability criteria it yields) not only allows for construction of versatile multipartite detection criteria for different kinds of entanglement, it also contains novel separability criteria for bipartite systems. Since these do not detect more entanglement than common criteria (such as the PPT criterion) and are rather more complicated to use, they do not offer any real advance in entanglement theory. However, these criteria (like related the multipartite ones) are rather easily implementable in experiments due to the comparatively low number of different measurements required. For sake of completeness, two examples for such criteria shall be presented here.\\

\section{Introducing Criteria}
As a starting point, consider the elementary bipartite separability criterion introduced and discussed in section \ref{sec_hmgh_example}:
\beq \sqrt{\bra{\Phi}\rho^{\otimes 2}\underline{\mathcal{P}}\ket{\Phi}} - \sqrt{\bra{\Phi}\mathcal{P}^\dagger_A\rho^{\otimes 2}\mathcal{P}_A\ket{\Phi}} \leq 0 \label{eq_app_ineq1} \eeq

\subsection{\texorpdfstring{$m$}--linear Inequality}
The above inequality can be straightforwardly generalised as follows.

\btm The inequality
\beq \label{eq_app_mlin} \sqrt{\Re e(\bra{\Phi}\rho^{\otimes m}\underline{\mathcal{P}^{\rightarrow}}\ket{\Phi})} - \sqrt{\bra{\Phi}\mathcal{P}_A^{\rightarrow\dagger}\rho^{\otimes m}\mathcal{P}_A^\rightarrow\ket{\Phi}} \leq 0 \eeq
is satisfied for all separable bipartite states $\rho \in \mathcal{H}$, for all fully separable states $\ket{\Phi} = \bigotimes_{i=1}^m\ket{\alpha_i}\otimes\bigotimes_{i=1}^m\ket{\beta_i} \in \mathcal{H}^{\otimes m}$ and for all $m \in \mathbbm{N}$, where the cyclic permutation operators $P_i^\rightarrow$ on the $m$-fold copies of the respective subsystems of $\rho$ are defined such that
\beq P_A^\rightarrow \ket{\alpha_1}\otimes\ket{\alpha_2}\otimes\cdots\otimes\ket{\alpha_m} = \ket{\alpha_2}\otimes\ket{\alpha_3}\otimes\cdots\otimes\ket{\alpha_m}\otimes\ket{\alpha_1} \eeq
Note that for $m=2$, this is equivalent to inequality (\ref{eq_app_ineq1}), where for $m \neq 2$ the permutation operators $\mathcal{P}^\rightarrow_i$ are different from the operators $\mathcal{P}_i$ used in other criteria constructed from the HMGH-framework, and in particular do not satisfy $\mathcal{P}_i^{\rightarrow\dagger} = \mathcal{P}_i^\rightarrow$.\etm

\bpf The inequality is equivalent to the inequality
\beq \bra{\Phi}\mathcal{P}_A^{\rightarrow\dagger}\rho^{\otimes m}\mathcal{P}_A^\rightarrow\ket{\Phi} - \frac{1}{2}(\bra{\Phi}\rho^{\otimes m}\underline{\mathcal{P}^\rightarrow}\ket{\Phi}+\bra{\Phi}\underline{\mathcal{P}^\rightarrow}^\dagger\rho^{\otimes m}\ket{\Phi}) \geq 0 \eeq
Since this inequality is not convex, it has to be proven for mixed states $\rho = \sum_i \ket{\phi_i}\bra{\phi_i}\otimes\ket{\chi_i}\bra{\chi_i}$ explicitly (where $\ket{\phi_i}$ and $\ket{\chi_i}$ are subnormalised states, i.e. $\bra{\phi_i}\phi_i\rangle = \bra{\chi_i}\chi_i\rangle = \sqrt{p_i}$, with $\sum_i p_i = 1$), which is equivalent to showing that

\beq \vec{X}_{A}^\ast\cdot\vec{X}_{A}- \frac{1}{2}\left(\vec{X}^\ast\cdot\vec{X}_{AB}+\vec{X}_{AB}^\ast\cdot\vec{X}\right) \geq 0 \eeq
where
\beq 
\left[\vec{X}\right]_{p_1\cdots p_nq_1\cdots q_n} = && \prod_{i=1}^m \langle\alpha_i|\phi_{p_i}\rangle \prod_{i=1}^m \langle\beta_i|\chi_{q_i}\rangle \nonumber \\
\left[\vec{X}_A\right]_{p_1\cdots p_nq_1\cdots q_n} = && \prod_{i=1}^m\langle\alpha_i|\phi_{p_{i\oplus1}}\rangle \prod_{i=1}^m\langle\beta_i|\chi_{q_i}\rangle \\ \nonumber
\left[\vec{X}_{AB}\right]_{p_1\cdots p_nq_1\cdots q_n} = && \prod_{i=1}^m\langle\alpha_i|\phi_{p_{i\oplus1}}\rangle \prod_{i=1}^m\langle\beta_i|\chi_{q_{i\oplus1}}\rangle \eeq
where $\oplus$ is the addition modulo $m$. Since 
\beq \vec{X}^\ast\cdot\vec{X} = \vec{X}_A^\ast\cdot\vec{X}_A = \vec{X}_{AB}^\ast\cdot\vec{X}_{AB} \eeq
this simplifies to 
\beq \frac{1}{2}\left|\vec{X}^\ast\cdot\vec{X}_{AB} - \vec{X}_{AB}^\ast\cdot\vec{X}\right|^2\geq 0 \eeq 
\qed \epf

\subsection{Rank-\texorpdfstring{$m$}--Determinant}
In order to obtain a different bipartite separability criterion, observe that inequality (\ref{eq_app_ineq1}) can also be written as
\beq \det\begin{pmatrix} \rho_{i_1j_1i_1j_1} & \rho_{i_1j_2i_2j_1} \\ \rho_{i_2j_1i_1j_2} & \rho_{i_2j_2i_2j_2} \end{pmatrix} \geq 0 \eeq
where $\rho_{ijkl} = \bra{ij}\rho\ket{kl}$ with $\ket{\Phi} = \ket{i_1j_1i_2j_2}$.\\
Starting from this observation, the criterion can also be generalised in a different way.

\btm The inequality
\beq \label{eq_app_psdet} \det\begin{pmatrix} \rho_{i_1j_1i_1j_1} & \rho_{i_1j_2i_2j_1} & \cdots & \rho_{i_1j_mi_mj_1} \\ \rho_{i_2j_1i_1j_2} & \rho_{i_2j_2i_2j_2} & \cdots & \rho_{i_2j_mi_mj_2} \\ \cdots & \cdots & & \cdots \\ \rho_{i_mj_1i_1j_m} & \rho_{i_mj_2i_2j_m} & \cdots & \rho_{i_mj_mi_mj_m}\end{pmatrix} \geq 0 \eeq
is satisfied for all separable states $\rho = \sum \rho_{ijkl} \ket{i}\bra{k}\otimes\ket{j}\bra{l}$, for all $m \in \mathbbm{N}$ and for all $\{i_\alpha\} \in \mathbbm{N}^{m}$ and $\{j_\alpha\}\in \mathbbm{N}^{m}$ (i.e. each $i_\alpha$ and $j_\alpha$ is an integer between 0 and $d_1$ or $d_2$ (respectively), and $\alpha = 1,2,\cdots,m$).\\
Note that for $m=2$, this is equivalent to inequality (\ref{eq_app_ineq1}). Also observe that for $m > \mathrm{rank}(\rho)$ equality holds, i.e. the inequality can never be violated. The same is true if any two $i_\alpha$ or $j_\alpha$ are chosen equal. \etm

\bpf To prove this inequality, remember that 
\beq \det A = \epsilon_{j_1j_2\cdots j_m} A_{1j_1} A_{2j_2} \cdots A_{mj_m} \eeq
Since every density matrix element of a separable state can be written as
\beq \rho_{ijkl} = \sum_{\alpha} p_\alpha a^\alpha_i b^\alpha_j a^{\alpha *}_k b^{\alpha *}_k \eeq
and by abbreviating 
\beq c^{\alpha_1 \alpha_2 \cdots \alpha_m}_{1, 2, \cdots, m} = a_{i_1}^{\alpha_1}b_{j_1}^{\alpha_1 *} a_{i_2}^{\alpha_2} b_{j_2}^{\alpha_2 *} \cdots a_{i_m}^{\alpha_m} b_{j_m}^{\alpha_m *} \eeq
we arrive at
\beq \Xi_m = \sum_{\alpha_1 \cdots \alpha_m} p_{\alpha_1}\cdots p_{\alpha_m} \epsilon_{k_1k_2\cdots k_m} c^{\alpha_1 \alpha_2 \cdots \alpha_m}_{i_1i_2\cdots i_m} c^{\alpha_1 \alpha_2 \cdots \alpha_m *}_{j_{k_1}j_{k_2}\cdots j_{k_m}} \eeq
Since the $c^{\alpha_1 \alpha_2 \cdots \alpha_m}_{j_1j_2\cdots j_m}$ are symmetric w.r.t. interchange of index pairs $\{\alpha_r, j_r\}$ and since the whole expression has to be fully symmetric in $\{\alpha_1, \alpha_2, \cdots, \alpha_m\}$, it follows that
\beq \Xi_m && = \sum_{\alpha_1 \cdots \alpha_m} p_{\alpha_1}\cdots p_{\alpha_m} \epsilon_{k_1k_2\cdots k_m} c^{\alpha_{i_1} \alpha_{i_2} \cdots \alpha_{i_m}}_{1, 2, \cdots, m} c^{\alpha_{j_{k_1}} \alpha_{j_{k_2}} \cdots \alpha_{j_{k_m}} *}_{1, 2, \cdots, m} \nonumber \\
&& = \sum_{\alpha_1 \cdots \alpha_m} p_{\alpha_1}\cdots p_{\alpha_m} c^{\alpha_{i_1} \alpha_{i_2} \cdots \alpha_{i_m}}_{1, 2, \cdots, m} c^{[\alpha_{j_1} \alpha_{j_2} \cdots \alpha_{j_m}]*}_{1, 2, \cdots, m} \\
&& = \sum_{\alpha_1 \cdots \alpha_m} p_{\alpha_1}\cdots p_{\alpha_m} \left|c^{[\alpha_{1} \alpha_{2} \cdots \alpha_{m}]}_{1, 2, \cdots, m}\right|^2 \geq 0 \nonumber \eeq
where $[1, 2, \cdots, m]$ is the antisymmetrisation. \qed\epf

\section{Discussion}
The main advantages of the HMGH-framework lie in the convex structure of the inequalities, which allows for discrimination of different kinds of partial separability. Since this concept is not present in the bipartite scenario, the bipartite entanglement detection criteria which can be constructed from the framework lack advantages over other bipartite separability criteria. Furthermore, they are disadvantageous in several ways, as e.g they require an optimisation over all local-unitary transformations in order to be implemented most efficiently. Also, they detect entanglement strictly worse than the (comparatively quite simple) PPT criterion.\\
In spite of the parameter $m$ representing a maximal number of (local) dimensions in which entanglement can be detected by means of the criteria (\ref{eq_app_mlin}) and (\ref{eq_app_psdet}), the criteria cannot straightforwardly be used to detect genuine multidimensional entanglement either, since already bipartite entanglement may violate the inequalities for arbitrary $m$. However, there might be other entanglement properties which could be related to the different detection capabilities of the criteria.\\
\\
Note that the detection power of the criteria formulated via the $m$-linear inequalities (\ref{eq_app_mlin}) decreases with increasing $m$, since the two sides of the respective inequalities (with $\ket{\Phi} = \ket{\phi_1^A}\otimes\ket{\phi_1^B}\otimes\ket{\phi_2^A}\otimes\ket{\phi_2^B}\otimes\cdots\otimes\ket{\phi_m^A}\otimes\ket{\phi_m^B}$) read
\beq \Re e\left(\prod_{i=1}^m \bra{\phi_i^A,\phi_i^B}\rho\ket{\phi_{i+1}^A,\phi_{i+1}^B}\right) \leq \prod_{i=1}^m \bra{\phi_{i+1}^A,\phi_i^B}\rho\ket{\phi_{i+1}^A,\phi_i^B} \eeq
where the index $m+1$ is identified with the index 1. Observe that both sides of the inequality are essentially geometric means of density matrix elements. Consequently, if the inequality is violated, there necessarily has to be a smaller number of density matrix elements on each side (forming the $m'-$linear inequality for some $m' < m$), such that the inequality is violated by a larger value (or at the equal value) than the $m$-linear inequality. Due to the freedom of choice in $\ket{\Phi}$, any phase which might be necessary for the violation (since only the real part of the product on the left hand side affects the criterion) can always be reproduced with $m'$ factors as well (in particular, for $m=2$ phases do not enter the criterion at all, as the left hand side of the inequality is just the absolute value of a density matrix element).\\

\chapter{Entanglement in Unstable Systems}
In unstable quantum systems (i.e. systems composed of radionuclids or other decaying particles), density matrices are often incorrectly obtained by only considering the undecayed part of the system. However, this implies a kind of post-selection, which leads to correlations in the obtained data, which do not reflect the information present in the system (see e.g. Ref.~\cite{popescu}). In order to obtain a correct description of the state in question (containing exactly the amount of information available in the physical scenario), the decay products have to be taken into account as well. This, in turn, is a very complicated task.\\
In other words: While the problem of determining a state's entanglement properties, given the density matrix, is - although highly nontrivial - a well defined one, finding the density matrix which correctly describes a system involves subtle complications which are only seldom mentioned.\\
For sake of simplicity, this concept shall be discussed here on the basis of (at most) bipartite entangled qubit systems (as the principle idea is the same for higher dimensional cases).\\

\section{Measurements}
Consider a single particle with two degrees of freedom, i.e. a qubit. If the particle is stable, any measurement (characterised by a measurement direction $\vec{a}$, e.g. in the Bloch-representation) can yield either of two outcomes:
\begin{itemize}
	\item The particle is in the state corresponding to the direction $\vec{a}$.
	\item The particle is in the state corresponding to the direction orthogonal to $\vec{a}$.
\end{itemize}
If the particle is unstable, this is not the case, since the particle may have decayed before the measurement. Therefore, measurements have to be adapted to incorporate this possibility. This can be done by explicitly allowing for more different outcomes, such as
\begin{itemize}
	\item The particle is in the state corresponding to the direction $\vec{a}$.
	\item The particle is in the state corresponding to the direction orthogonal to $\vec{a}$.
	\item The particle has decayed via channel 1.
	\item The particle has decayed via channel 2.
	\item et cetera.
\end{itemize}
where the different channels represent the different possible ways of decay for the particle (with different decay products, times of decay and/or properties of the decay products, e.g. momenta, spins, et cetera). This is a rather unfeasible way of measuring, as it not only requires complete knowledge over everything that happens in the experiment (including all involved particles), but also implies that both directions, $\vec{a}$ and its orthogonal, can be identified with certainty (which in experimental situations often is not the case).\\
A more experimentally suitable way of approaching this problem is by discriminating not between all possible results, but only subsets. An intuitive and practical choice of possible outcomes is e.g.
\begin{itemize}
	\item The particle is in the state corresponding to the direction $\vec{a}$.
	\item The particle is not in this state.
\end{itemize}
In this case, while a negative measurement result contains rather little information, a positive one can be used for statistical analysis, thus yielding valuable data.\\
An operational framework based on this approach was introduced in Ref.~\cite{hiesmayr_kaons}. In this approach, operators in the Heisenberg picture (i.e. time dependant operators acting on time independant states) are constructed, such that the probability for a positive measurement result decreases exponentially with time (corresponding to the decay rate of the particle). These effective operators can be written as
\beq \label{eq_app_effop} O^{eff}(\alpha,\phi,t) = (1-|\vec{n}|) \id + \vec{n}\cdot \vec{\sigma} \eeq
where $\vec{\sigma}$ is the vector of the three Pauli matrices, $t$ is the time parameter which together with $\alpha$ and $\phi$ parametrises the measurement direction
\beq \vec{n} = e^{-\Gamma t} \left(\begin{array}{c} \cos(t+\phi)\sin(\alpha) \\
\sin(t+\phi)\sin(\alpha) \\ \sinh(\Delta\Gamma t) + \cosh(\Delta\Gamma t)\cos(\alpha) \end{array}\right) \eeq
with $\Delta\Gamma = \frac{\Gamma_1-\Gamma_2}{2}$ and $\Gamma = \frac{\Gamma_1+\Gamma_2}{2}$ being the difference and mean value (respectively) of the decay widths of the two eigenvectors of $\sigma_z$, which is, without loss of generality, chosen to be the eigenbasis of decay width (if the latter is not constant, in which case obviously $\Delta\Gamma=0$ and $\Gamma$ is the usual decay width of the particle). Note that this formalism is contains an additional assumption, namely that the particles decay exponentially and obeying the law of decay.\\

\section{Bell Inequalities}
As an example, consider the problem of constructing experimentally suitable Bell-Inequalities for bipartite neutral kaon systems. These systems are typically entangled in the degrees of freedom corresponding to the quantum number of strangeness and are produced in a singlet state:
\beq \ket{\Psi^-}=\frac{1}{\sqrt{2}}\left(\ket{K^0}\otimes\ket{\bar{K^0}} - \ket{\bar{K^0}}\otimes\ket{K^0}\right) \eeq
where $\ket{K^0}$ and $\ket{\bar{K^0}}$ are the strangeness eigenstates (particle and antiparticle, respectively). For unstable systems like this one, Bell inequalities can - again, assuming the law of decay - be constructed in a more effective way by observing that the bounds for the expectation value of the Bell operator $\mathcal{B}$ may be time-dependant \cite{hiesmayr_bell}. That is, while at $t=0$, the bounds have fixed values (e.g. for the CHSH-ineuqality $|\tr(\rho\mathcal{B})|\leq2$), these bounds may change as time passes, because the probability for the particles to already have decayed increases.\\
In general, a Bell inequality is of the form
\beq \min_{\sigma \ \mathrm{is} \ \mathrm{local-realistic}} \tr(\sigma \mathcal{B}) \leq \tr(\rho \mathcal{B}) \leq \max_{\sigma \ \mathrm{is} \ \mathrm{local-realistic}} \tr(\sigma \mathcal{B}) \eeq
which only has to be modified such that the maximum has to be computed individually for each moment in time. By assuming that the set of all states which can be described by a local realistic theory is a convex one, the optimisation reduces to pure states only. By using that for pure states, non-local-realism and entanglement are equivalent (see e.g. \cite{gisin_bellent}), the optimisation can be performed over all separable pure states only, which is much more easily computable and well-defined:
\beq B_-= \min_{\sigma \in \mathcal{S}} \tr(\sigma\mathcal{B}) \leq \tr(\rho\mathcal{B}) \leq \max_{\sigma \in \mathcal{S}} \tr(\sigma\mathcal{B}) = B_+ \eeq
where the bound $B$ depends on the (possibly different) times of measurement contained in the measurement of $\mathcal{B}$. Bell inequalities of this kind are capable of detecting quantum nonlocality in systems, which cannot straightforwardly be accessed by standard quantum informational tools, such as the bipartite kaon system. In this case (since the kaon system is unstable), the Bell operator may be composed out of effective operators of the form (\ref{eq_app_effop}), such that e.g. a CHSH-type Bell operator assumes the form
\beq \nonumber \mathcal{B} = O^{eff}(\alpha_1^A,\phi_1^A,t_1^A)\otimes(O^{eff}(\alpha_1^B,\phi_1^B,t_1^B)+O^{eff}(\alpha_2^B,\phi_2^B,t_2^B)) \\
+O^{eff}(\alpha_2^A,\phi_2^A,t_2^A)\otimes(O^{eff}(\alpha_1^B,\phi_1^B,t_1^B)-O^{eff}(\alpha_2^B,\phi_2^B,t_2^B)) \eeq
Note that this sort of Bell inequality cannot be formulated in the Schrödinger picture, since measurements at different times cannot be contained in a single Schrödinger operator. This illustrates that the commonly used formalism of quantum information theory (based on the Schrödinger formalism) does not contain the full potential necessary for yielding a complete framework (and, consequently, a complete characterisation and understanding) of these phenomena.\\

\chapter{Mathematica Source Code}
Many of the results presented in this work were obtained by means of symbolic calculation and programming via the software Wolfram Mathematica. In order to facilitate reproduction of these results as well as any possible further study of the subject, the used Mathematica code shall be provided here (along with a brief documentation).\\
In the upcoming section, the plain source code will be presented (sorted alphabetically by name of the defined functions), such that in can be directly used for calculation. After that, in section \ref{sec_code_doc} the use of this code will be explained .\\

\section{Code}

\verb#Cgme[\[Psi]_List,d_Integer:2]:=Module[{n,\[Rho]red,j},#\\
\verb#  n=Log[d,Max[Dimensions[\[Psi]]]];#\\
\verb#  Min[Re[Table[\[Rho]red=PartialTrace[VecToDM[Normalize[\[Psi]]],#\\
\verb#    UniquePartitions[2,n][[j]][[1]],d];#\\
\verb#  Sqrt[2(1-Tr[\[Rho]red.\[Rho]red])],{j,1,Length[UniquePartitions[2,n]]}]]]];#\\
\\
\verb#DickeState[n_Integer:3,m_Integer:1,d_Integer:2]:=Module[{j},#\\
\verb#  Normalize[VecToDM[Sum[Sum[vd[Subsets[Range[n],{m}][[i]],n,d,j],#\\
\verb#    {i,1,Binomial[n,m]}],{j,0,d-2}]],Tr]];#\\
\\
\verb#DoubleClass[\[Sigma]_List]:=Module[{n,i,j,k},#\\
\verb#  n=Log[2,Dimensions[\[Sigma]][[1]]];#\\
\verb#  Sum[If[i!=j,#\\
\verb#    Re[MatrixElement[\[Sigma],UnitVector[n,i+1],#\\
\verb#      UnitVector[n,j+1]]+(-1)^(n+1) MatrixElement[\[Sigma],#\\
\verb#      1-UnitVector[n,i+1],1-UnitVector[n,j+1]]]#\\
\verb#  -(MatrixElement[\[Sigma],Table[If[MemberQ[{i,j},k],1,0],{k,0,n-1}]]#\\
\verb#    +MatrixElement[\[Sigma],Table[If[MemberQ[{i,j},k],0,1],{k,0,n-1}]]),#\\
\verb#  -(n-2)(MatrixElement[\[Sigma],UnitVector[n,i+1]]#\\
\verb#    +MatrixElement[\[Sigma],1-UnitVector[n,i+1]])#\\
\verb#  ],{i,0,n-1},{j,0,n-1}]#\\
\verb#  -n(n-1)/2(MatrixElement[\[Sigma],Table[0,{n}]]#\\
\verb#    +MatrixElement[\[Sigma],Table[1,{n}]])];#\\
\\
\verb#Flip[i_List,d_Integer:2]:=Module[{flipop,j},#\\
\verb#  flipop=KP[Table[Sum[{UnitVector[d,j]}\[ConjugateTranspose].#\\
\verb#    {UnitVector[d,d-j+1]},{j,1,d}],#\\
\verb#    {Log[d,If[Length[Dimensions[i]]==1,Dimensions[i][[1]],#\\
\verb#      Max[Dimensions[i][[1]],Dimensions[i][[2]]]]]}]];#\\
\verb#  If[Length[Dimensions[i]]==1,({i}.flipop)[[1]],#\\
\verb#    If[Dimensions[i][[1]]==1,i.flipop,#\\
\verb#      If[Dimensions[i][[2]]==1,flipop.i,#\\
\verb#        If[Dimensions[i][[1]]==Dimensions[i][[2]]==#\\
\verb#          Dimensions[flipop][[1]]==Dimensions[flipop][[2]],#\\
\verb#            flipop.i.flipop\[ConjugateTranspose],#\\
\verb#            Print["ERROR: Invalid input state."]]]]]];#\\
\\
\verb#GHZState[n_Integer:3,d_Integer:2]:=Module[{i},#\\
\verb#  Normalize[VecToDM[Sum[vd[{},n,d,i],{i,0,d-1}]],Tr]];#\\
\\
\verb#KP[i\_List]:=Module\[\{kprod,j\},#\\
\verb#  kprod={{1}};#\\
\verb#  Do[kprod=KroneckerProduct[kprod,i[[j]]],{j,1,Length[i]}];#\\
\verb#  kprod\];#\\
\\
\verb#MatrixElement[\[Sigma]_List,bra_List,d_Integer:2]:=#\\
\verb#  MatrixElement[\[Sigma],bra,bra,d];#\\
\verb#MatrixElement[\[Sigma]_List,bra_List,ket_List,d_Integer:2]:=#\\
\verb#  Module[{n,branum,ketnum},#\\
\verb#  n=Log[d,Dimensions[\[Sigma]][[1]]];#\\
\verb#  branum=Sum[bra[[n-i+1]] d^(i-1), {i,1,n}]+1;#\\
\verb#  ketnum=Sum[ket[[n-i+1]] d^(i-1), {i,1,n}]+1;#\\
\verb#  Return[\[Sigma][[branum]][[ketnum]]]];#\\
\\
\verb#nClass[\[Sigma]_List]:=Module[{n,\[Alpha]},#\\
\verb#  n=Log[2,Dimensions[\[Sigma]][[1]]];#\\
\verb#  \[Alpha]=If[n==3,3/2,If[n==4,1,If[n>4,1/2,#\\
\verb#    Print["nClass: Invalid particle number!"]]]];#\\
\verb#  Re[MatrixElement[\[Sigma],Table[0,{n}],Table[1,{n}]]]#\\
\verb#  -\[Alpha](1-MatrixElement[\[Sigma],Table[0,{n}]]#\\
\verb#  -MatrixElement[\[Sigma],Table[1,{n}]])];#\\
\\
\verb#nm1Class[\[Sigma]_List]:=Module[{n,i,j,k},#\\
\verb#n=Log[2,Dimensions[\[Sigma]][[1]]];#\\
\verb#Sum[If[i!=j,Re[MatrixElement[\[Sigma],UnitVector[n,i+1],UnitVector[n,j+1]]]#\\
\verb#-(n-2) MatrixElement[\[Sigma],Table[If[k==i || k==j,1,0],{k,0,n-1}]],#\\
\verb#-(n-2)MatrixElement[\[Sigma],UnitVector[n,i+1]]],{i,0,n-1},{j,0,n-1}]#\\
\verb#-n(n-1)/2 MatrixElement[\[Sigma],Table[0,{n}]]];#\\
\\
\verb#PartialTrace[\[Rho]_List,sys_Integer,d_Integer:2]:=#\\
\verb#  Module[{n,\[Sigma],k,DigitList,temp},#\\
\verb#  n=Log[d,Max[Dimensions[\[Rho]]]];#\\
\verb#  \[Sigma][i_Integer,j_Integer]:=Sum[#\\
\verb#    MatrixElement[\[Rho],Insert[IntegerDigits[i,d,n-1],k,sys],#\\
\verb#      Insert[IntegerDigits[j,d,n-1],k,sys],d],{k,0,d-1}];#\\
\verb#  Sum[{UnitVector[d^(n-1),i+1]}\[ConjugateTranspose].#\\
\verb#    {UnitVector[d^(n-1),j+1]}\[Sigma][i,j],#\\
\verb#    {i,0,d^(n-1)-1},{j,0,d^(n-1)-1}]];#\\
\verb#PartialTrace[\[Rho]_List,sys_List,d_Integer:2]:=Module[{matr,trsys},#\\
\verb#  matr=\[Rho];trsys=sys;#\\
\verb#  Do[matr=PartialTrace[matr,trsys[[1]]-i+1,d];#\\
\verb#  trsys=Drop[trsys,{1}];#\\
\verb#  ,{i,1,Length[sys]}];matr];#\\
\\
\verb#Partitions[2,st_List]:=Module[{i},#\\
\verb#  Table[{Subsets[st,{1,Length[st]-1}][[i]],#\\
\verb#  Complement[st,Subsets[st,{1,Length[st]-1}][[i]]]},#\\
\verb#  {i,1,Length[Subsets[st,{1,Length[st]-1}]]}]];#\\
\verb#Partitions[k_Integer,st_List]:=Module[{km1part,i,j},#\\
\verb#  km1part=Partitions[k-1,st];#\\ 
\verb#  Flatten[Table[{Join[Partitions[2,km1part[[i]][[1]]][[j]],#\\
\verb#    Table[km1part[[i]][[l]],{l,2,Length[km1part[[i]]]}]]},#\\
\verb#    {i,1,Length[km1part]},#\\
\verb#    {j,1,Length[Partitions[2,km1part[[i]][[1]]]]}],2]];#\\
\verb#Partitions[k_Integer,n_Integer]:=Partitions[k,Range[n]];#\\
\\
\verb#Q0[\[Sigma]_List,d_Integer:2,f_Integer:2]:=#\\
\verb#  Module[{n,Parts,k,l,i,j,s},#\\
\verb#  n=Log[d,Dimensions[\[Sigma]][[1]]];#\\
\verb#  Parts=UniquePartitions[2,n];#\\
\verb#  Sum[If[k==l,0,#\\
\verb#    (Abs[MatrixElement[\[Sigma],Table[l,{n}],Table[k,{n}],d]])-#\\
\verb#    (Sum[Sqrt[#\\
\verb#      MatrixElement[\[Sigma],Table[If[MemberQ[Parts[[i]][[1]],j],l,k]#\\
\verb#        ,{j,1,n}],d] MatrixElement[\[Sigma],#\\
\verb#        Table[If[MemberQ[Parts[[i]][[1]],j],k,l],{j,1,n}],d]]#\\
\verb#      ,{i,1,Length[Parts]}])] ,{k,0,f-1},{l,0,f-1}]];#\\
\\
\verb#Qk[\[Sigma]_List,k_Integer,d_Integer:2,f_Integer:2]:=#\\
\verb#  Module[{n,\[Gamma],i,kvec1,kvec2},#\\
\verb#  n=Log[d,Dimensions[\[Sigma]][[1]]];#\\
\verb#  \[Gamma]=UniquePartitions[k,n];#\\
\verb#  Sum[2 Abs[MatrixElement[\[Sigma],Table[i,{n}],#\\
\verb#    Table[i+1,{n}],d]]#\\
\verb#  -Sum[Product[(#\\
\verb#    MatrixElement[\[Sigma],Table[If[MemberQ[\[Gamma][[l]][[j]],#\\
\verb#      m],i,i+1],{m,1,n}],d] MatrixElement[\[Sigma],#\\
\verb#    Table[If[MemberQ[\[Gamma][[l]][[j]],m],i+1,i],{m,1,n}],d])#\\
\verb#  ,{j,1,k}]^(1/(2k)),{l,1,Length[\[Gamma]]}],{i,0,f-2}]];#\\
\\
\verb#Qm[\[Sigma]_List,m_Integer:1,d_Integer:2,f_Integer:2]:=#\\
\verb#  Module[{n,\[Omega],\[Mu],DickeSetsA,DickeSetsB,DickeSets,#\\
\verb#    dvec1,dvec2,\[Delta],i,j,k,l},#\\
\verb#  n=Log[d,Dimensions[\[Sigma]][[1]]];#\\
\verb#  If[m>n/2,\[Omega]=Flip[\[Sigma],d];\[Mu]=n-m,#\\
\verb#    \[Omega]=\[Sigma];\[Mu]=m];DickeSetsA=Subsets[Range[n],{\[Mu]}];#\\
\verb#  DickeSetsB[i_List]:=#\\
\verb#    Flatten[Table[Sort[Join[Delete[i,j],#\\
\verb#      Subsets[Complement[Range[n],i],{1}][[k]]]],#\\
\verb#      {j,1,\[Mu]},{k,1,n-\[Mu]}],1];#\\
\verb#  DickeSets=Flatten[Table[{DickeSetsA[[i]],#\\
\verb#    DickeSetsB[DickeSetsA[[i]]][[j]]},{i,1,Binomial[n,\[Mu]]},#\\
\verb#    {j,1,\[Mu] (n-\[Mu])}],1]; #\\
\verb#  (Sum[Sum[#\\
\verb#    Abs[MatrixElement[\[Omega],#\\
\verb#      Table[If[MemberQ[DickeSets[[i]][[1]],s],k+1,k],{s,1,n}],#\\
\verb#      Table[If[MemberQ[DickeSets[[i]][[2]],s],k+1,k],{s,1,n}],d]]#\\
\verb#    -Sqrt[#\\
\verb#      MatrixElement[\[Omega],Table[If[MemberQ[DickeSets[[i]][[1]],j],#\\
\verb#        If[MemberQ[DickeSets[[i]][[2]],j],k+1,k],k],{j,1,n}],d]#\\
\verb#      MatrixElement[\[Omega],Table[If[MemberQ[DickeSets[[i]][[1]],j],#\\
\verb#        k+1,If[MemberQ[DickeSets[[i]][[2]],j],k+1,k]],{j,1,n}],d]]#\\
\verb#    ,{k,0,f-2}]#\\
\verb#    +2Sum[\[Delta]=Subsets[Union[Complement[Range[n],#\\
\verb#      DickeSets[[i]][[2]]],DickeSets[[i]][[1]]],{1,n-2}];#\\
\verb#      Abs[MatrixElement[\[Omega],#\\
\verb#        Table[If[MemberQ[DickeSets[[i]][[1]],s],k+1,k],{s,1,n}],#\\
\verb#        Table[If[MemberQ[DickeSets[[i]][[2]],s],l+1,l],{s,1,n}],d]]#\\
\verb#    +Sum[-Sqrt[#\\
\verb#      MatrixElement[\[Omega],#\\
\verb#        Table[If[MemberQ[\[Delta][[h]],j],#\\
\verb#          If[MemberQ[DickeSets[[i]][[2]],j],l+1,l],#\\
\verb#          If[MemberQ[DickeSets[[i]][[1]],j],k+1,k]],{j,1,n}],d]#\\
\verb#      MatrixElement[\[Omega],Table[If[MemberQ[\[Delta][[h]],j],#\\
\verb#        If[MemberQ[DickeSets[[i]][[1]],j],k+1,k],#\\
\verb#        If[MemberQ[DickeSets[[i]][[2]],j],l+1,l]],{j,1,n}],d]]#\\
\verb#      ,{h,1,Length[\[Delta]]}]#\\
\verb#    ,{k,0,f-2},{l,0,k-1}],{i,1,Length[DickeSets]}]#\\
\verb#  -\[Mu] (n-\[Mu]-1)(f-1)Sum[MatrixElement[\[Omega],#\\
\verb#    Table[If[MemberQ[DickeSetsA[[i]],s],k+1,k],{s,1,n}],d],#\\
\verb#    {i,1,Length[DickeSetsA]},{k,0,f-2}])/\[Mu]];#\\
\\
\verb#UniquePartitions[2,st_Integer]:=UniquePartitions[2,Range[st]];#\\
\verb#UniquePartitions[2,st_List]:=Module[{start,out},#\\
\verb#  start=Subsets[st,{1,Length[st]-1}];#\\
\verb#  out=Table[If[MemberQ[start[[i]],st[[1]]],#\\
\verb#    {start[[i]],Complement[st,start[[i]]]}],{i,1,Length[start]}];#\\
\verb#  out=DeleteCases[out,Null];#\\
\verb#  out]#\\
\verb#UniquePartitions[k_,st_]:=Module[{km1part,temp,additems},#\\
\verb#  km1part=UniquePartitions[k-1,st];#\\
\verb#  out={};#\\
\verb#  Do[temp=UniquePartitions[2,km1part[[i]][[j]]];#\\
\verb#    If[temp!={}, additems=Table[Sort[Join[temp[[l]],#\\
\verb#      Complement[km1part[[i]],{km1part[[i]][[j]]}]]],{l,1,Length[temp]}];#\\
\verb#      out=Join[out,additems]]#\\
\verb#  ,{i,1,Length[km1part]},{j,1,k-1}];#\\
\verb#  out=Tally[out];#\\
\verb#  out=Table[out[[i]][[1]],{i,1,Length[out]}];#\\
\verb#  out];#\\
\\
\verb#vd[i_List,n_Integer,d_Integer:2,f_Integer:0]:=Module[{j},#\\
\verb#  KP[Table[If[MemberQ[i,j],{UnitVector[d,f+2]},#\\
\verb#    {UnitVector[d,f+1]}],{j,1,n}]]];#\\
\\
\verb#VecToDM[i_List]:=Module[{vectr},#\\
\verb#  If[Length[Dimensions[i]]==1,vectr={i},vectr=i];#\\
\verb#  If[Dimensions[vectr][[1]]==1,#\\
\verb#    vectr\[ConjugateTranspose].vectr,#\\
\verb#    vectr.vectr\[ConjugateTranspose]]];#\\

\section{Documentation\label{sec_code_doc}}
\texttt{Cgme[$\psi$,d]} computes the gme-concurrence (\ref{eq_cgme}) of a pure $n$ qudit state $\psi$. If no dimension $d$ is entered, $\psi$ is considered to be a multi-qubit-state.\\
\\
\texttt{DickeState[n,m,d]} generates a generalised $n$-qudit-$m$-Dicke state density matrix, as defined in \cite{spengler_gmd}. In particular, for $m=1$, this is the state (\ref{eq_quditw}). All three parameters are optional, the default values are $n=3$, $m=1$ and $d=2$.\\
\\
\texttt{DoubleClass[$\sigma$]} computes the value of the classification inequality for double states (\ref{eq_doubleineq}) for the multi-qubit density matrix $\sigma$.\\
\\
\texttt{Flip[i,d]} performs a generalised bit-flip on the multi-qudit state $i$, which can be either a state vector or a density matrix. If not entered, $d$ is set to 2.\\
\\
\texttt{GHZState[n,d]} generates an $n$-qudit GHZ-state density matrix (\ref{eq_quditghz}). By default, $n=3$ and $d=2$.\\
\\
\texttt{KP[i]} is a short notation for the subsequent multiple tensor product \linebreak[4] (\texttt{KroneckerProduct}) of all elements in the list $i$.\\
\\
\texttt{MatrixElement[$\sigma$,bra,ket,d]} picks the transition element $\bra{bra}\sigma\ket{ket}$ of the $n$-qudit matrix $\sigma$, where both $bra$ and $ket$ are lists of indices corresponding to numbers of unit vectors on the respective subsystems, following the notation $\ket{ijk} = \ket{i}\otimes\ket{j}\otimes\ket{k}$. E.g. the off-diagonal density matrix element $\bra{000}\rho_{GHZ}\ket{111}$ of a three-qubit GHZ-state $\rho_{GHZ}$ is given by \linebreak[4] \texttt{MatrixElement[$\rho_{GHZ}$,\{0,0,0\},\{1,1,1\},2]}.\\
The parameters $ket$ and $d$ are optional - if not entered, $d$ is set to 2 and $ket$ is set equal to $bra$, such that the function returns the corresponding diagonal matrix element.\\
\\
\texttt{nClass[$\sigma$]} yields the value of the $n$-tuple state classification inequality (\ref{eq_ntupleineq}) of the multi-qubit density matrix $\sigma$.\\
\\
\texttt{nm1Class[$\sigma$]} yields the value of the $(n-1)$-tuple state classification inequality, as defined in \cite{huberbruss}, of the multi-qubit density matrix $\sigma$.\\
\\
\texttt{PartialTrace[$\rho$,sys,d]} performs the partial trace over the multi-qudit matrix $\rho$ with respect to the system $sys$ (if $sys$ is an integer) or all systems whose number is contained in $sys$ (if $sys$ is a list of integers). By default, $d=2$.\\
\\
\texttt{Partitions[k,st]} gives a list of all $k$-partitions (including all permutations) of the set $st$ (if $st$ is a list), or a list of all $k$-partitions of the set $\{1,2,3,...,st\}$ (if $st$ is an integer). Unlike the function \texttt{UniquePartition}, this function treats different permutations of the same partition as different, i.e. returns all possible permutations of all partitions, instead of just one permutation per partition. Consequently, the returned set of partitions is $k!$-fold degenerate.\\
\\
\texttt{Q0[$\sigma$,d,f]} evaluates the genuine multipartite entanglement detection criterion (\ref{eq_gme_ghz}) (if $f=2$) with $\ket{\Phi} = \ket{0}^{\otimes n}\otimes\ket{1}^{\otimes n}$, or the criterion for genuinely $f$-dimensional genuine multipartite entanglement (\ref{eq_q0}) (if $f>2$) with $\ket{\Phi} = \ket{0}^{\otimes n}\otimes\ket{1}^{\otimes n}$ of the multi-qudit density matrix $\sigma$.\\
The parameters $d$ and $f$ are optional, both default values are 2.\\
\\
\texttt{Qk[$\sigma$,k,d,f]} computes the value of the criterion for $k$-separability (\ref{eq_ksep}) of the multi-qudit density matrix $\sigma$ with $\ket{\Phi} = \ket{0}^{\otimes n}\otimes\ket{1}^{\otimes n}$ (if $f=2$), or a multidimensional generalisation thereof for $f>2$. The default values for the optional parameters are $d=2$ and $f=2$.\\
\\
\texttt{Qm[$\sigma$,m,d,f]} yields the value of the genuine multipartite entanglement detection inequality for $m$-Dicke-like states (\ref{eq_gme_dicke}) for the multi-qudit density matrix $\sigma$ (if $f=2$) or the value of the criterion for multidimensional genuine mutlipartite entanglement for generalised qudit $m$-Dicke states (\ref{eq_qm}) (if $f>2$).\\
Default values are $d=2$ and $f=2$.\\
\\
\texttt{UniquePartitions[k,st]} gives a list of all $k$-partitions of the set $st$ (if $st$ is a list), or a list of all $k$-partitions of the set $\{1,2,3,...,st\}$ (if $st$ is an integer). Unlike the function \texttt{Partition}, this function treats different permutations of the same partition as equal, i.e. only returns one representative partition instead of all possible permutations.\\
\\
\texttt{vd[i,n,d,f]} returns the $n$-qudit product state vector of $\ket{f+1}$ in all subsystems whose labels are contained in the set $i$ and $\ket{f}$ else. The optional parameters are $d$ (default value 2) and $f$ (default value 0).\\
\\
\texttt{VecToDM[i]} yields the density matrix corresponding to the state vector $i$.

\end{appendix}

\end{document}